\documentclass[useAMS,usenatbib]{mn2e}
\bibpunct{(}{)}{;}{a}{}{,}
\usepackage{graphicx, longtable, lscape}
\usepackage{float}
\usepackage[figuresright]{rotating}
\usepackage{color}

\newcommand{\lya}{Ly$\alpha$~}

\newcommand{\be}{\begin{equation}}
\newcommand{\ee}{\end{equation}}
\newcommand{\ba}{\begin{eqnarray}}
\newcommand{\ea}{\end{eqnarray}}
\newcommand{\brr}{\begin{array}}

\newcommand{\err}{\end{array}}
\newcommand{\bc}{\begin{center}}
\newcommand{\ec}{\end{center}}

\newcommand{\kms}{km s$^{-1}$}

\newcommand{\spose}[1]{\hbox to 0pt{#1\hss}}
\newcommand{\simlt}{\mathrel{\spose{\lower 3pt\hbox{$\mathchar"218$}}
     \raise 2.0pt\hbox{$\mathchar"13C$}}}
\newcommand{\simgt}{\mathrel{\spose{\lower 3pt\hbox{$\mathchar"218$}}
     \raise 2.0pt\hbox{$\mathchar"13E$}}}

\definecolor{purple}{rgb}{0.63, 0.36, 0.94}

\DeclareMathAlphabet{\mathsc}{OT1}{cmr}{m}{sc}
\def\testbx{bx}%
\DeclareRobustCommand{\ion}[2]{%
\relax\ifmmode
\ifx\testbx\f@series
{\mathbf{#1\,\mathsc{#2}}}\else
{\mathrm{#1\,\mathsc{#2}}}\fi
\else\textup{#1\,{\mdseries\textsc{#2}}}%
\fi}

\title[The bimodality of \ion{C}{iii} absorbers at $z \sim 2.5$]
{Evidence of bimodal physical properties of intervening, optically-thin \ion{C}{iii} 
absorbers at $z \sim 2.5$}
\author[Kim, Carswell \& Ranquist]{
T.-S. Kim$^{1, 2}$\thanks{E-mail: kim@oats.inaf.it}, R. F. Carswell$^{3}$ \&
D. Ranquist$^{2,4,5}$ \\
$^{1}$Osservatorio Astronomico di Trieste, Via G. B. Tiepolo, 11, 34143, Trieste, Italy\\
$^{2}$Department of Astronomy, University of Madison-Wisconsin, WI 53706, USA\\
$^{3}$Institute of Astronomy, University of Cambridge, Madingley Road, Cambridge CB3 0HA\\
$^{4}$Department of Astrophysics \& Planetary Sciences, University of Colorado, Boulder,
CO 80309, USA\\
$^{5}$Department of Physics and Astronomy, Brigham Young University, Provo, UT 84602, USA}

\date{Accepted; Received}

\begin{document}   
\maketitle
\begin{abstract}
We present the Voigt profile analysis of 132
intervening \ion{C}{iv}+\ion{C}{iii} components 
associated with optically-thin
\ion{H}{i} absorbers at $2.1 < z < 3.4$ in the 19 high-quality 
UVES/VLT and HIRES/Keck QSO spectra.
For $\log N_{\mathrm{\ion{C}{iv}}} \in [11.7, 14.1]$, 
$N_{\mathrm{\ion{C}{iii}}} \propto N_{\mathrm{\ion{C}{iv}}}^{1.42\pm0.11}$ 
and
$<\!N_{\mathrm{\ion{C}{iii}}}/N_{\mathrm{\ion{C}{iv}}}\!> \, = 1.0\pm0.3$ with
a negligible redshift evolution.
For 54 \ion{C}{iv} components tied (aligned) with \ion{H}{i} 
at $\log N_{\mathrm{\ion{H}{i}}} \in [12.2, 16.0]$ and $\log N_{\mathrm{\ion{C}{iv}}} 
\in [11.8, 13.8]$, 
the gas temperature $T_{b}$ estimated from absorption line widths
is well-approximated to a Gaussian peaking at
$\log T_{b} \sim 4.4\pm 0.3$ for $\log T_{b} \in [3.5, 5.5]$, 
with a negligible non-thermal contribution.
For 32 of 54 tied \ion{H}{i}+\ion{C}{iv} pairs,
also tied with \ion{C}{iii}
at $\log N_{\mathrm{\ion{C}{iii}}} \in [11.7, 13.8]$,
we ran both photoionisation equilibrium (PIE) and non-PIE (using a fixed
temperature $T_{b}$) Cloudy models for the Haardt-Madau 
QSO+galaxy 2012 UV background. 
We find evidence of bimodality in observed and derived physical
properties.
High-metallicity branch absorbers have a carbon abundance 
[C/H]$_{\mathrm{temp}} \ge -1.0$, a line-of-sight length $L_{\mathrm{temp}} \le 20$\,kpc
and a total (neutral and ionised) hydrogen volume
density $\log n_{\mathrm{H, \, temp}} \in [-4.5, -3.3]$
and $\log T_{b} \in [3.9, 4.5]$. Low-metallicity branch absorbers have
[C/H]$_{\mathrm{temp}} \le -1.0$, $L_{\mathrm{temp}} \in [20, 480]$\,kpc
and $\log n_{\mathrm{H, \, temp}} \in [-5.2, -4.3]$
and $\log T_{b} \sim 4.5$. 
High-metallicity branch absorbers seem to be originated from extended disks,
inner halos or outflowing gas of intervening galaxies,
while low-metallicity absorbers are produced by
galactic halos or the surrounding IGM filament.
\end{abstract}


\section{Introduction}
\label{sec1}

Observational constraints on the interaction between galaxies
and their surrounding medium are important for improving our understanding of
how galaxies obtain their gas supply, grow over cosmic times and enrich their surroundings. 
In the current standard, simplified
$\Lambda$-CDM (cold dark matter) picture, the distribution of CDM grows
hierarchically out of the initial density fluctuation by gravity, forming
intersecting filamentary and sheet-like structures \citep{navarro96, bullock01}.
The intergalactic medium (IGM), mostly consisting of hydrogen
and helium, follows the underlying dark matter distribution.
Eventually,
the IGM cools down to form galaxies at the center of CDM halos located
at the density peaks in the filamentary structure. In return,
galaxies and quasars photoionise the IGM 
and eject processed materials into the halo and the IGM through galactic winds.
Galactic winds enrich the IGM, suppress star formation in galaxies and heat the gas
around them, while the intergalactic gas continuously falls into the CDM halos
and the large-scale gaseous filaments. This ongoing process results in
a complicated cosmic web of gas, galaxies and underlying CDM
\citep{cen94, dave99, keres05, oppenheimer09, steidel10, 
wiersma10, cen11, voort11, shen13}.

Observational constraints on the relationship between
the properties of surrounding gas and galaxies thus provide important tests for models  
of structure formation, galaxy formation/evolution, and past and ongoing
star formation.
The IGM is a diffuse, warm photoionised plasma, and the properties we may infer are
the physical gas density, temperature, metallicity and line-of-sight length. These properties
as a function of impact parameters, the projected distance from a galaxy to
the absorbing gas, are one of the observational goals in the IGM study. Nonetheless, 
even without time-intensive deep galaxy surveys around QSO sightlines to estimate
impact parameters, obtaining the IGM gas properties 
is still in its own right well worthwhile to constrain
theoretical predictions.

The IGM is  
well observed through rich absorption lines blueward the Ly$\alpha$ emission line 
in the spectra of background QSOs, therefore often referred as ``absorbers" or
``absorption lines". The majority of these lines are due to the resonance line transition 
by diffuse intervening neutral hydrogen \ion{H}{i} 
with the \ion{H}{i} column density at $\log N_{\mathrm{\ion{H}{i}}} 
\in [12, 17]${\footnote{The logarithmic
value of a quantity Q is expressed unitless as being divided by its unit before
converted into the logarithmic value, such as
$\log N_{\mathrm{\ion{H}{i}}} = \log (N_{\mathrm{\ion{H}{i}}} / {\mathrm{cm}^{-2}})$.
When the unit is noted for clarity, it is listed inside the bracket.}},
often referred as 
the Ly$\alpha$ forest regardless of its metal
association. There is also a small 
contribution from metal transitions, 
which is mostly associated with \ion{H}{i} absorbers at 
$\log N_{\mathrm{\ion{H}{i}}} \ge 14$ \citep{cowie95, songaila98, ellison00}. 
Due to its high temperature and a low density, the Ly$\alpha$ forest does not have 
in-situ star formation. Therefore, metals associated with the low-density
forest should have been transferred from somewhere. The most promising
candidate with the most abundant observational support is 
galactic-scale outflows driven by supernovae.

Bipolar galactic outflows are ubiquitous both at high and low redshifts \citep{strickland98,
martin05, shapley06, strickland09, rupke13}. High-$N_{\mathrm{\ion{H}{i}}}$ systems with 
$\log N_{\mathrm{\ion{H}{i}}} \ge 17$ are directly associated
with intervening galaxies at both high and low redshifts
\citep{bergeron86a, steidel92, lebrun97, fynbo10, werk13}, while the association 
with galaxies of the Ly$\alpha$ forest is less clear \citep{wakker09, rudie13, liang14}. 
Despite the large scatter, absorption strengths of both \ion{H}{i} and metals are 
anti-correlated with impact parameters \citep{lanzetta90, lanzetta95,
chen01, wakker09, steidel10, tumlinson11, bordoloi14, liang14}. The \ion{H}{i} gaseous
halo is larger than metal-enriched halo, with the latter being about $\sim 200$\,kpc
\citep{brooks11, cen11, shen13}.
In addition, \citet{bouche12} found
that their 11 \ion{Mg}{ii} absorber-galaxy pairs at $z \sim 0.1$ follow a different
absorption strength--impact parameter relation between the absorbers closer to
the minor axis and the major axis, evidence of an outflow origin of \ion{Mg}{ii} absorbers.
Similarly, \citet{lehner13}
found a bimodal metallicity distribution in the 28 absorbers with $\log N_{\mathrm{\ion{H}{i}}}
\in [16.2, 18.5]$ at $z < 1$ 
and interpreted the bimodality due to galactic winds and inflowing gas, respectively.

In general, cosmological numerical
simulations have successfully reproduced the observations. 
They predict that bipolar outflows originated in the disk disperse a
metal-enriched gas into the halo 
along the minor axis of star-forming galaxies at larger distance
than the major axis, since the interstellar medium (ISM) density in the disk is
lower along the minor axis. When the line of sight passes through closer to galaxies, 
i.e. at small
impact parameters, the absorption system has a higher metallicity, with more 
metal species at a wider range of ionisation states and greater absorption strengths
including \ion{H}{i}
\citep{aguirre01, cen11, voort11, oppenheimer12, shen13, barai13}.
These simulations also suggest that
some fraction of metals escapes from halos of parents galaxies into the intergalactic
space. The escaped metals become associated with 
low-$N_{\mathrm{\ion{H}{i}}}$ absorbers 
often at $\ge 2$ virial radii, and even in
the IGM filaments without any galaxies within 0.5--1\,Mpc. The details
of this process depend on when and where galactic outflows occur 
and how galactic outflows and infalling gas interact with each other.

For an absorbing gas produced by a single element,
i.e. the majority of the IGM \ion{H}{i} gas, there exists no robust 
observational constraint
on the physical properties, except for an upper limit on the gas temperature from the
absorption line width. When more than two elements 
having a very different mass display a similar profile with each other,
implying they are produced by the same gas,
the temperature and non-thermal motion of the gas can be directly estimated
from comparing their absorption line widths. When there exist more than 2 ionic transitions from 
the same metal element, such as \ion{C}{iv}, \ion{C}{iii} and \ion{C}{ii}, 
a simple assumption on the ionisation mechanism such as photoionisation
and collisional ionisation can be applied to derive other gas properties, 
if a radiation field to which absorbers are exposed can be provided.
Note that only a range of the carbon abundance can
be estimated from the photoionisation assumption if only \ion{H}{i} and \ion{C}{iv} 
are available.

The most commonly used tracer of metals associated with the Ly$\alpha$
forest is  \ion{C}{iv} doublet
$\lambda\lambda$1548.204, 1550.778
\citep{cowie95, schaye03}{\footnote{While the most
abundant cosmic metal element is oxygen, the \ion{O}{vi} doublet 
$\lambda\lambda$1031.926, 1037.616 occurs inside the dense forest region, 
therefore it is difficult to detect a clean \ion{O}{vi} associated with 
low-$N_{\mathrm{\ion{H}{i}}}$ absorbers at $z \sim 3$.
In addition, no other transitions of oxygen such as \ion{O}{iv} 
$\lambda$787.711 are covered at $z \sim 3$
in the ground-based optical spectra.}}, since its rest-frame wavelength longer
than \ion{H}{i} Ly$\alpha$ $\lambda$1215.670 places \ion{C}{iv} 
redward the Ly$\alpha$ emission line of the QSO, free of the forest \ion{H}{i} blends.
Although \ion{C}{ii} $\lambda$1334.532 can be outside the Ly$\alpha$
forest, it is usually detected with high-$N_{\mathrm{\ion{H}{i}}}$ absorbers 
\citep{cowie95, boksenberg15}.
While \ion{C}{iii} $\lambda$977.020  is commonly found for 
low-$N_{\mathrm{\ion{H}{i}}}$ 
absorbers, its much shorter
rest-frame wavelength often locates it in the wavelength regions severely
contaminated by high-order \ion{H}{i} Lyman lines and of less reliable 
continuum placement.  

The observational difficulty to detect \ion{C}{iii} and a lack of \ion{C}{ii}
of the forest \ion{H}{i} often leave
\ion{C}{iv} the sole metal ion used to infer the physical
structure of the low-$N_{\mathrm{\ion{H}{i}}}$ forest. 
The carbon abundance based on
the \ion{H}{i} and \ion{C}{iv} profile fitting at
$\log N_{\mathrm{\ion{H}{i}}} \ge 14.5$ has been estimated to be $\sim 10^{-2.5}$ 
solar with a large scatter in photoionisation equilibrium (PIE) \citep{cowie95, hellsten97, 
rauch97b, simcoe04}. 
On the other hand, the statistical pixel optical depth analysis of \ion{H}{i} and \ion{C}{iv} 
probing lower-$N_{\mathrm{\ion{H}{i}}}$ ends than the profile fitting 
analysis estimates that the median forest carbon abundance
is $\sim 10^{-3.5}$ solar with a over-density and $z$ dependence
\citep{schaye03}.

Theories indicate that 
\ion{C}{iv} associated with the forest 
is produced by a radiatively cooling non-equilibrium gas once
shock-heated by supernovae-driven galactic
outflows and/or hot stars, and that it is now exposed
to the external UV background radiation produced by QSOs and galaxies
\citep{wiersma09, cen11, voort11, oppenheimer13}. 
When an initially hot gas cools radiatively
below $\sim 10^{6.7}$\,K, the gas is no longer in 
collisional ionisation equilibrium (CIE) and it remains over-ionised
due to cooling occurring
faster than recombination. Presence of radiation to a radiatively
cooling non-CIE gas reduces a gas cooling rate at the gas temperature
$T \sim 10^{4}-10^{5}$\,K by a large
factor, which changes the thermal
and ionisation states of the gas
\citep{wiersma09, voort11, oppenheimer13}. 
By comparison, for a low-density gas in CIE 
without any external
radiation, \ion{C}{iv} peaks at 
$T \sim 10^{5}$\,K. The ionisation fraction of a given metal element is only a function 
of the gas temperature and 
the ratio of \ion{C}{iii} and \ion{C}{iv} is about 1 at
$T \sim 10^{5}$\,K \citep{gnat07}.
For a low-density PIE gas,
the \ion{C}{iv} fraction rapidly increases
from $T \sim 10^{5.4}$\,K to $T \sim 10^{4.9}$\,K, then becomes rather
independent of $T$ at $T \le 10^{4.9}$\,K. The ratio of \ion{C}{iii} and
\ion{C}{iv} is $\sim 1$ at $T \sim 10^{4.2}-10^{5.0}$\,K 
\citep{oppenheimer13}. 

Here we present the results from a detailed Voigt profile analysis and photoionisation
modelling of optically-thin \ion{H}{i} absorbers associated with \ion{C}{iv} and \ion{C}{iii} 
at $2.1 < z < 3.4$. 
The selected \ion{C}{iv} components have $\log N_{\mathrm{\ion{C}{iv}}} \in [11.7, 14.0]$,
which is a much lower $N_{\mathrm{\ion{C}{iv}}}$ range than the galaxy-galaxy pairs study 
by \citet{steidel10} at $z \sim 2.3$, 
$\log N_{\mathrm{\ion{C}{iv}}} \ge 13.5$ and most studies based on the
absorption strength--impact parameters.
Being optically thin, i.e. $\log N_{\mathrm{\ion{H}{i}}} \le 17.2$, 
these \ion{C}{iii}-selected absorbers do not require any complicated radiative 
transfer effect as is the case for optically thick Lyman limit absorbers.
Comparing the predicted column densities based on photoionisation modelling 
with the observed ones enables us to 
test the basic assumption on the IGM physics such as photoionisation and
hydrostatic equilibrium. It also provides
a more robust estimate of the predicted physical 
properties of the \ion{C}{iv}-enriched low-$N_{\mathrm{\ion{H}{i}}}$ gas, which
is less well constrained observationally and can 
shed a light on the enrichment mechanism in the low-density universe. 
We have found evidence that optically-thin \ion{H}{i} absorbers 
aligned (or tied in the profile fitting analysis to be specific)
both with \ion{C}{iv} and \ion{C}{iii} can be
classified into two populations as high- and low-metallicity branch absorbers
at the boundary of carbon abundance of one tenth of solar.
Each branch follows its own different scaling relation between various observed and
derived physical parameters.

This paper is organised as follows. Section~\ref{sec2} describes the 
sample, and gives details of the
 Voigt profile fitting analysis. The direct observables of the
\ion{C}{iii} absorbers including the gas temperature are presented in Section~\ref{sec3}. 
The extensive photoionisation modellings are summarised in Section~\ref{sec4},
using the photoionisation code Cloudy version c13.03  \citep{ferland13}.
The main findings on bimodality in observed/derived
physical parameters of optically-thin \ion{H}{i}+\ion{C}{iv}+\ion{C}{iii} absorbers
are presented and discussed in Sections~\ref{sec5} and \ref{sec6}, and 
summarised in Section~\ref{sec7}.

\begin{table}
\caption{Analysed QSOs}
\label{tab1}
\begin{tabular}{llccc}
\hline
\noalign{\smallskip}
QSO & $z_{\mathrm{em}}^{\mathrm{a}}$ & $z_{\mathrm{\ion{C}{iii}}}^{\mathrm{b}}$ 
   & S/N$^{\mathrm{c}}$ & Inst. \\
\noalign{\smallskip}
\hline
\noalign{\smallskip}

Q0055--269       & 3.656 & 2.663--3.390 &  32, 53, 20 & UVES\\
PKS2126--158   & 3.280 & 2.669--3.208  & 90, 168, 61 & UVES \\
HS1425+6039    & 3.180 & 2.865--3.110  & 85,  68, 15 & HIRES \\
Q0636+6801     & 3.175 & 3.000--3.105 & 70, 70, 7 & HIRES \\
Q0420--388      & 3.115 & 2.850--3.045 &  120, 120, 52 & UVES \\
HE0940--1050  & 3.082 & 2.776--3.014 & 80, 110, 52 & UVES \\
HE2347--4342  & 2.873 & 2.257--2.710 & 100, 96, 50 & UVES \\
                          &          & 2.770--2.809  &  &  \\
HE0151--4326  & 2.781$^{\mathrm{d}}$& 2.180--2.710 &  81, 150, 45 & UVES \\
Q0002--422    & 2.768 & 2.159--2.705 & 87, 137, 38 & UVES \\
PKS0329--255  & 2.704 & 2.248--2.643  &  39, 70, 20  & UVES \\
Q0453--423    & 2.657 & 2.136--2.595 &  65, 115, 20 & UVES \\
HE1347--2457  & 2.612$^{\mathrm{d}}$ & 2.135--2.552 & 69, 83, 30 & UVES \\
Q0329--385      & 2.435 &  2.151--2.378 & 45, 75, 13 & UVES \\
HE2217--2818  & 2.413 & 2.128--2.355 & 86, 120, 11 & UVES \\
Q0109--3518    & 2.405 & 2.149--2.348 & 68, 139, 22 & UVES \\
HE1122--1648  & 2.404 & 2.162--2.346  & 118, 227, 15 & UVES \\
HE0001--2340  & 2.264 & 2.144--2.211 &  96, 74, 11 & UVES \\               
PKS0237--23    & 2.222 & 2.122--2.167  &  138, 169, 13 & UVES \\
PKS1448--232  & 2.219  & 2.122--2.168 &  77, 122, 7  & UVES \\

\noalign{\smallskip}
\hline
\end{tabular}
\begin{list}{}{}
\item[$^{\mathrm{a}}$]
The redshift is measured from the observed \lya emission line of the QSOs.
\item[$^{\mathrm{b}}$]
Only the region with $S/N \ge 5$ per pixel is analysed.
\item[$^{\mathrm{c}}$]
The first, second and third number is the S/N per pixel in the 
central parts of the analysed Ly$\alpha$ forest, \ion{C}{iv}
and \ion{C}{iii} regions, respectively.
\end{list}
\end{table}

\section{QSO sample}
\label{sec2}

\subsection{Data}

A total of 19 QSO spectra were used to search for \ion{C}{iii},
with the 17 spectra taken with the UVES on the VLT and the remaining
2 spectra with the HIRES on Keck. 
These spectra were chosen from the larger sample analysed in \citet{kim15},
to include only the spectra covering a longer wavelength region 
useful for the \ion{C}{iii} search. 
The \citet{kim15} sample was chosen to study the low-density
IGM, i.e. containing no strong damped Ly$\alpha$ systems
and not many Lyman limit systems in one sightline, from the UVES and
HIRES archives and includes only the QSO spectra with the high 
signal-to-noise ratio (S/N)
and the long, continuous wavelength coverage to cover high-order \ion{H}{i}
Lyman series.
The details on the data treatment can be found in \citet{kim15}. 

The resolution 
is $\sim 6.7 $~\kms\/ (or $R \sim 45,000$) and the wavelength 
is in the heliocentric velocity frame.
To avoid the proximity effect, 
wavelength regions within 5000\,\kms\/ of
the QSO Ly$\alpha$ emission line 
were excluded in each case. No further exclusion was done,
e.g. in the regions near a sub-damped Ly$\alpha$ absorber.
As the \ion{C}{iii} rest-frame wavelength 977.020\,\AA\/ is around Ly$\gamma$ 
$\lambda$972.536, it is located in the shorter wavelength region usually
having a lower S/N.
Therefore, only the wavelength region with $S/N \ge 5$ per pixel was included in the
\ion{C}{iii} search. 
Table~\ref{tab1} gives the basic properties of the 
spectra and the redshift range searched for \ion{C}{iii}. For the rest of this study,
the listed S/N is per pixel. All 
the atomic wavelengths are in the rest frame, unless stated otherwise.

\subsection{Voigt profile fitting}

\subsubsection{Basic assumptions}

The most physically useful approach where high resolution, high S/N spectra of QSOs are available
is to decompose absorption profiles into several discrete components,
assuming Voigt profiles for each component.
The profile fitting analysis provides redshifts, column 
densities    in cm$^{-2}$
and line widths in \kms\/ (as the Doppler parameter or the $b$ parameter). For
a Maxwellian velocity distribution, the $b$ parameter ($=\sqrt{2}\sigma$, 
where $\sigma$ is the standard deviation)
is related to the temperature and the non-thermal motion of the gas:

\begin{equation}
b^{2} = \frac{2 k T}{m_{\mathrm{M}}} + b_{\mathrm{nt}}^{2},
\label{eq1}
\end{equation} 

\noindent where $k$ is the Boltzmann constant, $T$ is the thermal gas temperature
in K, $m_{\mathrm{M}}$ is the atomic mass of the element
M and $b_{\mathrm{nt}}$ is the Gaussian non-thermal
line width in km s$^{-1}$. The thermal line with is 
$b_{\mathrm{th}}^{2} = 2 k T/m_{\mathrm{M}}$.
When two different elements give rise to lines which have a similar profile shape and 
so the absorbing material in which they arise is probably co-spatial, 
Eq.~\ref{eq1} can be used to estimate a thermal and 
non-thermal contribution in the gas motion \citep{carswell12, dutta14}.

\subsubsection{The search and the profile fitting of \ion{C}{iii} $\lambda$\,977}
\label{sec2.2.2}

To search for \ion{C}{iii} $\lambda$\,977 and to measure the line parameters,
we used the same line lists of 19 QSOs analysed by \citet{kim15}.
These lists are complete in the redshift range for which high-order Lyman lines
of higher-$z$ \ion{H}{i} absorbers affect the \ion{C}{iii} region of interest. 
In order for further physical analysis to be meaningful, such as photoionisation
modelling, \ion{C}{iii} and \ion{C}{iv}
have to be produced in the same gas cloud. We assumed
that the ion components having the same velocity structure are co-spatial.

For every \ion{C}{iv} component associated with
optically-thin \ion{H}{i}, we searched for narrow lines at
the expected \ion{C}{iii} $\lambda$\,977 positions based on the velocity structure of  
\ion{C}{iv}.
If the expected \ion{C}{iii} candidates occur inside saturated lines or are 
severely blended 
with other lines, these \ion{C}{iii} candidates were discarded.
Only when \ion{C}{iii} candidates are clean and/or blends around the candidates
can be securely estimated from other identified and fitted lines, these
were flagged as potential \ion{C}{iii} components.
This means that all the \ion{C}{iii} components corresponding
to a multi-component \ion{C}{iv} absorber are not necessarily clean. 
Even when part of multi-component \ion{C}{iii} profiles
is clean, the line parameters of clean components could be still obtained and
useful. 

The \ion{C}{iii} candidates were then re-fitted with VPFIT version 
10.2 \citep{carswell14},
using its default
rest-frame wavelengths and the oscillator strengths. 
To satisfy a condition of being co-spatial,
when the \ion{C}{iii} and \ion{C}{iv} components have similar redshifts
(within 1.5\,\kms\/ of each other), their 
redshifts and Doppler parameters were constrained to be the same. 
For velocity differences larger than that value,
it was not generally possible to get good fits for sharp lines
if the redshifts were constrained in that way.

The redshifts and Doppler parameters of any \ion{H}{i} components
clearly corresponding to \ion{C}{iii} and \ion{C}{iv}
within 15\,\kms\/ were also 
tied to their values, but 
for \ion{H}{i} the Doppler parameter was constrained so that the temperature 
$T\ge 0$ and  $b_{\mathrm{nt}}\ge 0$ (see Eq.~\ref{eq1}). 
Note that tying one \ion{H}{i} component with the corresponding
\ion{C}{iv} sometimes significantly changes the line parameter of
other nearby, tied/untied \ion{H}{i} components in multi-component \ion{H}{i} absorbers,
i.e. a consequence of the Voigt profile fitting results not being unique, nor objective
\citep{kirkman97, kim13}.
In addition, not all \ion{H}{i} components within 15\,\kms\/ from \ion{C}{iv} can be
tied successfully.
Where an appropriate \ion{H}{i} component is measured, 
the best-fit values of $T$ and  $b_{\mathrm{nt}}$ were determined by VPFIT.
The value 15\,\kms\/ was adopted for similar
reasons to the \ion{C}{iii}/\ion{C}{iv} difference limit. It is larger here because
the \ion{H}{i} lines are almost invariably broader than both the \ion{C}{iv}
and \ion{C}{iii} lines, and generally have a larger redshift error (the median
is $\sim$\,8\,\kms\/).
If a single \ion{H}{i} component
is aligned with a \ion{C}{iv} component within 15\,\kms\/, 
but there are another nearby, strong \ion{C}{iv} components within the associated 
\ion{H}{i} line width,
then it is not physically meaningful to include more \ion{H}{i} components to be
associated to all the \ion{C}{iv} components. 
Therefore, we did not tie \ion{H}{i} and \ion{C}{iv} in this case.

During the fitting process the continuum level was adjusted as necessary to 
obtain a good fit to the data.
The non-zero flux level of saturated lines often shown in the UVES
spectra was also taken into account.  
This line-fitting-and-continuum-adjustment iteration was repeated 
until a satisfactory fit was achieved with the normalised $\chi^{2}\simlt1.3$.

\begin{figure*}
\hspace{-0.3cm}
\includegraphics[width=45mm]{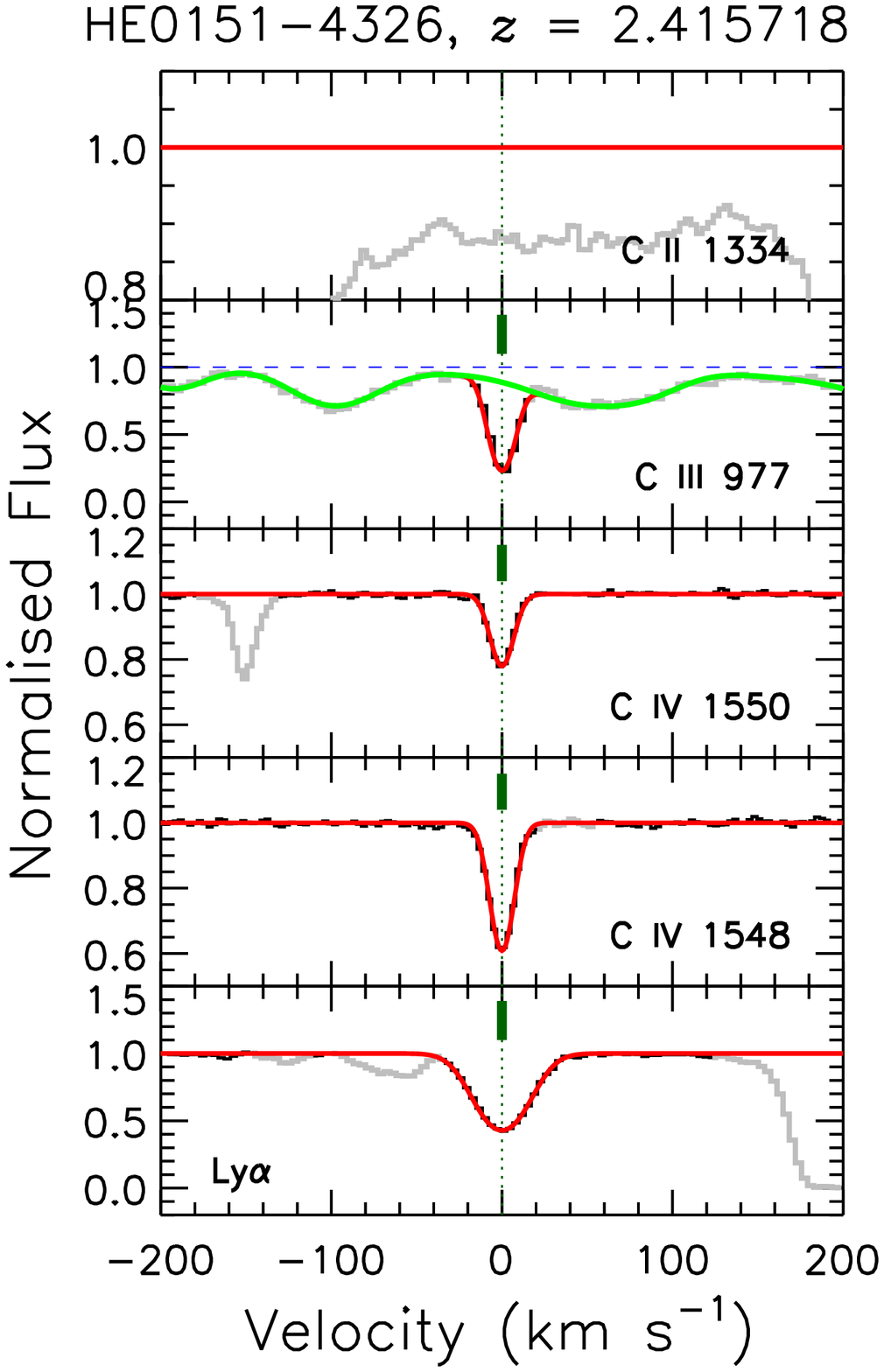}
\hspace{-0.3cm}
\includegraphics[width=45mm]{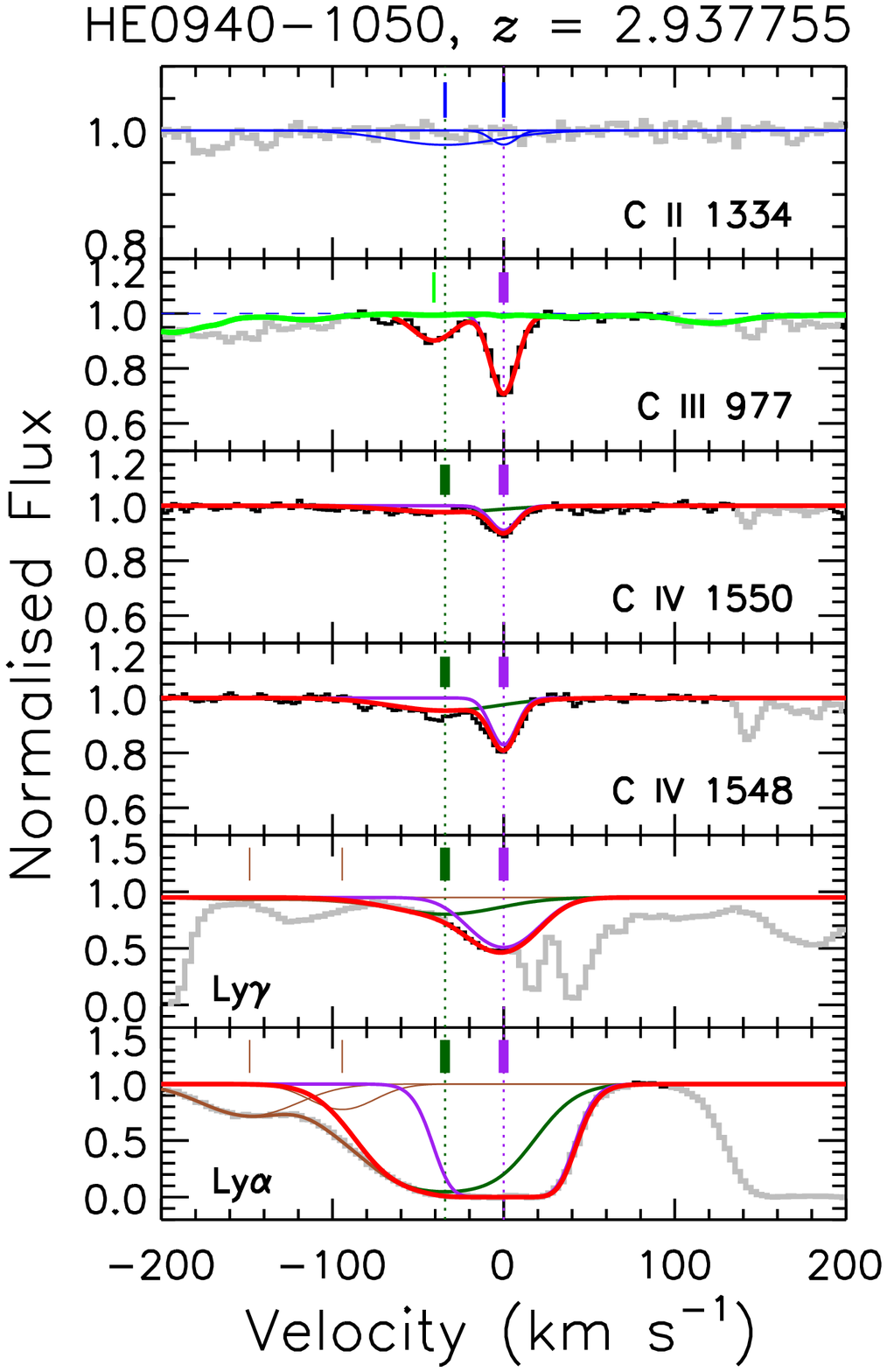}
\hspace{-0.3cm}
\includegraphics[width=45mm]{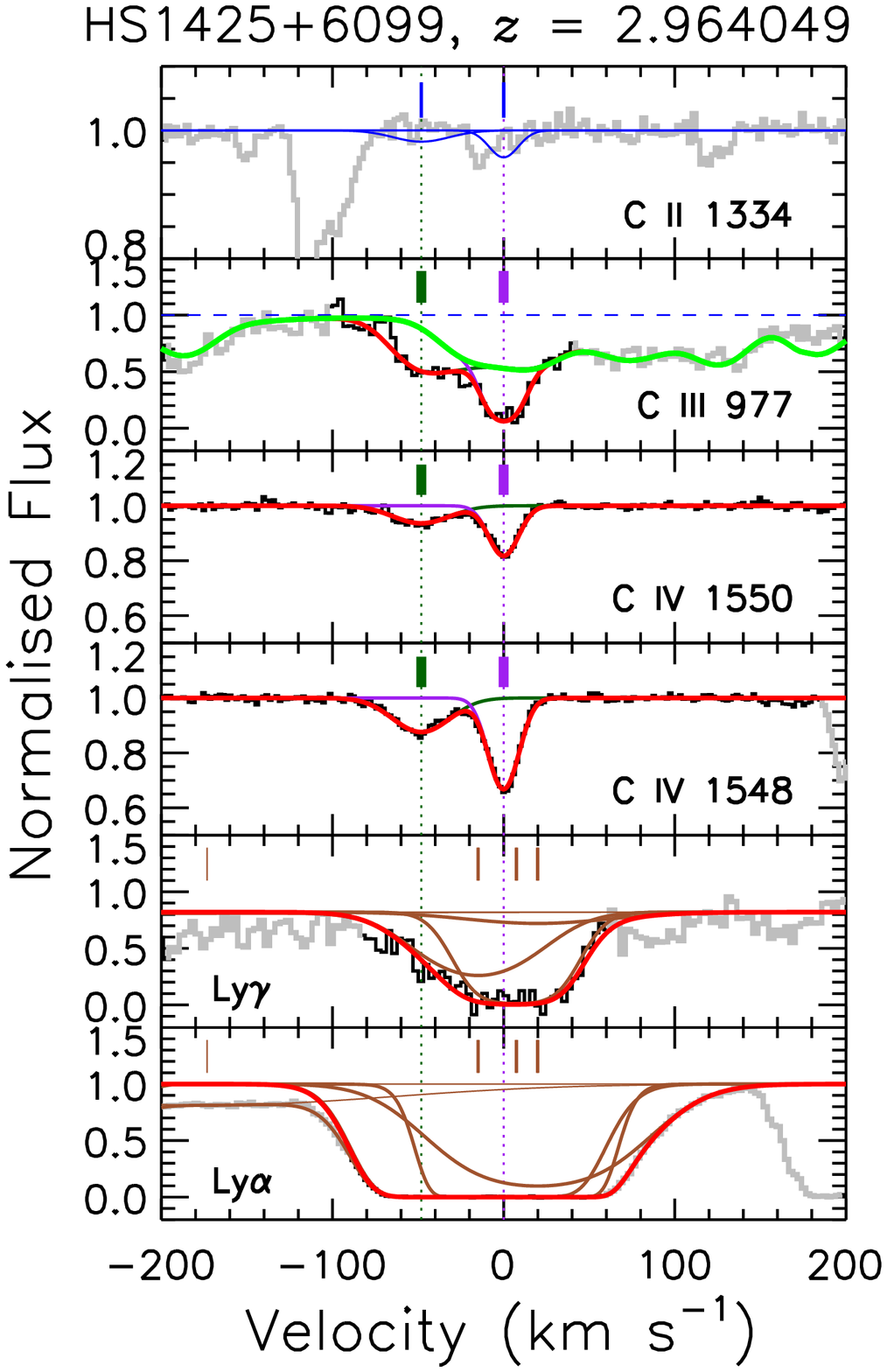}
\hspace{-0.3cm}
\includegraphics[width=45mm]{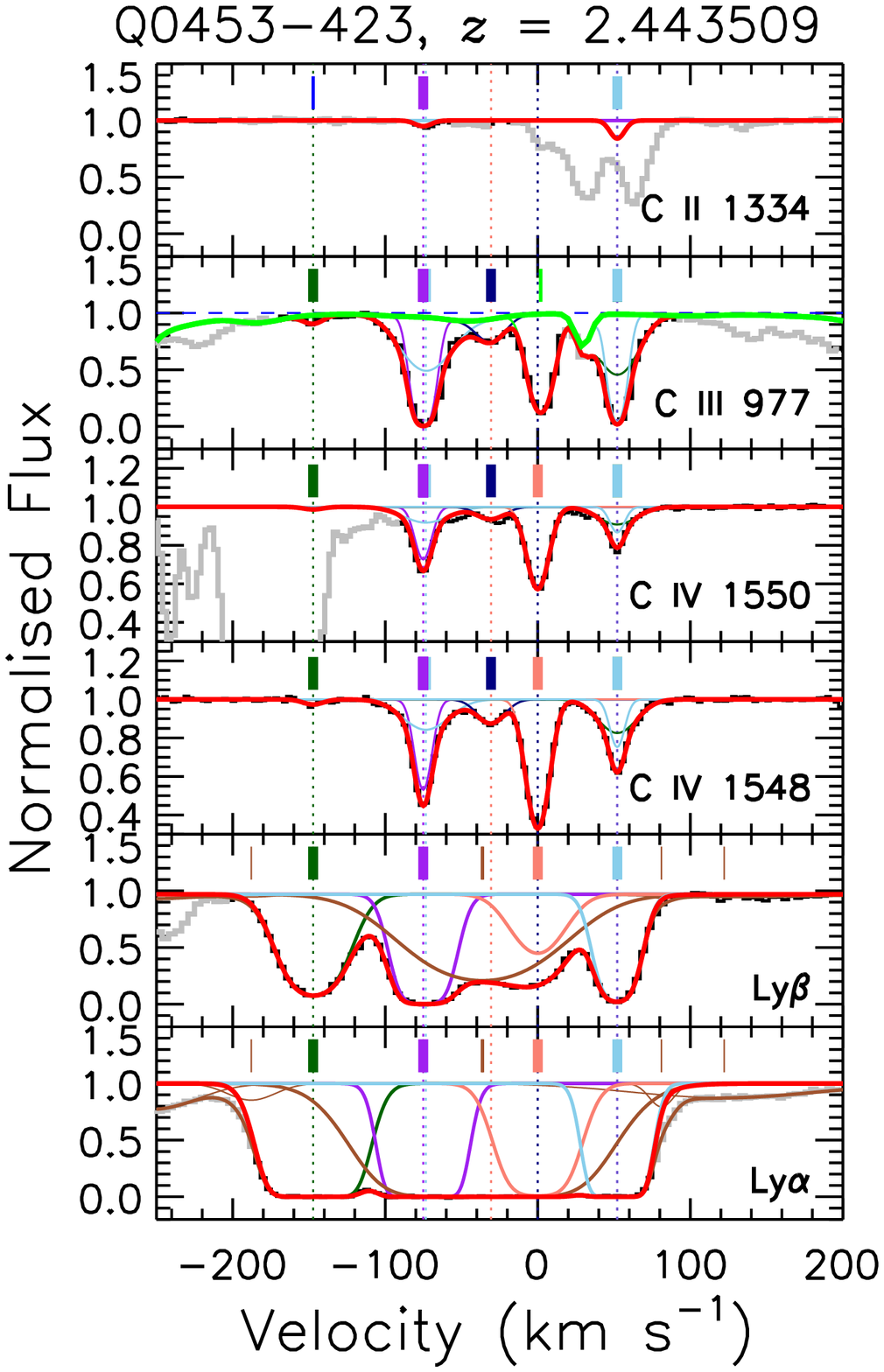}\\
\caption{The relative velocity $v$ versus normalised flux of 4 \ion{C}{iii} absorbers. 
The zero relative velocity is set to be at the
\ion{C}{iv} flux minimum, which is also set to be the redshift of a \ion{C}{iii} absorber.
The redshift and the QSO name are listed on top of 
each panel. The black histogram
is the observed spectrum, while the coloured, smooth profile is the generated
profile from the fitted line parameters. The centroid of each components marks
with thick vertical ticks.  Thick ticks of a same colour indicate the tied components.
The blue-coloured smooth profile indicates
upper limits. Uncertain \ion{C}{iii} components 
are indicated with a thin green tick. 
The thin brown \ion{H}{i} profile indicates
an individual \ion{H}{i} component near the region of interest, while the thick brown \ion{H}{i}
profile is the absorption profile from all the shown \ion{H}{i} components.
Metal lines are not shown to make a plot less complicated. 
The thick red \ion{H}{i} profile delineates only
the \ion{H}{i} components in the velocity range covered by \ion{C}{iv} and \ion{C}{iii}
in the plot. The thick green curve in the \ion{C}{iii} panel
shows the {\it working} continuum used in the VPFIT \ion{C}{iii} fit. This is a combination of the
final continuum, high-order Lyman lines, Lyman limit continuum depression
and metal lines, while in some cases 
\ion{H}{i} Ly$\alpha$ near the expected \ion{C}{iii} is also included in order to
remove a contribution of its broad wing, but never on top of \ion{C}{iii}.
{\it Left panel:} a certain, clean \ion{C}{iii} absorber. The redshift
and line width of \ion{H}{i}, \ion{C}{iv} and \ion{C}{iii} are tied. 
This absorber is suitable for the photoionisation modelling. 
{\it Second
panel:} \ion{C}{iii} at $v = 0$ \kms\/ has the same velocity structure with
\ion{H}{i} and \ion{C}{iv} at $v = 0$ \kms\/. However, \ion{C}{iii} at $v = -41$\,\kms\/
is shifted by $7\pm11$\,\kms\/ from \ion{C}{iv} at $v = -34$\,\kms\/ and has a narrower $b$.
{\it Third panel:} \ion{C}{iii} is blended with other absorptions.
Since the absorption from high-order Lyman lines is well constrained, indicated with 
the dark green profile, \ion{C}{iii} in red profile is also reasonably well measured. 
{\it Fourth panel:} the \ion{C}{iv} absorptions at $v = -75, {\mathrm and} \,+52$ \kms\/
display a narrow component on top of a much broader component, while
\ion{H}{i} at the same relative velocity does not suggest the same
component structure. The narrow \ion{C}{iv} component was assigned to the
seemingly single \ion{H}{i} component.
Refer to the online version for clarity.}
\label{fig1}
\end{figure*}

In addition to \ion{C}{iii} $\lambda$\,977, a search was made for corresponding
\ion{C}{ii} $\lambda\lambda$\,1334.532, 1036.336.
When detected these were fitted in the same
way as the \ion{C}{iii} line.
This generally gave only upper 
limits to the \ion{C}{ii} column density for systems where the Lyman limit is optically thin.

The detection limit for any line depends on the local S/N and the $b$ parameter. 
For \ion{C}{iii} $\lambda$\,977, it 
is more complicated since \ion{C}{iii} is located in heavily
blended regions with a uncertain continuum placement. Therefore, we use
a lower-end $N_{\mathrm{\ion{C}{iii}}}$ of fitted \ion{C}{iii} components
as a representative detection limit, though because of blending with Lyman forest lines 
the real limit will be higher in some individual cases. As for \ion{C}{ii}, we take the median 
upper limit of \ion{C}{ii} in absorption-free \ion{C}{ii} regions, as a majority of our
\ion{C}{iv}+\ion{C}{iii} pairs are not associated with \ion{C}{ii}.
Representative detection limits are
$\log N_{\mathrm{\ion{H}{i}, \, lim}} \sim 12.5$, 
$\log N_{\mathrm{\ion{C}{iv}, \, lim}} \sim 11.8$, 
$\log N_{\mathrm{\ion{C}{iii}, \, lim}} \sim 11.7$ and
$\log N_{\mathrm{\ion{C}{ii}, \, lim}} \sim 12.0$, although occasionally 
lower column density systems
are found in high S/N regions.

\subsection{Description of the \ion{C}{iii} sample}

\subsubsection{Examples of \ion{C}{iii} absorbers}

Figure~\ref{fig1} presents a velocity plot (the relative velocity $v$ versus 
the normalised flux) of four \ion{C}{iii} absorbers in our sample.
The left panel shows a clean, {\it certain} \ion{C}{iii} absorber whose 
component structure
is the same as the one of \ion{H}{i} and \ion{C}{iv}. The second panel illustrates 
an {\it uncertain} \ion{C}{iii} component at $v = -41$\,\kms\/. It is shifted by 
$7\pm11$\,\kms\/ from
the certain \ion{C}{iv} at $v = -34$\,\kms\/. The large velocity difference and
the different profile shape 
suggests that the component at $v = -41$\,\kms\/ is not likely to be \ion{C}{iii}.
Even if its identification is correct, the velocity shift implies that \ion{C}{iii} and \ion{C}{iv}
are not produced by the same gas, thus not suitable for the photoionisation 
modelling to derive any physical parameters.
The third panel illustrates a \ion{C}{iii} absorber blended with
high-order Lyman lines. The thick green profile indicates 
the contribution from the blended lines by high-order \ion{H}{i} lines and identified metal lines. 
The red profile delineates the entire \ion{C}{iii} absorption profile
around the region.
As blending in this case is well constrained from the high-order
\ion{H}{i} fit including all the identified metal lines, the \ion{C}{iii}
line parameters are also reasonably well measured. 

The fourth panel of Fig.~\ref{fig1} shows
an example of two \ion{C}{iv} components at the same relative velocity.
The \ion{C}{iv} absorption at $v = +52$\,\kms\/
displays a narrow component on top of a much broader component. On the contrary,
the \ion{H}{i} component at the same relative velocity does not indicate 
the same two-component structure. At $v = +52$\,\kms\/, the broader \ion{C}{iv} $b$ 
parameter is 16.6\,\kms\/, while the single \ion{H}{i} $b$ parameter 
is 13.0\,\kms\/.
For a gas producing both \ion{H}{i} and \ion{C}{iv} absorptions under the same 
physical condition, a \ion{C}{iv} $b$ parameter cannot be larger than a \ion{H}{i} $b$ 
parameter. A thermally broadened \ion{H}{i} produced
by the same \ion{C}{iv} gas with $b_{\mathrm{\ion{C}{iv}}} = 16.6$\,\kms\/ 
should have $b_{\mathrm{\ion{H}{i}}} = 57.5$\,\kms\/. 

Interestingly, the opposite type of absorption 
profile occurs occasionally at lower redshifts where blending is negligible so that the component
structure can be seen more clearly. For example, the $z = 0.22601$ \ion{O}{vi} absorber toward
HE0153--4520 has
a narrow \ion{H}{i} component on top of a broader component, with the latter being
associated with a single, broader \ion{O}{vi} \citep{savage11}. If the overall physical conditions
of metal-enriched intervening gas does not change significantly with redshifts,
in the case of our \ion{C}{iii} absorber,
a strong blending in \ion{H}{i} could have caused us to miss any broader \ion{H}{i}
coexisting with the broader \ion{C}{iv} component. However, the current data do not
require a broader \ion{H}{i}. Here
the narrow \ion{C}{iv}
component was assigned to the apparent, single \ion{H}{i} component. 
Similarly, the \ion{C}{iv} absorption at $v \sim -75$\,\kms\/ has the narrow and broader components 
almost on top of each other. The velocity difference is only $2\pm2$\,\kms\/, 
so forcing them to have the same redshift is easily possible.
Again, the narrow \ion{C}{iv} component was tied with the apparent, single \ion{H}{i} component.

\subsubsection{Selecting the clean \ion{C}{iii} components}
\label{sec2.3.3}

\ion{C}{iii} $\lambda$\,977 is a single line which falls in the 
Ly$\alpha$ forest region. Consequently, a careful choice of
\ion{C}{iii} components is required to compile a
sample where the Lyman forest contamination is less severe, and so reliable \ion{C}{iii} line 
parameters may be determined.

\begin{figure}
\hspace{-0.5cm}
\includegraphics[width=90mm]{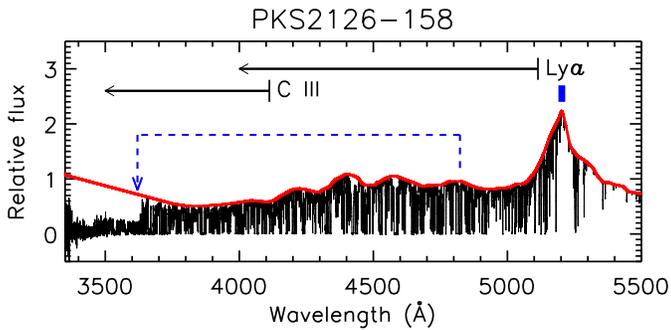}
\caption{The wavelength vs the relative flux of
PKS2126--158. The overlaid red curve is the final continuum, which was
extrapolated at $\le 3466$\,\AA\/.
The top arrows indicate the useful wavelength ranges for \ion{C}{iii} and \ion{H}{i}. For \ion{H}{i}
we exclude the proximity effect zone within 5,000\,\kms\/ of the
Ly$\alpha$ emission, which is indicated with the tick blue vertical tick at 5203\,\AA\/. 
\ion{C}{iii} corresponding to \ion{H}{i} is then at $\le 4112$\,\AA\/ .
In this object several high-$N_{\mathrm{\ion{H}{i}}}$ absorbers depress the continuum of
the QSO blueward their observed Lyman limit. Blue dashed line illustrates one such
absorber at $z \sim 2.967$, shown with
a vertical line at Ly$\alpha$ connected to the downward arrow at the 
Lyman limit depression at $\sim \! 3622$\,\AA\/.
}
\label{fig2}
\end{figure}

\begin{enumerate}

\item  Figure \ref{fig2} is
an example where the \ion{C}{iii} is in a wavelength region where the continuum of 
the background QSO is
depressed due to Lyman continuum absorption from high-column density \ion{H}{i} systems.
The continuum placement around \ion{C}{iii} regions below such depressions
is difficult to determine and is unreliable. If the continuum placement around the 
\ion{C}{iii} region is
highly uncertain despite the available high-order Lyman lines, 
these \ion{C}{iii} components were discarded. 

\item Any \ion{C}{iii} component displaying
a velocity difference larger than 1.5\,\kms\/
from the corresponding \ion{C}{iv} component 
was classified as uncertain 
and not used in the analysis. The velocity difference 
could be caused by many things, such as inaccurate wavelength calibration
in the low-S/N \ion{C}{iii} region, incorrect 
identification or unrecognised weak blending. It can also be due to a physical origin such as
the \ion{C}{iv}-producing gas not being co-spatial with 
the \ion{C}{iii}-producing gas, e.g. if  \ion{C}{iv} is produced in the outer transition region of
a hot-temperature gas cooling rapidly and \ion{C}{iii} is in a cool, dense core.
With the excellent wavelength calibration of UVES and HIRES spectra
within 1\,\kms\/ uncertainty,
inaccurate wavelength calibration is not a likely cause. 

\item When there exists no obvious absorption in the expected \ion{C}{iii} 
position, upper limits were calculated, following the procedure described in
\citet{jorgenson14}. \ion{C}{iii} components with 
upper limits were included in the further analysis.

\item Fitted \ion{C}{iii} components
were also checked for saturation, based on the flat-bottomed core of a line profile.
We note that $N_{\mathrm{\ion{C}{iii}}}$ of
a flat-bottomed-core profile is not necessarily a lower limit, 
as VPFIT looks for a best-fit $N_{\mathrm{\ion{C}{iii}}}$, unlike the apparent optical
depth analysis \citep{savage91}. However, when a \ion{C}{iii} component has a boxy shape,
we note its $N_{\mathrm{\ion{C}{iii}}}$ as a lower limit.

\item Since \ion{C}{iv} is usually found in high-S/N spectral regions and \ion{C}{iii} 
in low-S/N regions, sometimes a \ion{C}{iii} corresponding to a 
\ion{C}{iv} component in an absorption wing or in a broad profile is not 
required, i.e. no need of \ion{C}{iii} for a certain \ion{C}{iv}. Moreover, some
expected \ion{C}{iii} components occurs in severely blended regions. In
such cases, we did not estimate upper limits of \ion{C}{iii} components, since
upper limits are not very meaningful. Without a corresponding certain \ion{C}{iii},
these \ion{C}{iv} components were not included in the analysis.

\item Since the \ion{H}{i} absorbers 
were selected to be optically thin in the Lyman continuum,
most \ion{C}{iv} components in the sample do not have an associated
\ion{C}{ii}. When \ion{C}{ii} $\lambda\lambda$\,1334.53 or
1036.33 is not detected in 
absorption-free regions, upper limits for $N_{\mathrm{\ion{C}{ii}}}$ are also estimated.

\end{enumerate}

All the blending possibilities and the continuum placement uncertainties were 
taken into account when \ion{C}{iii} $\lambda$\,977 was selected and profile-fitted.
Unfortunately the \ion{C}{iii} line is a singlet, so there is no obvious way
to recognise weak \ion{H}{i} Ly$\alpha$ blending on top of \ion{C}{iii}.
The requirement that \ion{C}{iii} has the same redshift and Doppler parameter
as \ion{C}{iv} helps somewhat, but does not guarantee 
a clean or robust \ion{C}{iii} identification.
As a result, any \ion{C}{iii} column density in the sample could well be an upper limit only. 

With no other selection criterion except the apparent cleanness
of \ion{C}{iii}, our
\ion{C}{iii} selection might be considered biased in favour of narrower \ion{C}{iii} over broader.
We searched for a corresponding \ion{C}{iii} component candidate to a corresponding, 
detected \ion{C}{iv} component, regardless of the \ion{C}{iv} line width. With a ``tied" profile fit, 
a validity of \ion{C}{iii} component fits depend mostly on blending with \ion{H}{i} with
confusion introduced by continuum uncertainty and a low S/N. 

Blending by \ion{H}{i} should be
random in each analysed \ion{C}{iii} region,
since the $z$ and $N_{\mathrm{\ion{H}{i}}}$ distributions of 
\ion{H}{i} Ly$\alpha$ can be considered to be random over a small redshift range
\citep{hu95, kim01}. In addition, a strong \ion{H}{i} blending would not produce 
a similar \ion{C}{iii} profile shape as \ion{C}{iv}. If 
\ion{H}{i} blending does not change the \ion{C}{iii} profile shape,
it is likely to be weak, i.e. $N_{\mathrm{\ion{C}{iii}}}$ would not be
affected by a large factor. 
As our \ion{C}{iii} sample is mostly unsaturated and narrow, 
the fitted \ion{C}{iii} column density should not be far off from the {\it true} \ion{C}{iii} 
column density.

Continuum uncertainty and a low S/N are mostly
against a detection of weak and broad \ion{C}{iii} components. However, as long as 
\ion{H}{i} blending is weak, a ``tied" profile fit and a known \ion{C}{iv} velocity structure
enable us to obtain all the corresponding \ion{C}{iii} components, regardless of line
widths. We iterated the tied fits several times with a slightly different continuum 
adjustment each time and found a best-fit for a given \ion{C}{iii} absorber in order
to minimise the selection bias against weak and broad \ion{C}{iii}.
We note that not all analysed \ion{C}{iii} regions have a low S/N and a 
highly uncertain continuum placement.
The continuum placement uncertainty
in the \ion{C}{iii} region is generally up to 5 to 10\%, since we have fitted all 
the available \ion{H}{i} Lyman series and identified metal lines affecting the 
\ion{C}{iii} region.
When \ion{C}{iii} is fairly unblended or blending is well-characterised, 
the continuum uncertainty is less than 5\%. Any small
continuum adjustments at this level do not change the
fitted column density of \ion{C}{iii} more than $\sim$0.1\,dex. 
We discarded any \ion{C}{iii} 
located in a region having a large continuum uncertainty.

Weak \ion{H}{i} blending and continuum uncertainty would increase the scatter 
in any correlation involving with $N_{\mathrm{\ion{C}{iii}}}$,
but would not alter any obvious trends between real physical parameters.

\subsubsection{The final \ion{C}{iii} sample}
\label{sec2.3}

As our main interest
is the low-density absorbing gas, only \ion{C}{iv} systems associated with
Lyman continuum optically-thin \ion{H}{i} absorbers at 
$\log N_{\mathrm{\ion{H}{i}}} < 17.2$ were chosen to look for relatively clean
\ion{C}{iii}.
From 19 QSOs, a total of 53 \ion{C}{iv} systems was selected, 
consisting of 155 \ion{C}{iv} components. Out of 155, 23 \ion{C}{iv} components were
found to be associated with a blended or a shifted \ion{C}{iii}. Therefore, 
any analysis involving \ion{C}{iii}
is restricted to 132 tied (or aligned) \ion{C}{iii}+\ion{C}{iv} 
component pairs.
Of these, 104 have clean \ion{C}{iii} (not saturated, nor upper limits), 
in 4 it is saturated and in 24 only upper-limits
could be obtained.
In this work, we use the term
\ion{C}{iv} system in a conventional sense, {\it i.e.} an isolated, possibly complex,
\ion{C}{iv} absorption feature. Examples are shown in Fig.~\ref{fig1}. 
In general, a continuous \ion{C}{iv} absorption profile associated
with optically-thin \ion{H}{i} absorbers spans less than 500\,\kms\/. 
Components refer to the individually fitted absorption lines at some 
specific redshift as determined by VPFIT.

Although \ion{C}{iv} and
\ion{C}{iii} have associated \ion{H}{i}, these do not always show the same
velocity structure. In part this is due to the fact that \ion{H}{i} has a larger thermal
width than \ion{C}{iv} and \ion{C}{iii} for the same temperature, 
and it is more difficult to separate these broader lines into distinct components.
The number of the tied
\ion{H}{i}+\ion{C}{iv} component pairs from 155 \ion{C}{iv} components
in 53 \ion{C}{iv} systems is 
54, i.e. about 35\% of the analysed \ion{C}{iv} sample.
Recall that we tied components only if the velocity difference between the 
corresponding \ion{C}{iii} and \ion{C}{iv} (\ion{H}{i} and \ion{C}{iv}) components
is less than 1.5\,\kms\/ (15\,\kms\/), as explained Section~\ref{sec2.2.2}.

Out of 54 tied \ion{H}{i}+\ion{C}{iv} pairs, 
33 pairs also have a tied 
\ion{C}{iii} component, while 17 and 4 pairs have a upper-limit
and a blended/uncertain \ion{C}{iii} component, respectively.
Since the latter does not have a clean associated \ion{C}{iii} detected, they are
not in the sample of 132 \ion{C}{iv}+\ion{C}{iii} pairs, 
but included only in the analysis of the
temperature and non-thermal motion in Section~\ref{sec3}.

The tied \ion{H}{i}+\ion{C}{iv} component pairs are further
classified into 2 subclasses, certain and uncertain.
Even if
an isolated \ion{C}{iv} component is well-aligned with a cleanly
separable \ion{H}{i} component, the pair is labelled uncertain if
\ion{C}{iv} is located in the low-S/N spectral region since the \ion{C}{iv} line 
parameters are then not well-determined, e.g. 
the $z=2.572434$ pair toward HE2347$-$4342.
In addition, when a weak \ion{C}{iv}
component exists nearby with a strong \ion{C}{iv} component
clearly tied with a single \ion{H}{i} component, this pair is also classified
as uncertain. For example, the two \ion{H}{i}+\ion{C}{iv} pairs at $v = -75$ and
$+52$\,\kms\/ in the fourth
panel of Fig.~\ref{fig1} are both classified as uncertain. 
There are 40 certain and 14 uncertain tied \ion{H}{i}+\ion{C}{iv} pairs.

\begin{figure}
\includegraphics[width=85mm]{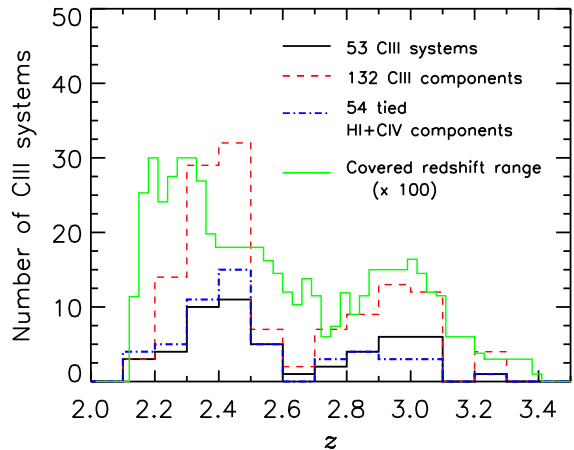}
\vspace{-0.3cm}
\caption{Redshift distributions of 53 \ion{C}{iii} systems, 132 \ion{C}{iii}
components including 24
upper limits and 108 detected \ion{C}{iii} components, 
and 54 tied \ion{H}{i}+\ion{C}{iv} pairs. The green histogram shows
the covered redshift range by 19 QSOs multiplied by 
100 times for clarity. Without incompleteness
correction for \ion{C}{iii} recovery rate mainly due to blending and low-S/N, the
number of \ion{C}{iii} absorbers is {\it not} a physically-meaningful detection rate.}
\label{fig3}
\end{figure}

Figure~\ref{fig3} displays the redshift distributions of 53 \ion{C}{iii} systems and 
132 \ion{C}{iv}+\ion{C}{iii} components in our sample. About 40\% of the 
\ion{C}{iii} absorbers is at
$2.3 < z < 2.5$. An apparent deficiency of \ion{C}{iii} systems
at $z \sim 2.6$ is an observational bias, caused by
a lower number of QSOs covered at that redshift range. We note
that there is no incompleteness correction for \ion{C}{iii} which might have been 
missed because of
heavy blending, low-S/N or unreliable continuum placement. Therefore,
the redshift distribution of analysed \ion{C}{iii} in Fig.~\ref{fig3} should
{\it not} be used as a physically meaningful number density evolution of
\ion{C}{iii} absorbers with redshift. 

Table~\ref{tab2} lists the line parameters of 
\ion{H}{i}, \ion{C}{iv}, \ion{C}{iii} and \ion{C}{ii} components in our 53 
\ion{C}{iii} systems, including
uncertain and upper-limit components. 
The entire Table~\ref{tab2} and the 
velocity plot of entire 53 \ion{C}{iii} systems similar to Fig.~\ref{fig1} 
are published electronically. 
The velocity plots also include uncertain \ion{C}{iii} components marked with a thin light
green tick. These have either
 a significantly shifted velocity centroid or are  located in the profile wing for which the fitted
line parameter is very uncertain. These are not included in the further analysis. 
Unless stated otherwise, all the analyses in this study are based on individual components.


\begin{table}
\caption[]{The line parameters of \ion{C}{iii} absorbers$\,^{\mathrm{a}}$}
\label{tab2}
\begin{tabular}{rlcclc}
\hline

\#$^{\mathrm{b}}$  & Ion & $\Delta v^{\mathrm{c}}$ & $z_{\mathrm{abs}}^{\mathrm{c}}$ & 
$b^{\mathrm{d}}$ & $\log N$ \\
  &  &  (km s$^{-1}$) &  & (km s$^{-1}$) &  \\
  
\noalign{\smallskip}
\hline

\multicolumn{6}{c}{    Q0055--269,    $z = $ 3.256208, $[   -40, +120]$\,km s$^{-1, \, {\mathrm{e}}}$} \\[0.07cm]

\hline \\[-0.25cm]
   x & \ion{C}{iv}   &    -17{\tiny $\pm$9}  &   3.255963 &     12.6{\tiny $\pm$2.0} &  12.92{\tiny $\pm$0.09} \\
   1 & \ion{C}{iv}   &      0{\tiny $\pm$3}  &   3.256208 &      9.0{\tiny $\pm$0.8} &  13.02{\tiny $\pm$0.07} \\
     & \ion{C}{iii}  &                       &            &      9.0                 &  13.13{\tiny $\pm$0.11} \\
   x & \ion{H}{i}    &      6{\tiny $\pm$17} &   3.256300 &     34.1{\tiny $\pm$3.0} &  15.18{\tiny $\pm$0.08} \\
   2 & \ion{C}{iv}   &     30{\tiny $\pm$7}  &   3.256628 &     10.8{\tiny $\pm$2.2} &  12.63{\tiny $\pm$0.09} \\
     & \ion{C}{iii}  &                       &            &     10.8                 &  12.45{\tiny $\pm$0.16} \\
   x & \ion{H}{i}    &     49{\tiny $\pm$16} &   3.256899 &     14.4{\tiny $\pm$3.5} &  14.49{\tiny $\pm$0.24} \\
   3 & \ion{C}{iv}   &     51{\tiny $\pm$3}  &   3.256940 &     11.4{\tiny $\pm$1.2} &  12.95{\tiny $\pm$0.04} \\
     & \ion{C}{iii}  &                       &            &     11.4                 &  13.01{\tiny $\pm$0.04} \\
{\bf 4} & \ion{C}{iv} &  81{\tiny $\pm$2}  &   3.257359 &     10.8{\tiny $\pm$0.8} &  12.76{\tiny $\pm$0.03} \\
     & \ion{H}{i}    &                       &            &     24.7{\tiny $\pm$0.9} &  13.93{\tiny $\pm$0.02} \\
     & \ion{C}{iii}  &                       &            &     10.8                 &  12.54{\tiny $\pm$0.06} \\
     &  \ion{C}{ii}  &                       &            &     25.6                 &  $\le 12.55$ \\[0.1cm]

\hline \\[-0.25cm]

\end{tabular}
\begin{list}{}{}
\item[$^{\mathrm{a}}$]
Only the beginning of the entire table is shown.
The full version of this table is available electronically on the MNRAS web page.
\item[$^{\mathrm{b}}$]
The tied component is grouped with a group number.  A component marked with ``x" indicates
a component with no other corresponding ions or a uncertain \ion{C}{iii} component with 
a certain \ion{C}{iv}.
\item[$^{\mathrm{c}}$]
The relative velocity is centred at the \ion{C}{iv} flux minimum. Only the $z$-reference-ion
\ion{C}{iv} has an error. 
\item[$^{\mathrm{d}}$]
For a tied fit, only the reference ion \ion{C}{iv} is provided with a 
fit error for the tied parameters $z$ and $b$.
\item[$^{\mathrm{e}}$]
The relative velocity ranged over which useful \ion{C}{iii} components are found.
\end{list}
\end{table}

\section{Column density and Doppler parameter distributions}
\label{sec3}

Using the fitted ion column densities and Doppler parameters
of \ion{H}{i}, \ion{C}{iv} and \ion{C}{iii} given in Table \ref{tab2}, 
we examine their distributions and look for any correlations between them.

\subsection{The \ion{H}{i}+\ion{C}{iv}+\ion{C}{iii} sample}
\label{sec3.1}

\begin{figure}
\hspace{-0.3cm}
\includegraphics[width=85mm]{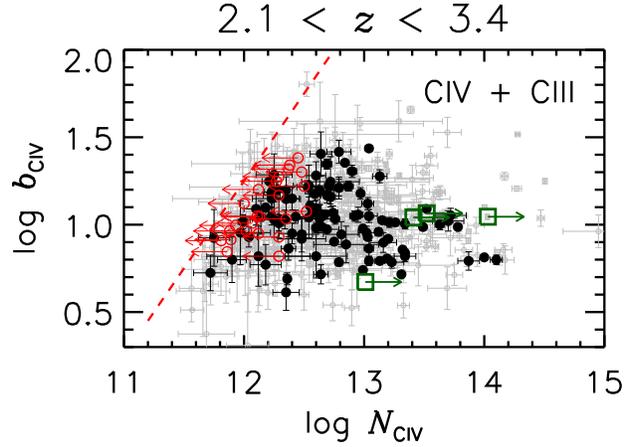}
\caption{The $\log N_{\mathrm{\ion{C}{iv}}}$--$\log b_{\mathrm{\ion{C}{iv}}}$ distribution.
Filled circles, red open circles and green open squares indicate \ion{C}{iv} associated
with 104 clean, 24 upper-limit and 4 lower-limit \ion{C}{iii} regardless of
association with a well-aligned \ion{H}{i}. Grey open 
circles show the distribution of
431 \ion{C}{iv} components from the 19 QSO spectra in the redshift range listed
in Table~\ref{tab1}. 
The dashed line indicates the \ion{C}{iv} detection limit. 
Refer to the online version for clarity.}
\label{fig4}
\end{figure}

First, in order to see whether or not our \ion{C}{iii} sample 
represent a typical set of IGM absorbers,
we consider the Doppler parameter vs column density distributions for 
\ion{C}{iv} and \ion{H}{i} for our selected samples
and compare these with the distributions for the populations as a whole.
Figure~\ref{fig4} shows the 
$\log N_{\mathrm{\ion{C}{iv}}}$--$\log b_{\mathrm{\ion{C}{iv}}}$ distribution 
of \ion{C}{iv} for those with associated \ion{C}{iii}, set against 
431 \ion{C}{iv} components with $\log N_{\mathrm{\ion{C}{iv}}} \in [11.5, 15.0]$
at $2.1 < z < 3.4$ from the 19 QSOs listed in Table~\ref{tab1} including \ion{C}{iv}
associated with optically-thick absorbers. 
As can be seen from the figure, those components for which only upper limits 
could be measured for \ion{C}{iii} tend
to correspond to lower \ion{C}{iv} column densities, while the few saturated 
\ion{C}{iii} lines correspond to higher \ion{C}{iv}
column densities.
Given that the current sample consists of 132 
\ion{C}{iii}+\ion{C}{iv} components pairs,
only about 31\% 
(132 out of 431) of all the \ion{C}{iv} components
are suitable for any further analysis. Note that
this is not an actual \ion{C}{iv}+\ion{C}{iii} association rate, however,
it may be taken as a lower limit. Unlike \ion{C}{iv}, blending of \ion{C}{iii} $\lambda$977
with Ly$\alpha$ forest lines reduces the effective S/N. Therefore, \ion{C}{iii} is not 
always be detected at the nominal limit for the S/N where it occurs.

\begin{figure}
\hspace{-0.1cm}
\includegraphics[width=85mm]{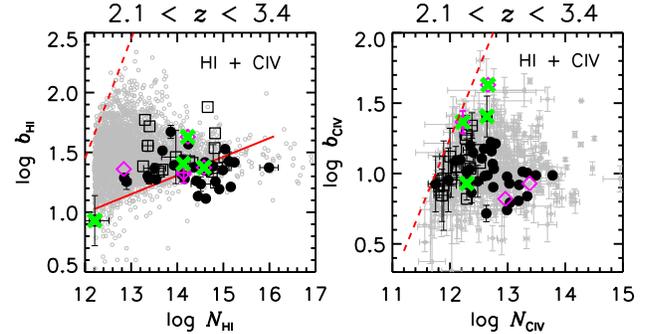}
\caption{{\it Left panel:} The Doppler parameter vs column density distribution of 
\ion{H}{i} for the same redshift range as in Fig.~\ref{fig4}.
Grey open circles indicate the
distribution for 3956 \ion{H}{i} components with $\log N_{\mathrm{\ion{H}{i}}} \in [12.2, 17.0]$ 
from the 19 QSOs, shown without errors for clarity.
These are from the high-order Lyman
fit using all the available high-order lines to obtain more robust line parameters 
of saturated lines used in the current study, including the tied parameter fits. 
Filled circles indicate
33 \ion{H}{i} components with well-aligned \ion{C}{iv} and \ion{C}{iii}
with errors where these are larger than the symbol size.
Open squares denote the 17 well-aligned \ion{H}{i}+\ion{C}{iv} 
components with \ion{C}{iii} upper limits, and 
magenta open diamonds are the 4 such components with blended \ion{C}{iii}. 
The dashed line indicates the \ion{H}{i} detection limit, while
the red solid line shows the cut-off $b$ as a function of $N_{\mathrm{\ion{H}{i}}}$
at $<z>\,\,=2.4$ taken from \citet{rudie12b}. 
Green crosses represent the non-thermally broadened components, which have
the gas temperature less than 5000 K. Some of the grey data points below 
the cutoff-$b$ line will be unidentified metal lines.
{\it Right panel:} $N_{\mathrm{\ion{C}{iv}}}$ vs $b_{\mathrm{\ion{C}{iv}}}$ of 
54 tied \ion{H}{i}+\ion{C}{iv} pairs overlaid with 431
\ion{C}{iv} components at $\log N_{\mathrm{\ion{C}{iv}}} \in [11.5, 15.0]$
as shown in Fig.~\ref{fig4}. The symbols
have the same meaning as in the left panel.
Refer to the online version for clarity.
}
\label{fig5}
\end{figure}

\begin{figure*}
\hspace{-0.2cm}
\includegraphics[width=180mm]{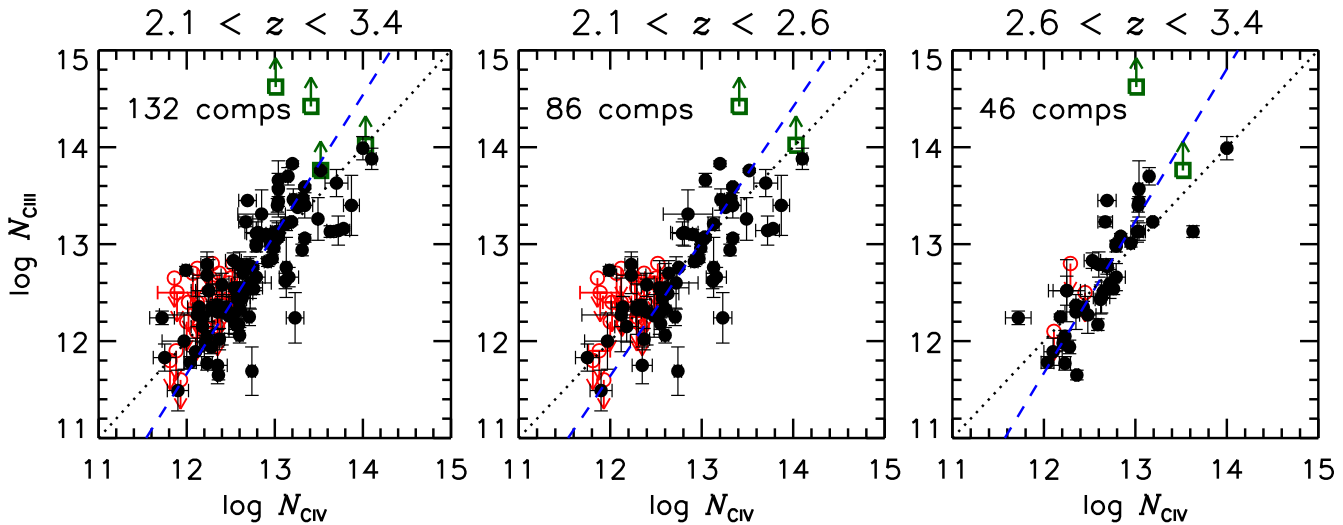}
\vspace{-0.5cm}
\caption{The $N_{\mathrm{\ion{C}{iv}}}$--$N_{\mathrm{\ion{C}{iii}}}$
relation at three redshift ranges. Filled circles, red open circles and green squares
indicate 104 clean, 24 upper-limit and 
4 lower-limit \ion{C}{iii} components, respectively.
Only errors larger than the symbol size are displayed.
Dotted lines indicate a one-to-one relation. Dashed lines represent a robust least squares
fit: $\log N_{\mathrm{\ion{C}{iii}}} = A + B \times \log N_{\mathrm{\ion{C}{iv}}}$,
while $A = (-5.61\pm1.44, -5.03\pm2.06, -7.16\pm2.44)$ and
$B = (1.44\pm0.11, 1.39\pm0.16, 1.57\pm0.19)$ at
$2.1 \le z \le 3.4$, $2.1 \le z \le 2.6$ and $2.6 \le z \le 3.4$, respectively.}
\label{fig6}
\end{figure*}

As with \ion{C}{iv}, we may examine the 
$N_{\mathrm{\ion{H}{i}}}$--$b_{\mathrm{\ion{H}{i}}}$ diagram for the
\ion{H}{i} components. This is shown in the left panel of Fig.~\ref{fig5},
with gray data points indicating 3956 \ion{H}{i} components in the 19 QSOs
with $\log N_{\mathrm{\ion{H}{i}}} \in [12.2, 17.0]$ at $2.1 < z < 3.4$ .
The \ion{H}{i} components highlighted here are those 54 components which have 
well-aligned associated \ion{C}{iv} with or without detected \ion{C}{iii}. 
For these, we estimate the temperature $T$ and 
turbulent Doppler parameter $b_{\rm nt}$ by constraining the redshifts to be 
the same and tying the Doppler parameters using Eq.~\ref{eq1}. 
In order to distinguish it from the gas temperature derived from 
a photoionisation model, we note the temperature calculated from the line 
widths by VPFIT as $T_{b}$ in this study.

Four green crosses represent a \ion{H}{i}+\ion{C}{iv} pair dominated by 
non-thermal broadening, as discussed in Section~\ref{sec3.3}.
They do not seem to occur in any preferred area, although
it could be caused by a small-number statistics. All of them have a low temperature at
$T_{b} \le 5000$\,K.

Only $\sim$1\% (54 out of 3956) of all the \ion{H}{i} components can be 
used for estimating the gas temperature and non-thermal 
motion in this study. This low fraction is mainly because a majority 
of low-$N_{\mathrm{\ion{H}{i}}}$, 
intervening \ion{H}{i} have no associated \ion{C}{iv} above 
the detection limit. In addition, the fraction is further reduced by the requirement 
that \ion{C}{iii} be 
measurable, since the tied 
\ion{H}{i}+\ion{C}{iv} sample is taken from
the \ion{C}{iv}+\ion{C}{iii} sample.

The lower cutoff-$b$ as a function of $N_{\mathrm{\ion{H}{i}}}$,
$\log b_{\mathrm{\ion{H}{i}}} = 1.244 + 0.156 \times (\log N_{\mathrm{\ion{H}{i}}} - 13.6)$ 
at $<z>\,\, = 2.4$ \citep{rudie12b}, is shown in the left panel of Fig. \ref{fig5}. 
This cutoff-$b_{\mathrm{\ion{H}{i}}}$ line is important as it is often used to estimate 
the IGM gas temperature
\citep{schaye00a, ricotti00, rudie12b, bolton14}. The \citet{rudie12b} 
minimum $N_{\mathrm{\ion{H}{i}}}$--$b_{\mathrm{\ion{H}{i}}}$ relation 
also appears to fit well for the general Ly$\alpha$ forest in this study.
Almost all of our tied \ion{H}{i}+\ion{C}{iv} pairs in the 
$N_{\mathrm{\ion{H}{i}}}$--$b_{\mathrm{\ion{H}{i}}}$
plane lie close to the cutoff-$b$ line, except a few pairs above the red solid line.
Except 4 \ion{H}{i}+\ion{C}{iv} pairs dominated by a non-thermal broadening (green
crosses),
the tied \ion{H}{i}+\ion{C}{iv} pairs without \ion{C}{iii} (open squares)
have a temperature at $\log T_{b} = 4.3$--5.5. For the \ion{H}{i}+\ion{C}{iv} pairs
with \ion{C}{iii}, the temperature spans a lower $T_{b}$ range at
$\log T_{b} = 3.9$--5.5, with 73\% having $\log T_{b} \le 4.5$.
This implies that some of \ion{H}{i}+\ion{C}{iv}
pairs with \ion{C}{iii} upper limits is due to a higher temperature. Indeed,
many outliers above the red solid line are 
$N_{\mathrm{\ion{C}{iii}}}$-upper-limit pairs.

Unlike the \ion{C}{iv}+\ion{C}{iii} pairs,
the \ion{H}{i}+\ion{C}{iv} pairs
do not sample a wide range of column density and $b$ value of \ion{H}{i}. Although there is
no well-defined one-to-one relation between $T$ and $b$ or between the volume density
and $N_{\mathrm{\ion{H}{i}}}$, a lack of tied
pairs with a larger $b$ or a smaller $N_{\mathrm{\ion{H}{i}}}$ implies that our tied
pairs do not include a higher-temperature gas and a lower density gas, but sample a gas
within a limited physical condition.
This is in part
caused by the observational bias, since a larger thermal width of \ion{H}{i} compared
to \ion{C}{iv} does not allow to recognise closely adjacent \ion{H}{i} components as
precisely as \ion{C}{iv} in a multi-component \ion{C}{iv} absorber.

Unlike \ion{H}{i} in the left panel, \ion{C}{iv} in the 
tied \ion{H}{i}+\ion{C}{iv}
pairs in the right panel of Fig.~\ref{fig5} 
displays a wider range of column density and $b$. Combined with 
the $N_{\mathrm{\ion{H}{i}}}$--$b_{\mathrm{\ion{H}{i}}}$ distribution, 
\ion{C}{iv} components of the tied pairs is not expected to show very
well-correlated relations with \ion{H}{i} components, 
i.e. $N_{\mathrm{\ion{C}{iv}}}$ is not a well-behaved 
function of $N_{\mathrm{\ion{H}{i}}}$.

\subsection{Correlations between $N_{\mathrm{\ion{H}{i}}}$, $N_{\mathrm{\ion{C}{iv}}}$ 
and $N_{\mathrm{\ion{C}{iii}}}$}
\label{sec3.2}

A straightforward quantity to consider is the 
$N_{\mathrm{\ion{C}{iv}}}$--$N_{\mathrm{\ion{C}{iii}}}$ relation. For the collisional
ionisation equilibrium (CIE), the ratio of two ions depends on the gas temperature. For
the photoionisation equilibrium (PIE), the ratio depends largely on the ionisation parameter,
the ionising photon flux divided by the gas density.

Figure~\ref{fig6} illustrates the $N_{\mathrm{\ion{C}{iv}}}$--$N_{\mathrm{\ion{C}{iii}}}$ relation
for three different redshift ranges, $2.1 \le z \le 3.4$, $2.1 \le z \le 2.6$, and $2.6 \le z \le 3.4$, 
along with least-squares fits for clean \ion{C}{iii}
detections only, i.e. excluding upper limits and lower limits.
The division redshift is simply chosen from Fig.~\ref{fig3}, which 
shows an apparent dip in the number distribution of \ion{C}{iii} systems at $z\sim 2.6$
due to the observational bias.
There are no indications of any redshift evolution of the 
$N_{\mathrm{\ion{C}{iv}}}$--$N_{\mathrm{\ion{C}{iii}}}$ 
relation. For a given $N_{\mathrm{\ion{C}{iv}}}$, the spread in $N_{\mathrm{\ion{C}{iii}}}$
is $\sigma \sim 0.3$\,dex, when fitted to a Gaussian.

For the full redshift range,
$N_{\mathrm{\ion{C}{iii}}} \propto N_{\mathrm{\ion{C}{iv}}}^{1.42\pm0.11}$
at $\log N_{\mathrm{\ion{C}{iv}}} \in [11.7, 14.1]$. 
However, at $\log N_{\mathrm{\ion{C}{iv}}} \sim 14$,
$N_{\mathrm{\ion{C}{iii}}}$ is lower than expected from the dashed
line, suggesting that the log-linear relation 
between $N_{\mathrm{\ion{C}{iii}}}$ and $N_{\mathrm{\ion{C}{iv}}}$ might well
break down at higher $N_{\mathrm{\ion{C}{iv}}}$. 
However, given only a few \ion{C}{iv} components at 
$\log N_{\mathrm{\ion{C}{iv}}} \ge 13.6$,
any firm conclusions cannot be drawn. At the same time, due to a lack of enough data at
high-$N_{\mathrm{\ion{C}{iv}}}$ ends, a one-to-one relation could be
a good approximation between 
$N_{\mathrm{\ion{C}{iv}}}$ and $N_{\mathrm{\ion{C}{iii}}}$. A lack of
high-$N_{\mathrm{\ion{C}{iv}}}$ components is in part due to the fact that 
the number density of higher-$N_{\mathrm{\ion{C}{iv}}}$ components 
is smaller \citep{kim13}.

Even though $N_{\mathrm{\ion{C}{iii}}}$ increases faster than 
$N_{\mathrm{\ion{C}{iv}}}$, $N_{\mathrm{\ion{C}{iii}}}$ is lower than
$N_{\mathrm{\ion{C}{iv}}}$ at $\log N_{\mathrm{\ion{C}{iv}}} \le 13$ and
is higher at $\log N_{\mathrm{\ion{C}{iv}}} \ge 13$. Therefore,
the mean $N_{\mathrm{\ion{C}{iii}}}/N_{\mathrm{\ion{C}{iv}}}$ is
$<\!N_{\mathrm{\ion{C}{iii}}}/N_{\mathrm{\ion{C}{iv}}}\!> \, = 1.0\pm0.3$ at $2.1 < z < 3.4$
for the \ion{C}{iv} components aligned with the clean \ion{C}{iii} components.
If the \ion{C}{iii}+\ion{C}{iv} gas is in CIE,  
$<\!N_{\mathrm{\ion{C}{iii}}}/N_{\mathrm{\ion{C}{iv}}}\!> \, \sim 1$ suggests
the gas temperature of $\sim \! 10^{5}$\,K \citep{gnat07}. For the PIE gas, 
$<\!N_{\mathrm{\ion{C}{iii}}}/N_{\mathrm{\ion{C}{iv}}}\!>$ indicates 
$T \sim 10^{4.2-5}$\,K \citep{oppenheimer13}. 

\begin{figure}
\hspace{-0.2cm}
\includegraphics[width=88mm]{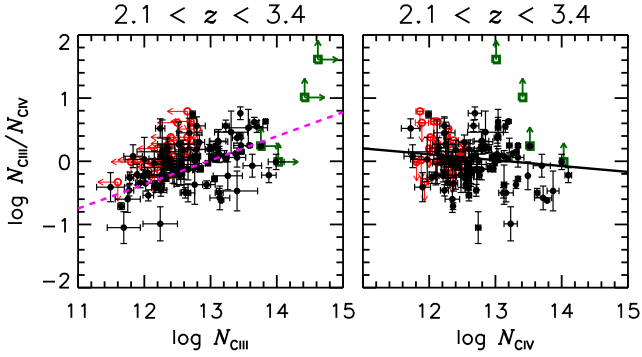}
\vspace{-0.5cm}
\caption{$N_{\mathrm{\ion{C}{iii}}}/N_{\mathrm{\ion{C}{iv}}}$ as a function
of $N_{\mathrm{\ion{C}{iii}}}$ (left panel) and of $N_{\mathrm{\ion{C}{iv}}}$
(right panel) at $2.1 < z < 3.4$. All the symbols are the same as in Fig.~\ref{fig6}.
The magenta dashed line in the left panel represents a least-square
fit of clean \ion{C}{iii} detections excluding lower and upper limits: 
$\log N_{\mathrm{\ion{C}{iii}}}/N_{\mathrm{\ion{C}{iv}}} = (-4.94\pm0.16)
+ (0.38 \pm 0.01) \times \log N_{\mathrm{\ion{C}{iii}}}$. In the right panel, 
the solid line is a least-square fit:
$\log N_{\mathrm{\ion{C}{iii}}}/N_{\mathrm{\ion{C}{iv}}} = (1.23\pm0.18)
+ (-0.09 \pm 0.01) \times \log N_{\mathrm{\ion{C}{iv}}}$.}
\label{fig7}
\end{figure}

\begin{figure}
\hspace{-0.2cm}
\includegraphics[width=85mm]{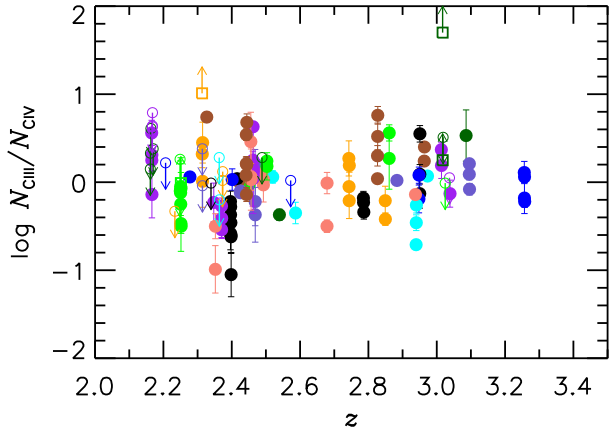}
\vspace{-0.4cm}
\caption{$N_{\mathrm{\ion{C}{iii}}}/N_{\mathrm{\ion{C}{iv}}}$ as a function
of $z$ at $2.1 < z < 3.4$. Filled symbols are the \ion{C}{iv}+\ion{C}{iii} pairs
with cleanly detected \ion{C}{iii}. Open symbols with upward and downward arrows
indicate the \ion{C}{iv}+\ion{C}{iii} pairs with lower-limit \ion{C}{iii} and
upper-limit \ion{C}{iii}, respectively. The pairs belonging to the same system are
coded with the same colour.
Refer to the online version for clarity.}
\label{fig8}
\end{figure}

Figure~\ref{fig7} displays $N_{\mathrm{\ion{C}{iii}}}/N_{\mathrm{\ion{C}{iv}}}$ as
a function of $N_{\mathrm{\ion{C}{iii}}}$ (left panel) and of $N_{\mathrm{\ion{C}{iv}}}$
(right panel), respectively.
In the left panel, $N_{\mathrm{\ion{C}{iii}}}/N_{\mathrm{\ion{C}{iv}}}$ 
increases with $N_{\mathrm{\ion{C}{iii}}}$ as expected, 
since $N_{\mathrm{\ion{C}{iii}}}$ increases faster 
than $N_{\mathrm{\ion{C}{iv}}}$.
On the other hand, $N_{\mathrm{\ion{C}{iii}}} / N_{\mathrm{\ion{C}{iv}}}$ seems to suggest a weak
decrease with $N_{\mathrm{\ion{C}{iv}}}$ in the right panel (solid line). 
This is mainly caused by the data points having a lower $N_{\mathrm{\ion{C}{iii}}}$ 
than the log-linear relation at $\log N_{\mathrm{\ion{C}{iv}}} \sim 14$ as seen in 
Fig.~\ref{fig6}. The same data points are scattered at  $\log N_{\mathrm{\ion{C}{iv}}} 
\in [13.0, 14.2]$ and $\log N_{\mathrm{\ion{C}{iii}}}/N_{\mathrm{\ion{C}{iv}}} \in [-1.0, 0.0]$
in the right panel.
This leads to a weak decrease of $N_{\mathrm{\ion{C}{iii}}}/N_{\mathrm{\ion{C}{iv}}}$
with $N_{\mathrm{\ion{C}{iv}}}$.

The redshift evolution of $N_{\mathrm{\ion{C}{iii}}}/N_{\mathrm{\ion{C}{iv}}}$ is
presented in Fig.~\ref{fig8}. Filled and open symbols indicate the 
\ion{C}{iv}+\ion{C}{iii} pairs with clean \ion{C}{iii} and \ion{C}{iii} limits, respectively.
There is a large spread of $\sim$1\,dex
in $N_{\mathrm{\ion{C}{iii}}}/N_{\mathrm{\ion{C}{iv}}}$
even within the same system, as commonly seen in the high-$N_{\mathrm{\ion{H}{i}}}$
absorbers, cf. \citet{boksenberg15}. This suggests that the \ion{C}{iv}-bearing gas 
components within a system have a range of internal conditions.

\begin{figure}
\hspace{-0.2cm}
\includegraphics[width=88mm]{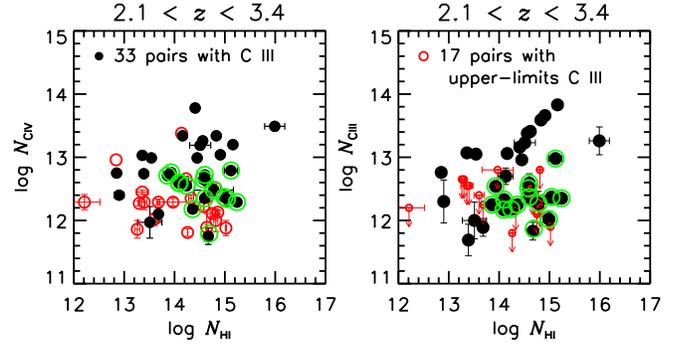}
\vspace{-0.4cm}
\caption{$N_{\mathrm{\ion{H}{i}}}$--$N_{\mathrm{\ion{C}{iv}}}$ 
(left panel) and $N_{\mathrm{\ion{H}{i}}}$--$N_{\mathrm{\ion{C}{iii}}}$ 
(right panel) of the tied \ion{H}{i}+\ion{C}{iv}+\ion{C}{iii}
components at $2.1 < z < 3.4$. Filled circles and open red circles are 
the 33 and 17 \ion{H}{i}+\ion{C}{iv} pairs with clean \ion{C}{iii} and 
upper-limit \ion{C}{iii},
respectively. No lower-limit \ion{C}{iii} is associated with a well-aligned
\ion{H}{i}. Data points embedded with a larger green open circle are
low-metallicity branch absorbers discussed in Section~\ref{sec5}.
Refer to the online version for clarity.
}
\label{fig9}
\end{figure}

\begin{figure}
\hspace{-0.2cm}
\includegraphics[width=88mm]{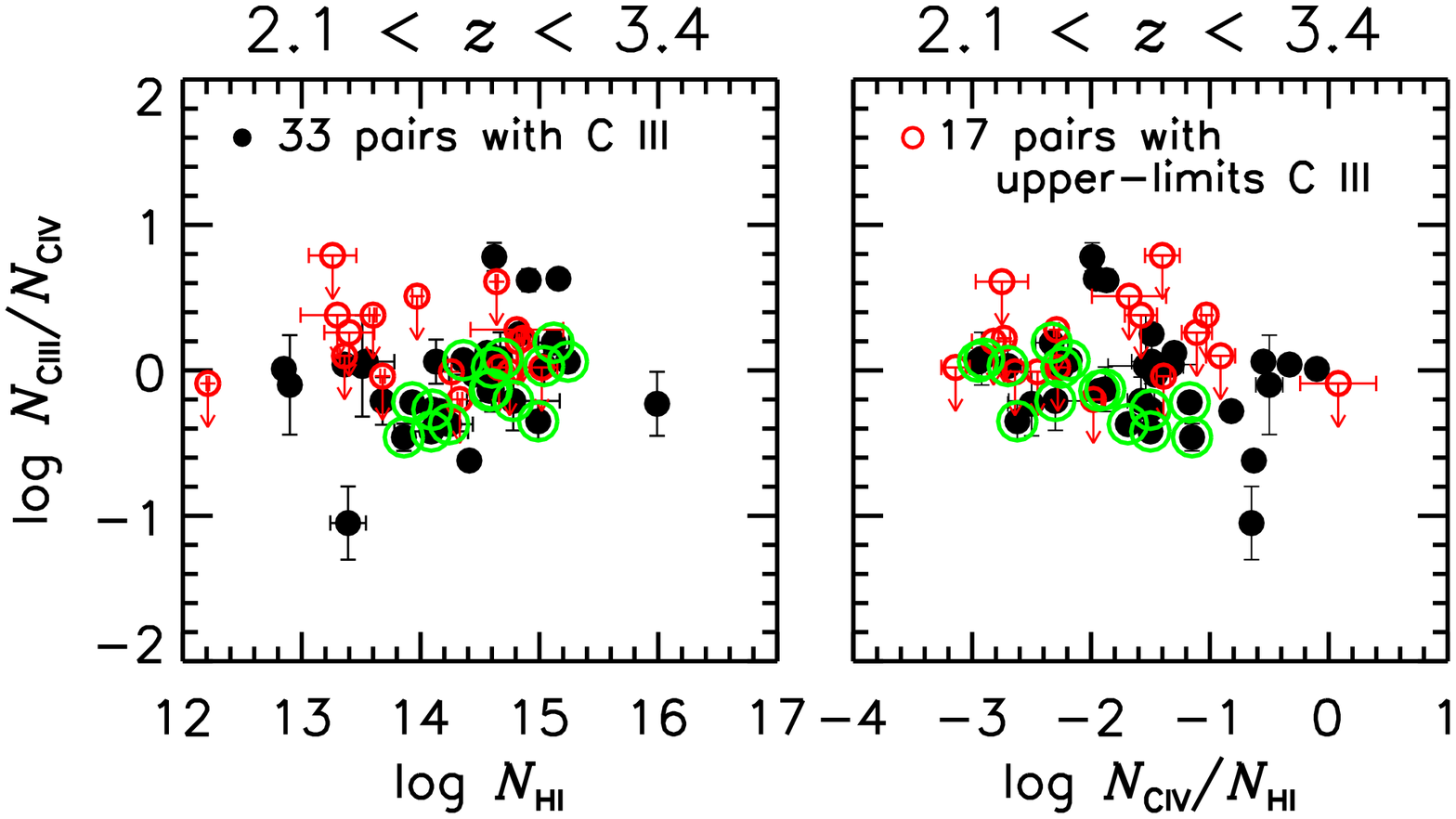}
\vspace{-0.5cm}
\caption{$N_{\mathrm{\ion{C}{iii}}}/N_{\mathrm{\ion{C}{iv}}}$ as a function
of $N_{\mathrm{\ion{H}{i}}}$ (left panel) and as a function of
$N_{\mathrm{\ion{C}{iv}}}/N_{\mathrm{\ion{H}{i}}}$ (right panel)
at $2.1 < z < 3.4$. All the symbols have the same meaning as in 
Fig.~\ref{fig9}. Refer to the online version for clarity.}
\label{fig10}
\end{figure}

There are only a small number of tied \ion{H}{i}+\ion{C}{iv}+\ion{C}{iii} 
components, so
the $N_{\mathrm{\ion{H}{i}}}$--$N_{\mathrm{\ion{C}{iv}}}$ 
and the $N_{\mathrm{\ion{H}{i}}}$--$N_{\mathrm{\ion{C}{iii}}}$ relations are poorly constrained 
and are effectively scatter plots (Fig.~\ref{fig9}). However, when tied \ion{C}{iii} trios 
shown as filled circles are grouped into 2 classes
as high-metallicity branch absorbers and low-metallicity branch absorbers 
introduced in Section~\ref{sec1} and discussed
in Section~\ref{sec5}, there seems a trend.
For high-metallicity branch absorbers (clean, filled circles),
$N_{\mathrm{\ion{H}{i}}}$ roughly increases with $N_{\mathrm{\ion{C}{iv}}}$
and $N_{\mathrm{\ion{C}{iii}}}$. On the other hand,
low-metallicity branch absorbers embedded with a larger green open circle
show no correlation.

In Fig.~\ref{fig10},
the detected data points on the 
$N_{\mathrm{\ion{H}{i}}}$--$N_{\mathrm{\ion{C}{iii}}}/N_{\mathrm{\ion{C}{iv}}}$ plane 
(left panel) and on the
$N_{\mathrm{\ion{C}{iv}}}/N_{\mathrm{\ion{H}{i}}}$--$N_{\mathrm{\ion{C}{iii}}}/N_{\mathrm{\ion{C}{iv}}}$
plane (right panel) suggest no 
correlation. Since the upper limits on both planes
have the values similar to the detected ones, the lack of correlation between these
quantities is reasonable.
This means that both $N_{\mathrm{\ion{C}{iv}}}$ and $N_{\mathrm{\ion{C}{iii}}}$ increase
with $N_{\mathrm{\ion{H}{i}}}$ in a roughly similar way, even with the larger scatter at higher
$N_{\mathrm{\ion{C}{iii}}}$ and $N_{\mathrm{\ion{C}{iv}}}$ seen 
in Fig.~\ref{fig7}.

\subsection{The gas temperature and the non-thermal motion}
\label{sec3.3}

\begin{figure}
\hspace{-0.2cm}
\includegraphics[width=86mm]{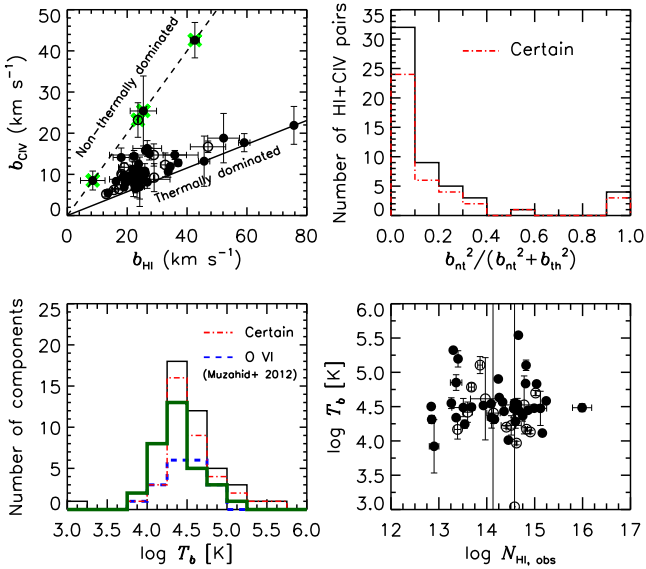}
\vspace{-0.6cm}
\caption{
Gas temperature and non-thermal $b$, $b_{\mathrm{nt}}$, of the 54
tied \ion{H}{i}+\ion{C}{iv} pairs. Filled and open circles indicate the 40 certain 
pairs and the 14 uncertain pairs as described in Section~\ref{sec2.3}. 
{\it Upper left:} $b_{\mathrm{\ion{H}{i}}}$ vs
$b_{\mathrm{\ion{C}{iv}}}$. Green crosses are the pairs as in Fig.~\ref{fig5}
and have $T_{b} \le 5000$\,K.
{\it Upper right:} the ratio between the non-thermal energy
and the total (thermal plus non-thermal) energy, 
$b_{\mathrm{nt}}^{2}/(b_{\mathrm{nt}}^{2}+b_{\mathrm{th}}^{2})$, for 
the full (solid histogram) and certain (red dot-dashed histogram) samples. 
{\it Lower left:} 
the gas temperatures
for the full (solid histogram) and certain (red dot-dashed histogram) samples. 
The thick green histogram is for the 33 pairs 
with a detected \ion{C}{iii} excluding lower limits.
The overlaid blue dashed histogram is taken from \citet{muzahid12} for
\ion{O}{vi} at $z \sim 2.3$. Only the \ion{H}{i}+\ion{O}{vi} pairs
whose velocity difference is less than 10\,\kms\/ are included.
{\it Lower right:} the gas temperature as a function 
of $\log N_{\mathrm{\ion{H}{i}}}$. The four non-thermally broadened
absorbers are not included.
}
\label{fig11}
\end{figure}

A comparison of the Doppler parameters of \ion{C}{iv} and \ion{H}{i} for the 54 
tied \ion{H}{i}+\ion{C}{iv} pairs is shown in Fig.~\ref{fig11} (upper left panel).
Table~\ref{tab_a1} lists the line parameters of individual tied
components, their derived gas temperature $T_{b}$ and 
non-thermal $b_{\mathrm{nt}}$ parameter.

A majority of the pairs are close to being thermally dominated,
while the 4 pairs are turbulently broadened.  
These 4 tied pairs are at $z = 2.937304$
toward HE0940--1050, $z= 2.787112$ toward HE2347--4342,
$z = 2.456208$ toward PKS0329--255 and
$z = 2.248307$ toward Q0329--385 with
the gas temperature of $\le 92$\,K, $\le 12830$\,K, $\le 538$\,K
and $\le 385$\,K, respectively. These pairs have the same $b$ value for
\ion{H}{i} and \ion{C}{iv}, therefore, the gas temperature
cannot be constrained at all. With $b = 42.6, 23.6, 25.4$ and 8.5\,\kms\/, respectively,
the non-thermally broadened absorbers seem to prefer a larger $b$, but
with only 4 absorbers, no firm conclusion cannot be drawn.

The non-thermal and thermal contributions to the \ion{H}{i}+\ion{C}{iv}
pairs are more clearly seen in the upper-right panel, which shows the
histogram of the ratio of the non-thermal energy to the total energy,
$b_{\mathrm{nt}}^{2}/(b_{\mathrm{th}}^{2}+b_{\mathrm{nt}}^{2})$.
For about 59\% of the \ion{H}{i}+\ion{C}{iv} pairs (32 out of 54), the 
non-thermal energy is less than 10\,\%. 
The median $b_{\mathrm{nt}}$ and $b_{\mathrm{th}}$ are 
7\,\kms and 22\,\kms\/, respectively.
When $b_{\mathrm{\ion{C}{iv}}} \ge 20$ \kms\/, 
there is a possibility that a seemingly single-component \ion{C}{iv} breaks into many
weaker components in a higher-S/N spectrum, in which case a
contribution by non-thermal broadening could be smaller.

The lower-left panel shows the 
distribution of the gas temperature for both the full and ``certain" samples.
While a broader \ion{H}{i} or \ion{C}{iv}
component could reveal a multi-component nature in much higher-S/N spectra,
$\log T_{b} \ge 5$ is also expected since intergalactic \ion{C}{iv} is  
produced by radiatively
cooling gas once shock-heated to $\log T_{b} \ge 7$ and now exposed
to the UVB \citep{cen11, shen13}. 
 
For the 54 tied pairs, the mean temperature of the full and
certain samples is
$<\!\log T_{b}\!> \,= 4.27\pm1.00$ and $<\!\log T_{b}\!> \,= 4.25\pm1.14$,
respectively. 
The median temperature is $\log T_{b, \mathrm{med}} = 4.47$ and 
$\log T_{b, \mathrm{med}} = 4.48$ for the
full and certain samples, respectively.
When excluded 4 non-thermally broadened pairs, for the full sample,
$<\!\log T_{b}\!> \,= 4.52\pm0.33$ and $\log T_{b, \mathrm{med}} = 4.48$.
Considering a large error, our mean temperature of the 54 pairs is 
consistent with the one, $<\log T_{b}> \,\,= 4.58$, found
by \citet{rauch96} from the 26 \ion{C}{iv}+\ion{Si}{iv} pairs at the similar redshifts. 
When excluded non-thermal pairs, our mean temperature 
of the 50 pairs is in better agreement
with the Rauch's mean temperature, although 
\ion{C}{iv}+\ion{Si}{iv} pairs typically sample a higher-$N_{\mathrm{\ion{H}{i}}}$
absorber than typical \ion{H}{i}+\ion{C}{iv} pairs without associated \ion{Si}{iv}.

Since our \ion{H}{i}+\ion{C}{iv} pairs do not have any unusual selection criterion
apart from being kinematically simple, we may
assume that this small number of
pairs is a representative of typical optically-thin \ion{C}{iv}-bearing gas found in the spectra
of high-$z$ QSOs.
The $T_{b}$ distribution is well-approximated by a Gaussian
peaking at $\log T_{b} \sim 4.43$ with 1$\sigma = 0.27$ and
$\log T_{b} \sim 3.5$--5.5. Considering that \ion{C}{iv} peaks at
$\log T \sim 5$ in CIE, the observed 
temperature distribution implies that the \ion{H}{i}+\ion{C}{iv} gas
samples both photoionised and collisionally ionised \ion{C}{iv}.
This \ion{C}{iv} temperature distribution is very similar to the simulated result for 
$\log N_{\mathrm{\ion{C}{iv}}} \in [12, 14]$ at $z = 2.6$ by \citet{cen11}.
The \ion{H}{i}+\ion{C}{iv}+\ion{C}{iii} trios excluding lower-limit 
$N_{\mathrm{\ion{C}{iii}}}$ components
account for only 61\% (33 out of 54) of the total sample, but show a similar 
$T_{b}$ distribution.

For comparison, the lower-left panel also shows the temperature estimates
from the observed $b$ parameters of aligned \ion{H}{i}+\ion{O}{vi} at $z \sim 2.3$
taken from \citet{muzahid12}.
Muzahid et al.'s \ion{O}{vi} temperature distribution 
is very similar to our \ion{C}{vi} temperature distribution.
This is not what
is expected from simulations, which predicts that \ion{O}{vi} is produced by higher
temperature gas than \ion{C}{iv} \citep{cen11}. However, the detection of high 
temperature, and hence broad, 
\ion{O}{vi} lines in the Ly$\alpha$ forest is difficult, since they are more likely to 
be misidentified as \ion{H}{i}. This bias may be responsible for part of the discrepancy. 

The lower-right panel of Fig.~\ref{fig11} displays 
$\log T_{b}$ as a function of $N_{\mathrm{\ion{H}{i}}}$. It shows that the
gas temperature does not correlate with $N_{\mathrm{\ion{H}{i}}}$, 
which is expected from the optically thin PIE gas.
We have found that there is no noteworthy correlations between $b_{\mathrm{nt}}$,
$T_{b}$, $z$, $N_{\mathrm{\ion{C}{iv}}}$ or $N_{\mathrm{\ion{C}{iii}}}$.

In summary, our data suggest that 
most of the tied low-$N_{\mathrm{\ion{H}{i}}}$ absorbers
located around the cutoff-$b$ line are thermally broadened, with a small
contribution from turbulent 
motion.  The result also confirms that the cutoff-$b$
line is a valid tool to probe the IGM temperature.
\ion{H}{i}+\ion{C}{iv} pairs display a wide range of the temperature at
$\log T_{b} \in [3.5, 5.5]$, with a few non-thermally broadened pairs with
$\log T_{b} \le 3.7$. The temperature distribution peaks at
$\log T_{b}  \sim 4.4$, implying that a majority of \ion{C}{iv} is 
photoionisation dominated.

\section{Photoionisation modelling}
\label{sec4}

\subsection{Basic assumptions}
\label{sec4.1}

The narrow line width of the low-density Ly$\alpha$ forest implies that
photoionisation is the dominant ionisation mechanism rather than collisional 
ionisation \citep{bergeron86b, haehnelt96}.
Theories reproducing the general observed properties of the 
\ion{H}{i} forest also predict that the forest is photoionised by the external 
UV background (UVB) contributed both from QSOs and galaxies,
assuming it in photoionisation equilibrium \citep{bergeron86b, cen94, 
haehnelt96, theuns98a, 
schaye00a, schaye01, dave10}. 

There are several critical pre-requirements for the photoionisation modelling to be valid. 
First, all the absorption lines of interest should be produced in the same gas. Thus,
for a constant gas temperature through the region,
ionic absorption components can be used only when their velocity structure is
the same. Second, the absorption lines should be properly 
resolved and deblended so that reliable 
column densities and line widths can be determined.
Third, there should exist a \ion{H}{i} component clearly associated with 
other ionic transitions of interest. The ratio of ionic column densities is insensitive
to the metallicity and $N_{\mathrm{\ion{H}{i}}}$ for the
ionization parameter $-3.0 \le \log U \le 0$, a $U$ range for typical Ly$\alpha$ absorbers.
However, as one of our scientific interests is to estimate the carbon abundance, 
$N_{\mathrm{\ion{H}{i}}}$ is necessary to break the degeneracy.
Fourth, more than two ionic transitions of the same
metal species should be present. 
In general, the metallicity and the gas density (or the ionisation parameter $U$)
are unknown. When only one ionic transition is available, a different combination
of the metallicity and density predicts the same column
density.  Additional ionic transitions break this degeneracy.

These requirements imply that photoionisation 
modelling should be applied only on 
clean, individual {\it components}, as opposed to total 
column densities obtained by adding up all the components of an absorption system.
If a physical condition of gas producing each component of a multi-component
system is similar, the total column densities of the system can be used, i.e. the
derived physical parameters of the system approximate the average of the individual
components. However, there is no solid 
physical reason to assume that each component is produced by gas under a similar
gas density, temperature and carbon abundance. In fact, the large spread in
$N_{\mathrm{\ion{C}{iii}}}/N_{\mathrm{\ion{C}{iv}}}$ within a given system as shown in
Fig.~\ref{fig8} hints that a physical condition might be different among components in
the same system.

\begin{figure}
\includegraphics[width=85mm]{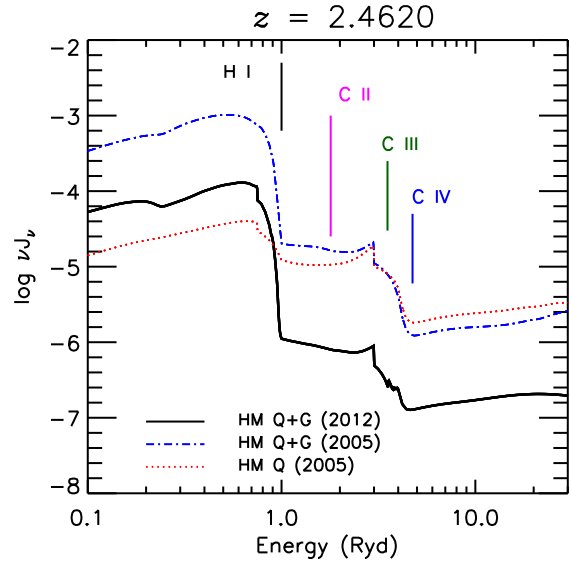}
\hspace{-0.8cm}
\caption{Examples of three UV background radiation fluxes $J_{\nu}$
as a function of the frequency $\nu$ in Rydbergs
at $z = 2.4620$ used for Cloudy modelling.
Our fiducial HM Q+G 2012 is
shown as the solid curve, while the Cloudy-default HM Q+G  2005
and the HM Q 2005 are the blue dot-dashed and the red
dotted curves, respectively (see Section \ref{sec4.2}). The ionisation potential of each ion of
interest is marked with a long vertical line: 1 Ryd (13.6 eV) for \ion{H}{i},
4.74 Ryd for \ion{C}{iv}, 3.52 Ryd for \ion{C}{iii} and 1.79 Ryd for \ion{C}{ii}. 
}
\label{fig12}
\end{figure}

Of the 155 \ion{C}{iv} components
in our sample, only the 54 components (35\%) have well-aligend \ion{H}{i},
and of these only the 33 components
have \ion{H}{i}, \ion{C}{iv} and detected \ion{C}{iii} (no lower limits) 
components at the same relative velocity.
Of the 33 trios, the $z=2.456208$ trios
toward PKS0329--255 is non-thermally broadened with $T_{b} \le 538$\,K. 
With a detected high-ion \ion{C}{iv} and such a low gas temperature, this absorber
is likely to be out of photoionisation equilibrium with the external UV background. 
Therefore, this trio is excluded in the further photoionisation analysis.
Four trios of the 32 also show 
clean, un-saturated \ion{C}{ii}. This provides 
more stringent constraints and a check on the validity of the photoionisation modelling: 
$z=2.462358$ (Q0002--422), 
$z=2.462044$ (Q0002--422),  $z=2.444109$ (Q0453--423)
and $z=2.442644$ (Q0453--423).
About one third of
\ion{C}{ii}-free components is associated with \ion{Si}{iv} (sometimes \ion{Si}{iii}
and \ion{Si}{ii} as well), while the rest is \ion{Si}{iv}-free, as listed in the
last columns of Table~\ref{tab_a2} and \ref{tab_a3}.

\subsection{Cloudy modelling}
\label{sec4.2}

\begin{figure*}
\includegraphics[width=186mm]{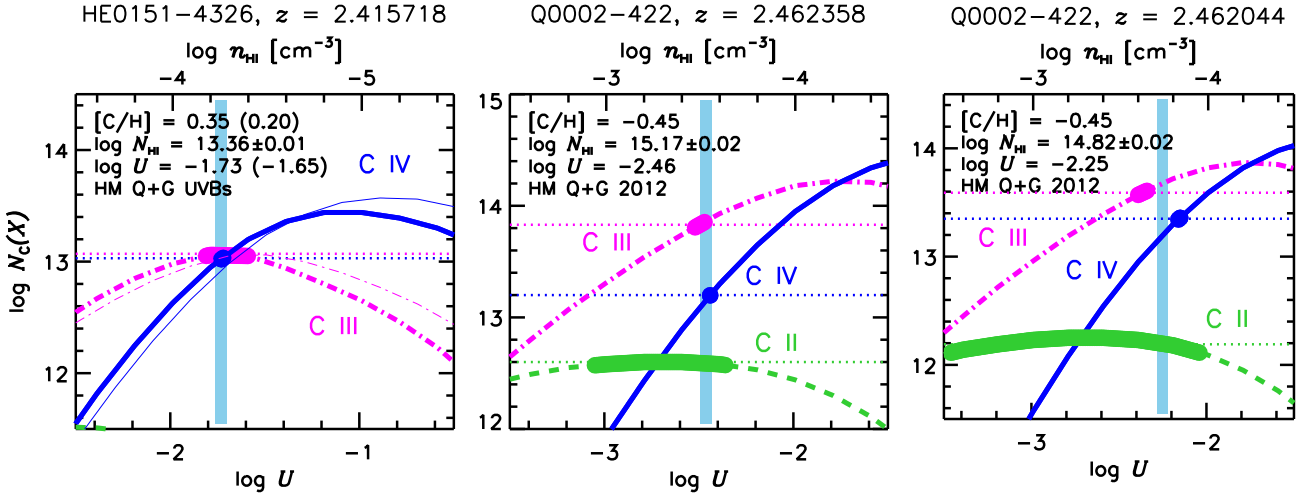}
\caption{
Three examples of Cloudy-predicted column densities $\log N{\mathrm{c}}(X)$ as a function of 
ionisation parameter $U$, where $N_{\mathrm{c}}(X)$ is the predicted column density of \ion{C}{iv},
\ion{C}{iii} and \ion{C}{ii} for different UV background radiations. 
Cloudy-predicted $N_{\mathrm{c}}(\ion{C}{iv})$, $N_{\mathrm{c}}(\ion{C}{iii})$
and $N_{\mathrm{c}}(\ion{C}{ii})$ are shown as solid blue, dot-dashed magenta
and dashed green curves, respectively.
The allowed column density range from the observations 
is indicated by thicker curves and the horizontal dotted lines with the
same colour as the curves. The sky-blue-shaded region is the best-fit $U$ value 
for observed column densities. 
{\it Left panel:} the 
$z = 2.415718$ absorber toward HE0151--4326. 
Cloudy-predicted column densities for the HM Q+G 2012 (HM Q+G 2005)
are shown as thick (thin) curves. 
{\it Middle panel:} the $z = 2.462358$ absorber toward Q0002--422 for
the HM Q+G 2012.
The best-fit $\log U = -2.46$ matches the observed 
$N_{\mathrm{\ion{C}{iv}}}$, $N_{\mathrm{\ion{C}{iii}}}$ and $N_{\mathrm{\ion{C}{ii}}}$ 
within the allowed error ranges simultaneously.
{\it Right panel:} the $z = 2.462044$ absorber toward Q0002--422, the worst
case in our sample.
No $U$ is found to match the observed carbon ions simultaneously.
}
\label{fig13}
\end{figure*}

We used the photoionisation code Cloudy version c13.03 \citep{ferland13}.
The gas was assumed to be a uniform slab in thermal and ionisation
equilibrium with a constant density
exposed to the external UVB. The Cloudy-default
solar abundance pattern is assumed, with a solar carbon abundance of 
(C/H)$_{\odot} = -3.61$. The carbon abundance is expressed as
${\mathrm{[C/H]}} = \log({\mathrm{C/H}}) - \log{\mathrm{(C/H)_\odot}}$. 

The fiducial UVB was chosen to be the latest
Haardt-Madau (HM) UVB 2012 version contributed both
by QSOs and galaxies (Q+G) \citep{haardt12}.
The main observational inputs for theoretical UVB models, such as
the relative contributions of QSOs and galaxies and the UV photon
escape fraction as a function of redshift, are still poorly known, 
and this leads to large uncertainties in the models \citep{bolton05, shapley06,
faucher08, siana10, haardt12, barger13}. Therefore, we
also used the Cloudy-default HM UVB 2005 version 
for QSOs-only (Q) and for Q+G for comparison. 

Based on the inversion method to reconstruct the spectral shape of the UVB from
metal absorbers at $1.8 < z < 2.9$,
\citet{agafonova07} claimed that the intergalactic \ion{He}{ii} Ly$\alpha$ absorption
modulates and fluctuates the UVB spectral shape from a QSO-dominated UVB without a need of   
additional sources such as starburst galaxies. However, their reconstruction
was mostly based on border-line Lyman limit systems at $\log N_{\mathrm{\ion{H}{i}}} \ge 16.7$,
where a radiative transfer effect by the internal \ion{He}{ii} Ly$\alpha$ absorption is important
with their estimate on $N_{\mathrm{\ion{He}{ii}}}$/$N_{\mathrm{\ion{H}{i}}}$ of 50--150.
By comparison, 
all of our sample but one have $\log N_{\mathrm{\ion{H}{i}}} \le 15.2$. In addition,
at $2.2 < z < 2.8$ where most of our sample belong to, 
$N_{\mathrm{\ion{He}{ii}}}$/$N_{\mathrm{\ion{H}{i}}}$ has been found to be
50--80 \citep{reimers97, shull10, syphers13}. 
Therefore, the internal 
\ion{He}{ii} Ly$\alpha$ absorption would not change the spectral shape of the UVB
significantly in our case.
Although the intensity and spectral shape of the UVB are likely to fluctuate,
we assume that a uniform UVB is a good approximation to our optically-thin
\ion{C}{iii} absorbers.
Moreover, \citet{bolton11} found 
that any realistic UVB fluctuations would not have a significant effect on the intergalactic
\ion{C}{iv} and \ion{Si}{iv} from a variety of toy UVB models.

Figure~\ref{fig12} 
shows the fiducial UVB 
and the two other UVBs at $z = 2.4620$. The intensity of the HM Q+G 2012 is
about 20 times lower than the HM Q+G 2005, but their overall
spectral shapes are similar. On the other hand, the 
HM Q+G 2005 and HM Q 2005 have different spectral shapes, 
especially at $\le 1$\,Ryd around the \ion{H}{i} ionisation edge, 
since the galaxy contribution is predominantly low-energy photons with energies $< 1$\,Ryd. 
Although the intensity of the HM Q+G 2012
is likely to be a factor of 3 to 5 too low at $z \sim 0$, it is consistent
with other measurements at $z \sim 2$--3, the redshift range of our interest
\citep{kollmeier14, shull15}.

The degree of ionisation depends on the intensity and spectral shape
of the UVB as well as the density of the gas. The ionisation level depends mainly on the
ionisation parameter $U = n_{\gamma} / n_{\mathrm{H}}$, where
$n_{\gamma}$ is the number density of (hydrogen) ionising photons
and $n_{\mathrm{H}}$ is
the total (neutral and ionised) hydrogen volume density.
For a chosen UVB and the observed $N_{\mathrm{\ion{H}{i}}}$ as the stopping
criterion of Cloudy, a determination of $U$ is sufficient 
to derive the hydrogen density.
In the low-density limit, the metallicity does not
strongly affect the ionisation structure, but becomes important for the thermal
balance since cooling is then dominated by the atomic line cooling rather than the 
recombination or bremstrahulung. In the low-metallicity limit,
the relative ionic ratio is independent of the metallicity \citep{osterbrock74}.

We generated a 
grid of models for each 
\ion{H}{i}+\ion{C}{iv}+\ion{C}{iii} component trio. We varied [C/H]
and $n_{\mathrm{H}}$ with a
logarithmic step size of 0.1 or 0.05. 
For each absorber, the observed column densities with their fit errors
and the Cloudy-predicted column densities were compared for a given 
[C/H] and ionisation parameter $U$ in order to find
the best-fit model. 

Figure~\ref{fig13} shows three such examples. 
The left panel shows a typical absorber in our sample, for which a well-measured 
single $U$ matches $N_{\mathrm{\ion{C}{iv}}}$ and $N_{\mathrm{\ion{C}{iii}}}$ 
simultaneously. All of our \ion{H}{i}+\ion{C}{iv}+\ion{C}{iii} sample without \ion{C}{ii}
have a well-determined $U$ and [C/H] regardless of a given UVB. 
Unsurprisingly, whenever \ion{C}{ii} is included,
the Cloudy solution is not as good as the \ion{C}{ii}-free trios 
since the ionisation is now set by two ratios rather than one, so a single 
ionisation parameter may not provide an adequate description,  if  
the spectral shape assumed for the ionising flux is incorrect, or if 
we are dealing with a multiphase medium.
The middle and right panels of Fig.~\ref{fig13} show two of such examples
under the HM Q+G 2012 UVB.
In the middle panel, a single $U$ matches 
the observed column density of \ion{C}{iv}, \ion{C}{iii} and \ion{C}{ii} within errors. 
On the other hand, the right panel shows the worst case in our
sample, in that no single best-fit $U$ is found to reproduce the observation
within errors. This absorber is fit slightly better for the HM Q 2005, but 
not significantly.

\subsection{Caveats}
\label{sec4.3}

In the presence of the UV background,
when a metal-enriched hot gas cools down or is mixed with the low-density \ion{H}{i}, 
these processes result in a change in the thermal
and ionisation states of the metal-enriched gas, a departure from
an equilibrium state. As a 
result, even for the same \ion{H}{i} column density ranges
at $\log N_{\mathrm{\ion{H}{i}}} \in [13, 15]$,
\ion{C}{iv}-enriched \ion{H}{i} is expected to have a higher gas temperature 
and a higher [C/H]
than the typical Ly$\alpha$ forest, 
whether \ion{C}{iv} is collisionally ionised or
photoionised \citep{cen11, shen13}. However, the degree of this
departure from photoionisation equilibrium in the metal-enriched
forest gas is not observationally constrained \citep{haehnelt96, oppenheimer12},
and this is one of our objectives in this study.

At $z \sim 2.4$ and at the cosmic mean density, $\log n_{\mathrm{H}} \sim 
-6.72 + 3 \times \log (1+z) \sim -5.13$ \citep{wiersma09}. For an absorber 
below this density, the Hubble expansion should be taken into account. 
The density of most \ion{C}{iii} absorbers in our sample
is $-5 < \log n_{\mathrm{H}} < -4$ as shown in Section~\ref{sec5.1}. 
For most \ion{C}{iii} absorbers, the Hubble expansion does not play
a significant role. 

In addition, 
for the HM Q+G 2012, the photoionisation time scale is 
$1/\Gamma_{\mathrm{\ion{H}{i}}} \sim 1.04 \times 10^{12}$\,sec $\sim 3.3 \times
10^{4}$\, yrs, where $\Gamma_{\mathrm{\ion{H}{i}}}$ is the \ion{H}{i}
photoionisation rate \citep{haardt12}. The dynamical time scale is $t_{\mathrm{dyn}} =
1/\sqrt{G\rho} \sim 1.16 \times 10^{10}$ yrs, where $G$ is the gravitational
constant and $\rho$ is the gas mass density \citep{schaye01}.
The photoionisation time scale is much shorter 
than the dynamical time scale.
Therefore, we ignore the Hubble expansion and any dynamical
evolution in the photoionisation model, i.e. assuming a static absorber.

Of 32 tied trios, 4 trios are found to be associated clearly with 
\ion{C}{ii} $\lambda$\,1334 or $\lambda$\,1036 or both.
While 20 \ion{C}{iii} absorbers have absorption-free \ion{C}{ii}
regions, therefore, a reliable $N_{\mathrm{\ion{C}{ii}}}$ upper limit,
8 \ion{C}{iii} absorbers have the \ion{C}{ii} region of $\lambda$\,1334 
and 1036 blended or unreliable, 
as noted in Table~\ref{tab_a2}. Four trios with \ion{C}{ii} have
highest $N_{\mathrm{\ion{C}{iii}}}/N_{\mathrm{\ion{C}{iv}}}$ values as well as the
lower gas temperature at $\log T_{b} \le 4.17$. There are no other clear distinctions 
between \ion{C}{ii} absorbers and upper-limit-$N_{\mathrm{\ion{C}{ii}}}$ absorbers.
Since 8 \ion{C}{iii} absorbers with a blended \ion{C}{ii} region have a lower 
$N_{\mathrm{\ion{C}{iii}}}/N_{\mathrm{\ion{C}{iv}}}$ and higher $T_{b}$,
we expect them to have $N_{\mathrm{\ion{C}{ii}}}$ similar to our assumed
$N_{\mathrm{\ion{C}{ii}}}$ detection limit of $\log N_{\mathrm{\ion{C}{ii}}} \sim 12.0$,
if they are associated with \ion{C}{ii}.

\subsection{Results and systematic uncertainties}
\label{sec4.4}

The Cloudy results of the 32 tied components are tabulated
in Table~\ref{tab_a2} for the HM Q+G 2012, including
the observed column densities with their fit errors. 

Yielding a formal 1$\sigma$ error of Cloudy-predicted parameters is not
trivial, since they are correlated without errors.
Instead, we use a rough acceptable ranges of [C/H] and $\log U$ 
as a [C/H] error and as a {\it relative} error substitute for other parameters.
As seen in Fig.~\ref{fig13},
the predicted column density has a upward parabola curve as a function of $\log U$.
When an observed column density error is small and the intercepting $U$ 
is located in the steeper side of the parabola curve, 
both [C/H] and $\log U$
can be estimated within 0.05\,dex. On the other hand, when
the observed column density error is large and the intercepting $U$ 
occurs at near the flatter top of the upward parabola curve, $\log U$ is less 
sharply estimated, with larger than 0.1\,dex for an acceptable $\log U$ range. 
In general, [C/H] is better estimated than $U$. 
About 47\% of the sample under the HM Q+G 2012 have
both [C/H] and $\log U$ estimates within 0.05\,dex. 

\begin{figure}
\hspace{-0.2cm}
\includegraphics[width=85mm]{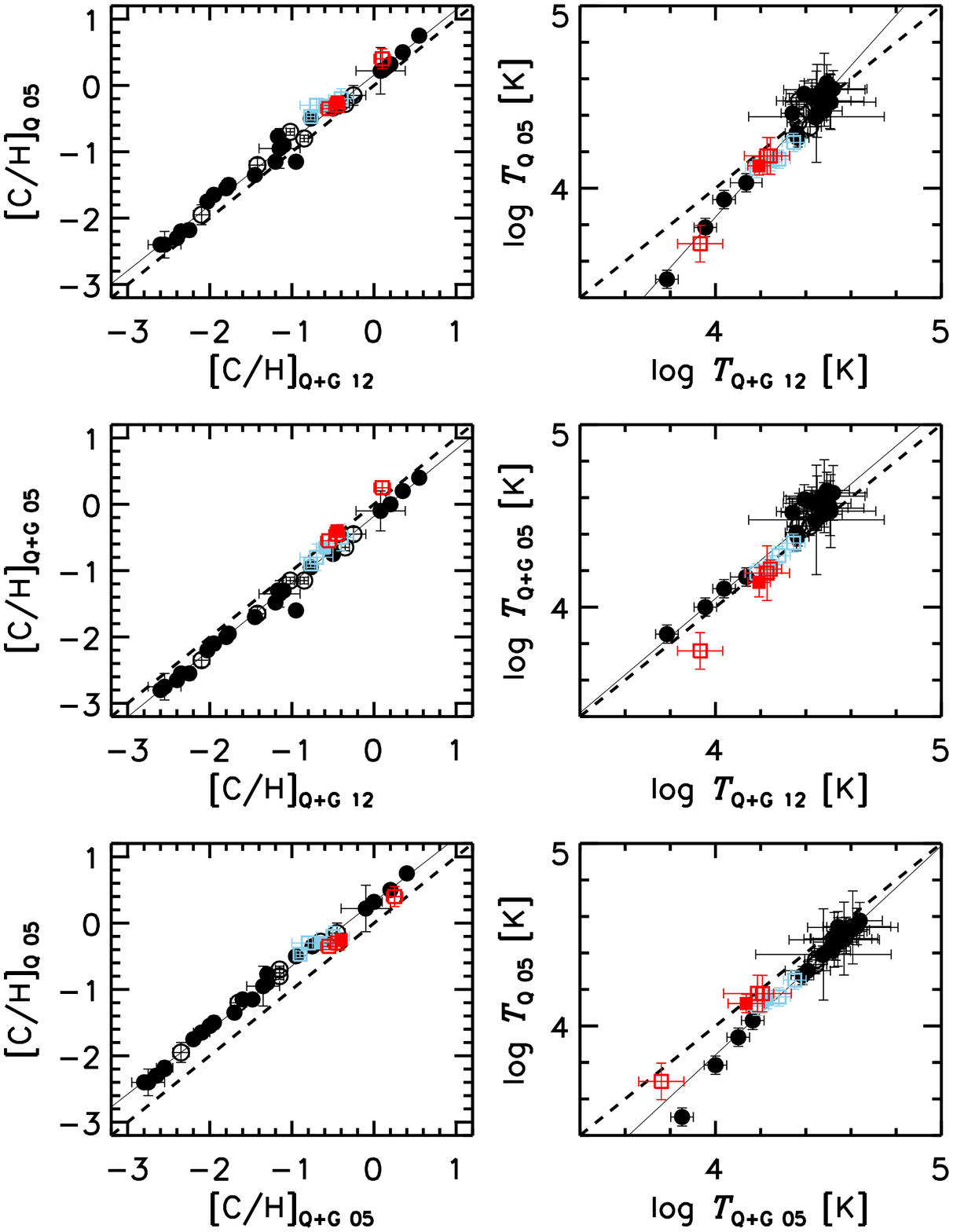}
\caption{Comparisons between the three HM UVBs
for the carbon abundance [C/H] (left panels) and
the temperature (right panels). Red squares indicate 
the 4 \ion{C}{iii} absorbers with \ion{C}{ii}, while
black circles are the \ion{C}{ii}-free absorbers. Sky-blue squares
indicate the 4 \ion{C}{ii} absorbers, when only \ion{C}{iv} and \ion{C}{iii}
are considered without \ion{C}{ii}.
Filled and open symbols are for
the certain and uncertain absorbers listed in Table~\ref{tab_a1}.
Only errors larger than the symbol size are plotted.
The dashed line represents equality in the quantities, and  the
thin solid line is a least-square fit to the results: 
[C/H]$_{\mathrm{Q \, 05}} =  0.16 + 0.98 \times {\mathrm{[C/H]_{Q+G \, 12}}}$,
$\log T_{\mathrm{Q \, 05}} = -1.69 + 1.38 \times {T_{\mathrm{Q+G \, 12}}}$,
[C/H]$_{\mathrm{Q+G \, 05}} = -0.18 + 1.00 \times {\mathrm{[C/H]_{Q+G \, 12}}}$,
$\log T_{\mathrm{Q+G \, 05}} = -0.13 + 1.05 \times {T_{\mathrm{Q+G \, 12}}}$,
[C/H]$_{\mathrm{Q \, 05}} = 0.32 + 0.97 \times {\mathrm{[C/H]_{Q+G \, 05}}}$,
$\log T_{\mathrm{Q \, 05}} = -0.74 + 1.15 \times {T_{\mathrm{Q+G \, 05}}}$,
respectively.
Refer to the online version for clarity.
}
\label{fig14}
\end{figure}

In the following analysis, the plotted error on the Cloudy-predicted
parameters except [C/H] are {\it not real} 1$\sigma$ errors, but just an illustration based
on the uncertainty on $\log U$, i.e. a larger acceptable $\log U$ range 
translates into a larger, {\it assumed} relative error.

\begin{figure*}
\includegraphics[width=175mm]{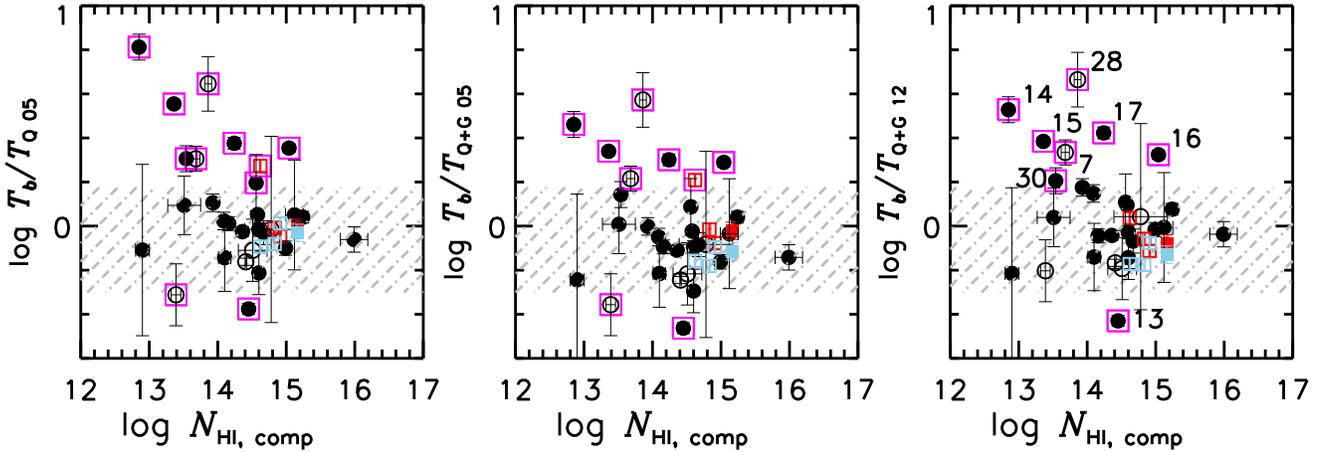}
\caption{Ratio of the VPFIT-derived temperature $T_{b}$ 
based on Eq.~\ref{eq1} to the Cloudy-predicted temperature
$T_{\mathrm{Q \, 05}}$, $T_{\mathrm{Q+G \, 05}}$ and $T_{\mathrm{Q+G \, 12}}$
for the HM Q 2005 (left panel), the HM Q+G 2005 (middle panel)
and the HM Q+G 2012 (right panel)
as a function of
$N_{\mathrm{\ion{H}{i}}}$ of a \ion{H}{i} component. 
Symbols are the same as in Fig.~\ref{fig14}.
Errors on the ratio are based only on the $T_{b}$ errors. 
The gray-shaded region indicates the difference
between $T_{b}$ and Cloudy-derived $T_{\mathrm{PIE}}$ less than 50\%.
Data points embedded in a magenta open square
are the absorbers which have more than 50\% difference between
the two temperatures. In the right panel, a number noted next to the
data points embedded in a magenta open square is an absorber number listed in 
Tables~\ref{tab_a2} and \ref{tab_a3}.} 
\label{fig15}
\end{figure*}

\subsection{The UV backgrounds} 
\label{sec4.5}

Figure~\ref{fig14} presents the comparisons of the predicted parameters [C/H] and 
$T$ between the three UVBs used. 
Any differences in the inferred carbon abundance or the temperature are 
effectively just a small rescaling, with the absorbers
having \ion{C}{ii} being somewhat discrepant in the 
HM Q+G 2005 and the HM Q 2005 cases. When considered only
\ion{C}{iv} and \ion{C}{iii} excluding \ion{C}{ii}, [C/H] and $\log U$ of the 4 
\ion{C}{iv}+\ion{C}{iii}+\ion{C}{ii} absorbers are lower by $\sim$0.2\,dex
and higher by $\sim$0.3\,dex, respectively. Their Cloudy-predicted parameters
(sky-blue squares) are also perfectly on the one-to-one
relation with other \ion{C}{ii}-free absorbers.
Other parameters such as the total hydrogen volume density
and the line-of-sight length also have a similar behaviour. In short, the
predicted physical parameters for one of the UVBs can be roughly
scaled from the ones for the other two UVBs. 
As long as each absorber is exposed to a similar UVB (one of our
assumptions), the evidence for bimodal distributions of 
observed and derived physical 
parameters does not depend on which of the three HM UVBs is used, 
though the absolute values of each derived parameter do.

We note that the 4 \ion{C}{ii} absorbers fit slightly better to the 
HM Q 2005 background than both HM Q+G UVBs, in particular,
the $z = 2.462044$ absorber toward Q0002--422.
However, this does not necessarily imply that all tied \ion{C}{iii}
absorbers are exposed to the UVB similar to the HM Q 2005. 
Considering that the shape of the UVB at high redshifts is not
observationally well constrained, 
all the assumed UVBs are equally adequate for our data within more realistic errors.

\subsection{Validity of the photoionisation modelling}
\label{sec4.6}

\begin{figure}
\vspace{0.5cm}
\hspace{-0.2cm}
\includegraphics[width=88mm]{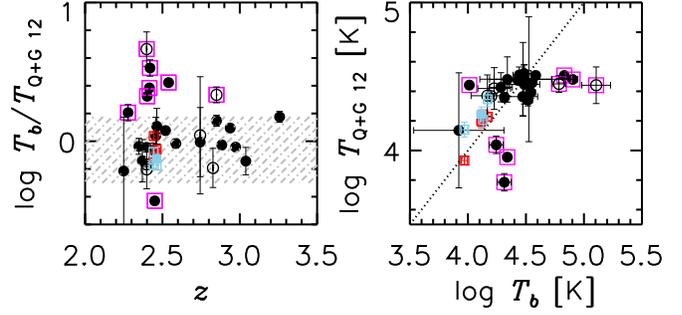}
\vspace{-0.5cm}
\caption{The ratio of $T_{b}$ and the Cloudy-predicted $T_{\mathrm{PIE}}$ 
as a function of $z$ (left panel) and of $T_{b}$ (right panel) for the fiducial HM Q+G 2012. 
All other symbols are the same as in Fig.~\ref{fig15}.}
\label{fig16}
\end{figure}

A way to check whether or not \ion{C}{iii} absorbers
are in photoionisation equilibrium (PIE) 
is to compare the VPFIT-derived gas temperatures $T_{b}$ from Eq.~\ref{eq1} and 
the Cloudy-predicted temperatures. Figures~\ref{fig15} and \ref{fig16}
present the ratio of $T_{b}$ and Cloudy-predicted temperatures $T_{\mathrm{PIE}}$ 
as a function
of $N_{\mathrm{\ion{H}{i}}}$, and of $z$ and $T_{b}$, respectively.

In Fig.~\ref{fig15}, most data points are reasonably close to the
$T_{b}/T_{\mathrm{PIE}} = 1$ line, but 
there are some outliers, with 
about 25\% of the \ion{C}{iii} absorbers having a
temperature difference $T_{b} - T_{\mathrm{PIE}}$ larger than
50\% of the Cloudy model temperature $T_{\mathrm{PIE}}$. 
The absorbers with a large temperature difference 
have a higher $N_{\mathrm{\ion{C}{iv}}}$
than the one with a similar $N_{\mathrm{\ion{H}{i}}}$.
For $\log N_{\mathrm{\ion{H}{i}}} \ge 14$,
in most cases $T_{b}/T_{\mathrm{PIE}} \sim 1$,
implying that the absorbers are roughly 
in PIE. It is not straightforward to draw
any clear conclusions at $\log N_{\mathrm{\ion{H}{i}}} \le 14.0$ as
there are only a few data points, but they
seem to show larger departures from PIE. 

In most cases, \ion{C}{iii} absorbers with a large departure from PIE show
a higher $T_{b}$ than the PIE-predicted $T_{\mathrm{PIE}}$. As the PIE temperature
is determined by photoionisation heating and atomic line cooling in the low-density
limit, a higher observed temperature $T_{b}$ compared to $T_{\mathrm{PIE}}$
strongly suggests that these absorbers are exposed to additional heating source
compared to other absorbers having $T_{b}$ more closer to their PIE 
temperature.

The median temperature ratio is $(T_{b}/T_{\mathrm{Q\, 05}})_{\mathrm{med}} = 1.03$,
$(T_{b}/T_{\mathrm{Q+G\, 05}})_{\mathrm{med}} = 0.92$ and
$(T_{b}/T_{\mathrm{Q+G\, 12}})_{\mathrm{med}} = 0.97$, respectively.
The mean temperature ratio is 
$<\!T_{b}/T_{\mathrm{Q\, 05}}\!>\,= 1.46 \pm 1.26$,
$<\!T_{b}/T_{\mathrm{Q+G\, 05}}\!>\,= 1.11 \pm 0.74$ and
$<\!T_{b}/T_{\mathrm{Q+G\, 12}}\!>\,= 1.31 \pm 0.90$, respectively. 
These values indicate that overall in our sample, the HM Q+G
2012 produces the predicted temperature similar to the observed $T_{b}$,
while the HM Q 2005 
has the largest number of the discrepant absorbers, 10 out of 32 absorbers.
This is why the HM Q+G 2012 was taken as a fiducial UVB. 
However, we note that the HM Q 2005
predicts a better match for the cluster of data points at 
$\log N_{\mathrm{\ion{H}{i}}} \in [14, 15]$.

The left panel of Fig.~\ref{fig16} shows that the VPFIT and Cloudy temperatures differ 
most often at lower $z$, though the small number of components at
$z > 2.6$ means that we cannot draw any firm conclusions.
The right panel of Fig.~\ref{fig16} implies that most absorbers have
$\log T_{\mathrm{Q+G\, 12}} \sim 4.5$ due to the imposed PIE condition at
low [C/H] (see Section~\ref{sec5.2}). Absorbers with a large temperature 
discrepancy tend to have a larger $T_{b}$.

The combining results from $T_{b}/T_{\mathrm{PIE}}$ as a function of  
$N_{\mathrm{\ion{H}{i}}}$ and $z$ suggest that
\ion{C}{iii} component trios
with $\log N_{\mathrm{\ion{H}{i}}} \le 14$ or at lower redshifts are
less likely to be in PIE.

\subsection{The photoionisation modelling with a fixed temperature}
\label{sec4.7}

Since the gas temperature derived from the $b$ parameter is a direct 
estimate from the observations, the Cloudy-predicted PIE temperature
may not be appropriate.
Therefore, another set of the Cloudy grid models was 
run with the fixed constant temperature $T_{b}$ for each \ion{C}{iii} absorber. 
We used only the HM Q+G 2012 for the fixed $T_{b}$ model,
since predicted parameters for a different HM UVB can be roughly
scaled from the HM Q+G 2012 and any correlations between parameters
are not altered significantly due to a different UVB among the three HM UVBs.

\begin{figure}
\hspace{-0.2cm}
\includegraphics[width=87mm]{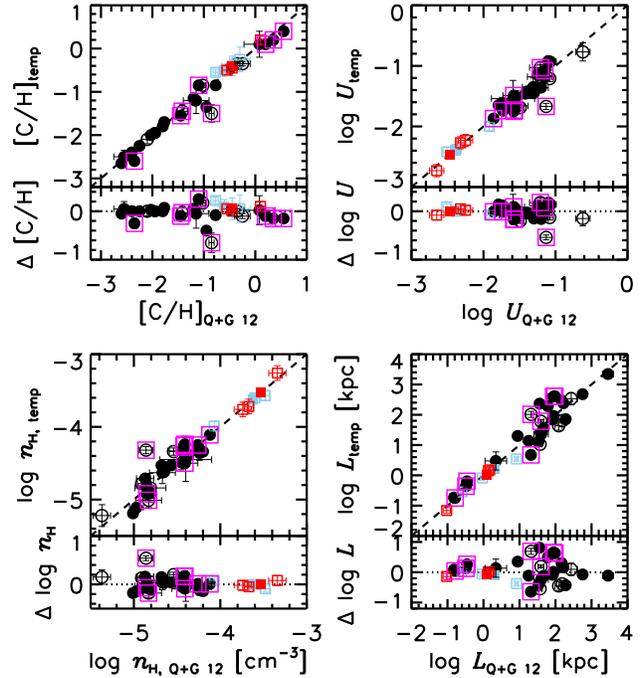}
\vspace{-0.5cm}
\caption{Comparisons between the PIE (x-axis) and the non-PIE (y-axis) models
for the fiducial HM Q+G 2012 UVB. The non-PIE model is set to have
a fixed temperature $T_{b}$
derived by VPFIT based on Eq.~\ref{eq1}. Filled and open symbols
represent the certain and uncertain absorbers. Data points embedded
with a magenta open square are the absorbers for which the Cloudy-predicted
$T_{\mathrm{Q+G \, 12}}$ is 50\% larger or smaller than $T_{b}$.
Plotted errors on [C/H]
are an error based on an allowed [C/H] range, while the ones on
other predicted parameters are based on an allowed $\log U$ range.
The difference between the non-PIE $T_{b}$ and the PIE models is illustrated
in the lower part of each panel. 
All other symbols are the same as in Fig.~\ref{fig14}. 
Refer to the online version for clarity.
}
\label{fig17}
\end{figure}

Figure~\ref{fig17} compares the carbon abundance [C/H], ionisation parameter $U$, 
hydrogen density $n_{\mathrm{H}}$
and line-of-sight length $L$ for PIE and non-PIE (with a fixed temperature $T_{b}$) models.
As expected, the 8 absorbers embedded in a magenta open square 
display larger discrepancies,
while the remaining 24 absorbers do not show any significant difference.
This is due to the fact that for a low-density, low-metallicity PIE gas,
the internal thermal balance is independent of $n_{\mathrm{H}}$ but
sensitive to the relative abundance of ions.

The mean departure from PIE is $<\!\Delta {\mathrm{[C/H]}}\!> \, = -0.04\pm0.16$,
$<\!\Delta \log U\!>\, = -0.02\pm0.14$, $<\Delta \log n_{\mathrm{H}}> \, = 0.02\pm0.14$,
and $<\!\Delta \log L\!>\, = 0.05\pm0.36$, respectively. While the mean departure from
PIE is negligible, the scatter is not. The line-of-sight length 
$L = N_{\mathrm{H}}/n_{\mathrm{H}}$ shows the largest
departure, as $L \propto T^{0.41}$ and other parameters have a smaller $T$ dependence
\citep{schaye01}.

\section{{\bf Derived} properties of the well-aligned 
\ion{H}{i}+\ion{C}{iv}+\ion{C}{iii} components}
\label{sec5}

In this section, we present our main results based
on the non-PIE Cloudy modelling with a fixed temperature $T_{b}$ 
for the HM Q+G 2012 UVB. Whenever necessary, the difference from the PIE model
is also discussed. Predicted parameters for the PIE assumption, i.e.
a free-fit temperature model, are subscripted with ``PIE", while parameters from
the non-PIE $T_{b}$ model are noted with a subscript ``temp". The column density
range for the 32 tied \ion{H}{i}+\ion{C}{iv}+\ion{C}{iii} component trios
is $\log N_{\mathrm{\ion{H}{i}}} \in [12.9, 16.0]$, 
$\log N_{\mathrm{\ion{C}{iv}}} \in [11.8, 13.8]$ and
$\log N_{\mathrm{\ion{C}{iii}}} \in [11.7, 13.8]$, respectively. 
For the 4 absorbers,
\ion{C}{ii} $\lambda$\,1334 or $\lambda$\,1036 or both 
is also detected at $\log N_{\mathrm{\ion{C}{ii}}} \in [12.2,12.7]$.

Note that apart from
the observed column density and $b$ value of \ion{H}{i} and \ion{C}{iv} along with the
gas temperature $T_{b}$, all of the parameters presented in this Section
are Cloudy-predicted. We also stress that despite all the caveats in the 
Cloudy-modelling and a choice of the UVB discussed in Section~\ref{sec4},
our main results are not altered by them.

\subsection{Carbon abundance, $N_{\mathrm{\ion{H}{i}}}$, 
$N_{\mathrm{\ion{C}{iv}}}$ and line-of-sight length}
\label{sec5.1}

\begin{figure*}
\includegraphics[width=170mm]{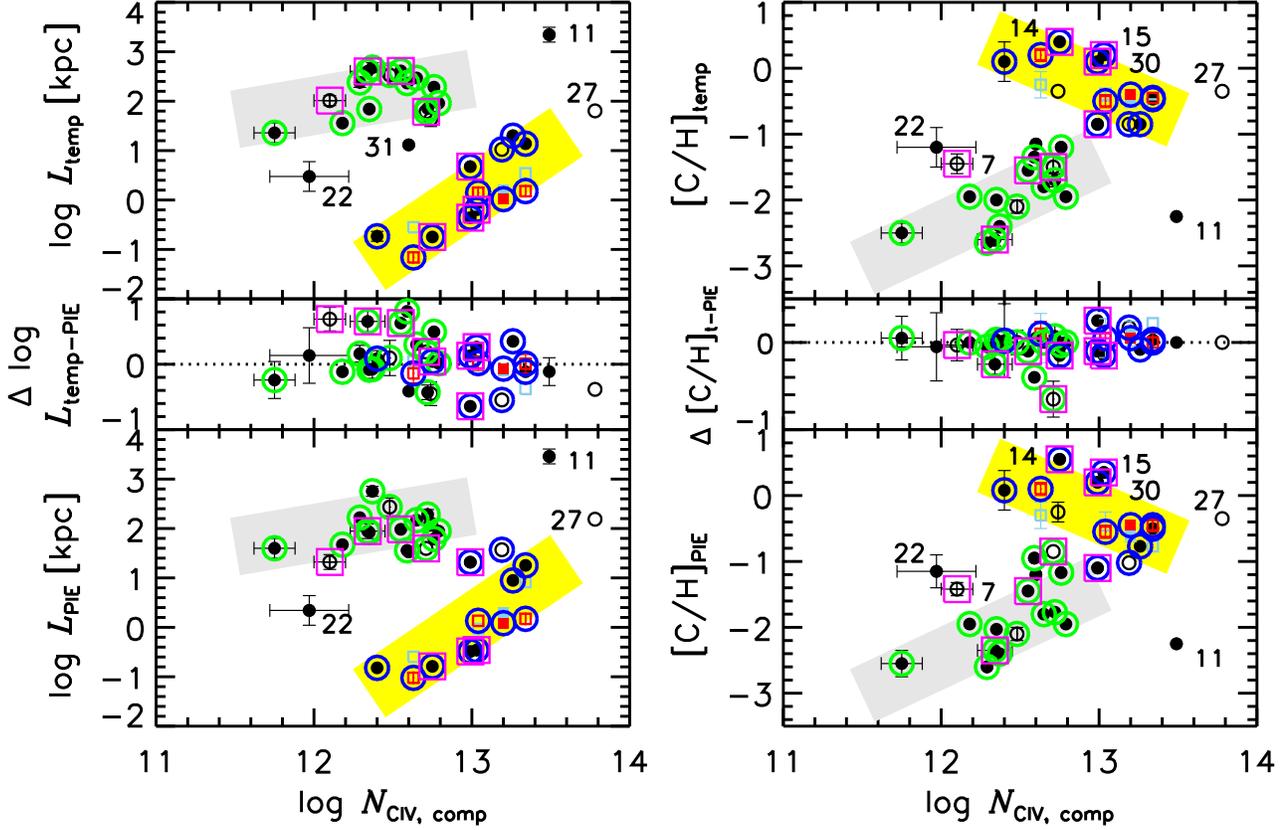}
\caption{{\it Upper panels:} 
the line-of-sight length $L_{\mathrm{temp}}$ (left) and
[C/H]$_{\mathrm{temp}}$ (right) as a function of 
$N_{\mathrm{\ion{C}{iv}}}$
of 32 tied components.
All the symbols are the same as in Fig.~\ref{fig17}.  
Errors on $\log L_{\mathrm{temp}}$ and [C/H]$_{\mathrm{temp}}$ are
the allowed $\log U$ and [C/H]$_{\mathrm{temp}}$ range, respectively,
and are shown only when they are larger than the symbols.
The yellow-shaded (gray-shaded) region represents
the fit to the data points embedded in a blue (green) open circle 
for the non-PIE model.
An absorber number next to the data points is the same one listed in 
Tables~\ref{tab_a2} and \ref{tab_a3}.
{\it Middle panels:} the difference between the non-PIE $T_{b}$ and PIE models
for the HM Q+G 2012.
{\it Lower panels:} for the PIE model. 
The yellow-shaded and gray-shaded regions  
are the same ones as in the upper panels.
Refer to the online version for clarity.
}
\label{fig18}
\end{figure*}

Among the correlations between various observed and predicted parameters,
two of the most unexpected correlations are presented in Fig.~\ref{fig18}.
In the upper left panel, the line-of-sight length,  
$L = N_{\mathrm{H}}/n_{\mathrm{H}}$, is shown as a function of
$N_{\mathrm{\ion{C}{iv}}}$ for the non-PIE model. The lower left panel presents 
the same parameter space for the PIE model, with
the middle left panel displaying the difference between the two models.
As $L$ is dependent on $T^{0.41}$ \citep{schaye01}, which is also dependent on
$U$ and [C/H], there is a larger difference between the non-PIE $T_{b}$ 
and PIE models shown in the middle left panel.

In both left panels, despite a few interlopers,
the $N_{\mathrm{\ion{C}{iv}}}$--$L$ plane can be clearly
divided into two regions, suggesting that our optically-thin
\ion{C}{iii} absorbers may consist of two distinct populations.
14 absorbers embedded in a green open circle have
$L_{\mathrm{temp}} = 20$--480\,kpc,
with the median $L_{\mathrm{temp}}$ of
233\,kpc. For these absorbers, $L_{\mathrm{temp}}$ is a weak
function of $N_{\mathrm{\ion{C}{iv}}}$. On the other hand,
12 absorbers embedded in a blue open circle
have $L_{\mathrm{temp}} \le 20$\,kpc, showing that 
$L_{\mathrm{temp}}$ increases rapidly 
with $N_{\mathrm{\ion{C}{iv}}}$ at  $\log N_{\mathrm{\ion{C}{iv}}} \in [12.2, 13.5]$.
The yellow-shaded (gray-shaded) region illustrates a least-square fit to 
the data points embedded in a blue (green) open circles, and 
the fit results are tabulated in Table~\ref{tab3}. 

A similar behaviour is seen in the [C/H]--$N_{\mathrm{\ion{C}{iv}}}$ relation in
the upper and lower right panels of Fig.~\ref{fig18}, where there is a
segregation of \ion{C}{iii} absorbers into two distinct regions as a high- or a low-metallicity
branch, divided
at [C/H]$_{\mathrm{temp}} \! \sim \! -1.0$. 
Again, the yellow-shaded (gray-shaded) region shows a
fit to the data points embedded in a blue (green) open circles as in the left panels, 
with the fit results listed in Table~\ref{tab4}. 

Most \ion{C}{iii} absorbers having $L_{\mathrm{temp}} \! \sim \! 200$\,kpc have
[C/H]$_{\mathrm{temp}} \le -1.0$, which increases with $N_{\mathrm{\ion{C}{iv}}}$.
As noted in Tables~\ref{tab_a2} and \ref{tab_a3},
none of these low-metallicity branch absorbers is associated with \ion{Si}{iv},
which is usually found with \ion{C}{iv} at  
$\log N_{\mathrm{\ion{C}{iv}}} \ge 13.0$ \citep{shen13}. These low-metallicity branch
absorbers have
a saturated \ion{H}{i} Ly$\alpha$ absorption profile with a weak \ion{C}{iv}. 
The \ion{C}{iii} components at
$v = 0$\,\kms\/ in the second panel of Fig.~\ref{fig1} provides a good
example of low-metallicity branch absorbers.

Absorbers with a higher [C/H]$_{\mathrm{temp}}$ show an opposite trend in that
[C/H]$_{\mathrm{temp}}$ anti-correlates with $N_{\mathrm{\ion{C}{iv}}}$.
High-metallicity branch absorbers further seem to be grouped into two subclasses,
1) part of a multi-component,
strong \ion{C}{iv} complex with a saturated \ion{H}{i} Ly$\alpha$ or 2) associated
with a strong \ion{C}{iv} and unsaturated \ion{H}{i} Ly$\alpha$. 
Most of the former subclass, complex high-metallicity branch absorbers, 
are associated with \ion{Si}{iv}. The \ion{C}{iii} components at
$v = +52$ and $v = -75$\,\kms\/ in the right panel of Fig.~\ref{fig1} are an 
example of high-metallicity branch absorbers, which is part of a \ion{C}{iv} complex.
A good example of the latter subclass, simple high-metallicity branch absorbers, 
is shown in the left
panel of Fig.~\ref{fig1}. However, on top of a small-number statistics, 
simple high-metallicity branch absorbers
sometimes occur in the vicinity of complex high-metallicity branch absorbers. 
Therefore, our data cannot clearly distinguish the two subclasses, 
however, suggest that high-metallicity branch absorbers could be
a mixed bag of more kinematically complex absorbers than low-metallicity branch ones.
 
In Fig.~\ref{fig18}, 4 sky-blue circles are \ion{C}{ii} absorbers
when $N_{\mathrm{\ion{C}{ii}}}$ is excluded, i.e. \ion{C}{ii} absorbers
are treated as \ion{C}{ii}-free absorbers. No significant
difference is found between the two subsamples, in that no \ion{C}{ii} 
high-metallicity branch absorbers become low-metallicity branch absorbers
even if \ion{C}{ii} was ignored. This implies that
the overall observational trends would hold, 
even if some of upper-limit-$N_{\mathrm{\ion{C}{ii}}}$ or blended-\ion{C}{ii}
absorbers are in fact associated with \ion{C}{ii}. Since a similar behaviour
is found for other correlations, no further consideration on possibly missed
\ion{C}{ii} above the \ion{C}{ii} detection limit is included. However, we
present the Cloudy-predicted parameters of 4 \ion{C}{ii} absorbers when
excluded \ion{C}{ii} as sky-blue symbols if appropriate.  

\begin{figure}
\hspace{-0.2cm}
\includegraphics[width=86mm]{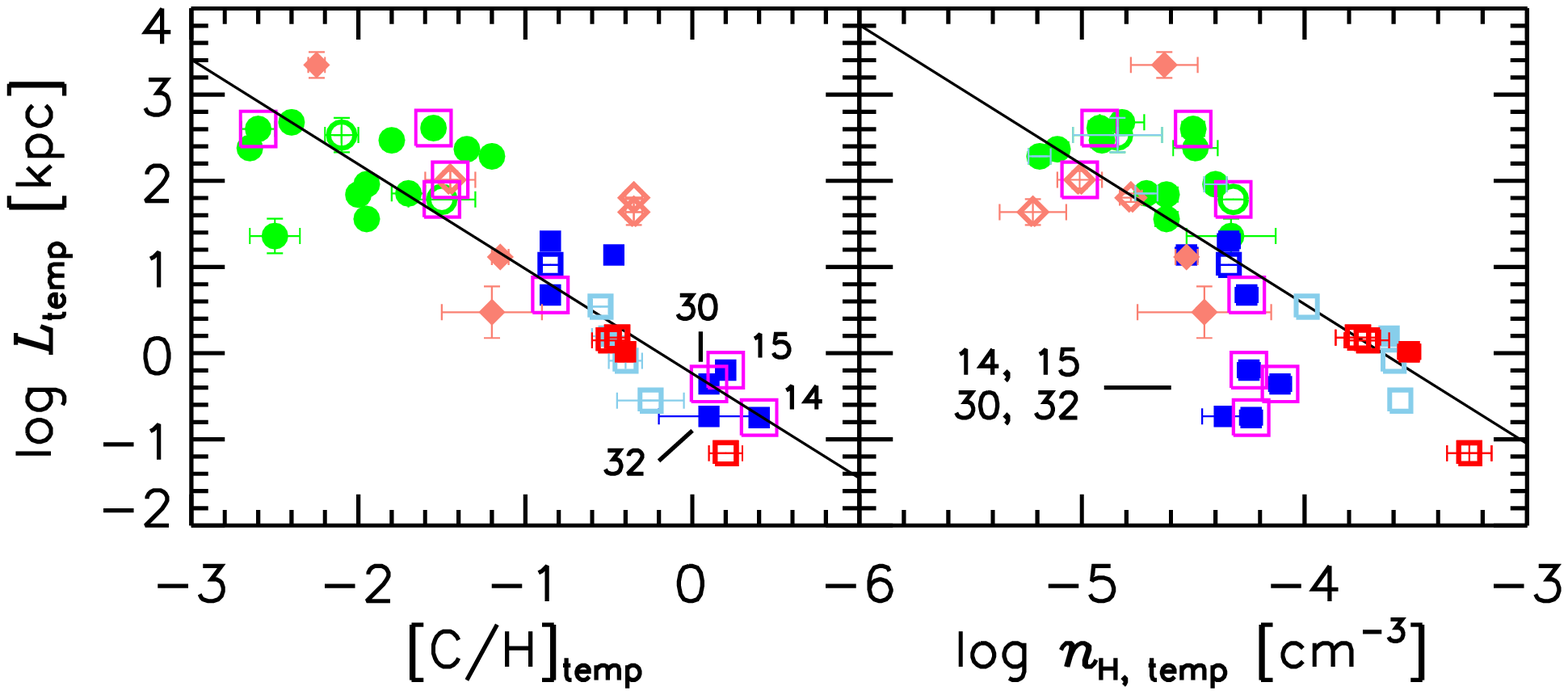}
\vspace{-0.5cm}
\caption{{\it Left panel:} the line-of-sight length $L_{\mathrm{temp}}$ 
as a function of [C/H]$_{\mathrm{temp}}$. 
Green circles and blue squares indicate low-metallicity and high-metallicity 
branch absorbers, with filled and open symbols being
for certain and uncertain absorbers. \ion{C}{iii} absorbers which do not clearly
belong to either branch are shown as filled and open orange diamonds. 
Red squares are for \ion{C}{ii} absorbers
(belonging to high-metallicity branch absorbers), 
while sky-blue squares are the
same \ion{C}{ii} absorbers when \ion{C}{ii} is excluded in the Cloudy modelling.
The solid line is 
a least-square fit to the full sample,
$\log L_{\mathrm{temp}} = -0.24 - 1.21 \times {\mathrm{[C/H]_{temp}}}$.
{\it Right panel:} $L_{\mathrm{temp}}$ as a function
of a total (neutral + ionised) hydrogen volume density
$n_{\mathrm{H, \, temp}}$. The least-square fit is
$\log L_{\mathrm{temp}} = -5.91 -1.62 \times \log n_{\mathrm{H, \, temp}}$.
Refer to the online version for clarity.
}
\label{fig19}
\end{figure}

For high-metallicity branch absorbers, as there are correlations
between [C/H]$_{\mathrm{temp}}$ and 
$N_{\mathrm{\ion{C}{iv}}}$ as well as
$L_{\mathrm{temp}}$ and $N_{\mathrm{\ion{C}{iv}}}$, a correlation between 
[C/H]$_{\mathrm{temp}}$ and
$L_{\mathrm{temp}}$ is also expected. This is shown in the left panel of
Fig.~\ref{fig19}. Although overall
an anti-correlation can be assumed, low-metallicity branch absorbers show 
a large spread of 1\,dex in $\log L_{\mathrm{temp}}$ for a given 
[C/H]$_{\mathrm{temp}}$ at [C/H]$_{\mathrm{temp}} \le -1.0$. Due to the large scatter,  
low-metallicity branch absorbers
at [C/H]$_{\mathrm{temp}} \le -1.0$ can be viewed as having almost a constant
$L_{\mathrm{temp}}$ at $L_{\mathrm{temp}} \sim 200$\,kpc, in good agreement 
with the conclusion 
from the $N_{\mathrm{\ion{C}{iv}}}$--$L_{\mathrm{temp}}$ relation.
The 4 \ion{C}{ii}-free super-solar-[C/H]$_{\mathrm{temp}}$ absorbers
including the 3 absorbers with a highest $T_{b} - T_{\mathrm{PIE}}$ value 
have a smallest $L_{\mathrm{temp}} \le 1$\,kpc. They are all simple
high-metallicity branch absorbers, characterised by a stronger \ion{C}{iv} than
typical \ion{H}{i} absorbers with a similar $N_{\mathrm{\ion{H}{i}}}$. 

The right panel of Fig.~\ref{fig19} presents $L_{\mathrm{temp}}$
as a function of total (neutral + ionised) hydrogen volume density
$n_{\mathrm{H, \, temp}}$. Low-metallicity branch absorbers have 
a lower $\log n_{\mathrm{H, \, temp}} \in [-5.2, -4.3]$ with a mean of
$<\!\log n_{\mathrm{H, \, temp}}\!> \, = -4.70\pm0.28$, slightly higher than the cosmic
mean number density at $z \sim 2.4$. High-metallicity branch absorbers have a higher 
$n_{\mathrm{H, \, temp}}$ ranged at $\log n_{\mathrm{H, \, temp}} \in [-4.5, -3.3]$.
Compared to the warm ionised gas in the Milky Way with a comparable temperature range,
our \ion{C}{iii} absorbers have about two orders of magnitude lower $n_{\mathrm{H, \, temp}}$
\citep{haffner09}.
The 4 \ion{C}{ii}-free
absorbers with $L_{\mathrm{temp}} \le 1$\,kpc, marked with an absorber number next
to the data points, have a smaller $L_{\mathrm{temp}}$ than
expected from the fit for $\log n_{\mathrm{H, \, temp}} \sim -4.2$.

\begin{table}
\caption{Least-square fit coefficients for the non-PIE $L_{\mathrm{temp}}$ as a function of
$N_{\mathrm{\ion{C}{iv}}}$ and $N_{\mathrm{\ion{H}{i}}}$:
$\log L_{\mathrm{temp}} = A + B \log N_{X}$. The parameters are
associated with the left panel of Figs.~\ref{fig18} and \ref{fig24}.}
\label{tab3}
\begin{tabular}{lrrl}

\hline
\noalign{\smallskip}
ion $X$ & $A$ & $B$ & Branch \\

\hline

\ion{C}{iv} (yellow) & $-28.11\pm1.88$ & $2.17\pm0.52$ & High-metallicity\\
\ion{C}{iv} (gray) & $-4.67\pm1.42$ & $0.55\pm0.40$ & Low-metallicity\\
\\[-0.2cm]

\ion{H}{i} (all) & $-10.71\pm0.95$ & $0.84\pm0.25$ & \\
\ion{H}{i} (yellow) & $-6.86\pm0.91$ & $0.50\pm0.24$ & High-metallicity\\
\ion{H}{i} (gray) & $-0.78\pm1.01$ & $0.20\pm0.26$ & Low-metallicity\\

\hline

\end{tabular}
\end{table}

\begin{table}
\caption{Least-square fit coefficients for the non-PIE [C/H]$_{\mathrm{temp}}$ 
as a function of
$N_{\mathrm{\ion{C}{iv}}}$ and $N_{\mathrm{\ion{H}{i}}}$:
[C/H]$_{\mathrm{temp}} = A + B \log N_{X}$.
The parameters are
associated with the right panel of Figs.~\ref{fig18} and \ref{fig24}.}
\label{tab4}
\begin{tabular}{lrrl}
\hline
\noalign{\smallskip}
ion $X$ & $A$ & $B$ & Branch \\

\hline

\ion{C}{iv} (yellow) & $13.19\pm1.30$ & $-1.04\pm0.36$ 
   & High-metallicity \\
\ion{C}{iv} (gray) & $-16.01\pm1.19$ & $1.13\pm0.34$ 
    &  Low-metallicity \\
\\[-0.2cm]

\ion{H}{i} (all) & $10.04\pm0.72$ & $-0.78\pm0.19$ & \\
\ion{H}{i} (yellow) & $5.39\pm0.50$ & $-0.40\pm0.13$ & High-metallicity \\
\ion{H}{i} (gray) & $11.39\pm0.59$ & $-0.91\pm0.15$ & Low-metallicity \\
\\[-0.2cm]

\hline

\end{tabular}
\end{table}

\begin{figure*}
\hspace{-0.2cm}
\includegraphics[width=180mm]{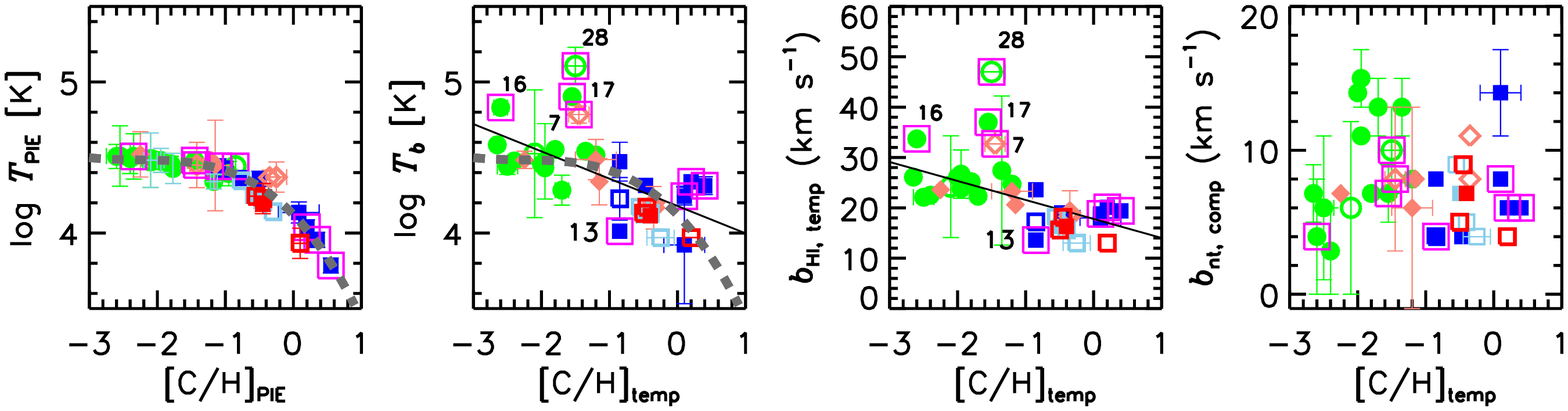}
\caption{{\it Left panel:} 
the PIE temperature $T_{\mathrm{PIE}}$ versus 
[C/H]$_{\mathrm{PIE}}$ for the fiducial HM Q+G 2012. The overlaid 
dark gray dashed
curve is a polynomial fit to the temperature vs metallicity,
$T_{\mathrm{PIE}} = 4.12 - 0.46 \times {\mathrm{[C/H]_{PIE}}}
-0.20 \times {\mathrm{[C/H]_{PIE}}}^{2} -0.03 
\times {\mathrm{[C/H]_{PIE}}}^{3}$.
Plotted errors on [C/H]$_{PIE}$
are an error based on an allowed [C/H] range, while the ones on
other predicted parameters are based on an allowed $\log U$ range.
All the symbols are the same as in Fig.~\ref{fig19}.
{\it Second panel:} the predicted carbon abundance [C/H]$_{\mathrm{temp}}$ 
for the non-PIE model with a fixed temperature $T_{b}$ versus $T_{b}$.
The solid line
is the least-square-fit to the data: $\log T_{b} = (4.18\pm0.06)  + (-0.18\pm0.04) 
\times \mathrm{[C/H]_{temp}}$.
The dark gray dashed curve is the PIE relation shown in the left panel.
{\it Third panel:} the observed total (thermal and non-thermal) line width of \ion{H}{i}
as a function of [C/H]$_{\mathrm{temp}}$. The solid line is the least-square
fit: $b_{\mathrm{\ion{H}{i}}} = (17.80\pm1.78)  +(-3.77\pm0.76) \times \mathrm{[C/H]_{temp}}$.
{\it Right panel:} the derived non-thermal line width $b_{\mathrm{nt}}$ 
as a function of [C/H]$_{\mathrm{temp}}$.
Refer to the online version for clarity.
}
\label{fig20}
\end{figure*}

\subsection{Carbon abundance, 
temperature, $b_{\mathrm{\ion{H}{i}}}$,
$b_{\mathrm{nt}}$, $N_{\mathrm{\ion{C}{iv}}}/N_{\mathrm{\ion{H}{i}}}$}  
\label{sec5.2}

Relations between the carbon abundance [C/H]$_{\mathrm{temp}}$ 
and various parameters are shown in Fig.~\ref{fig20}. The left panel
illustrates a PIE gas for the HM
Q+G 2012 background. For a low-density gas in the presence of a UVB, 
the ratio of ionic abundances does not depend 
on the metallicity at less than $10^{-1}$ solar, i.e. [C/H] does not play a 
significant role in the internal ionisation structure. 
A gas at higher temperature cools to an equilibrium temperature 
$T_{\mathrm{eq}}$ and a gas at lower temperature gets heated to 
$T_{\mathrm{eq}}$. This temperature for the HM Q+G 2012 is 
$\log T_{\mathrm{eq}} \sim 4.5$, as clearly seen in the left panel.
As expected, low-metallicity branch absorbers marked as green symbols
have a temperature around $T_{\mathrm{eq}}$.
When the metallicity is higher than $10^{-1}$ solar, the 
radiative metal line cooling becomes important, which
decreases $T_{\mathrm{eq}}$. A slight departure from the general
trend by \ion{C}{ii}-enriched absorbers is due to their higher $n_{\mathrm{H}}$.

In the second panel of Fig.~\ref{fig20}, the relation between 
[C/H]$_{\mathrm{temp}}$ and $T_{b}$ is not as clear as in the left panel,
due to the non-PIE condition. A large departure from the 
[C/H]$_{\mathrm{PIE}}$--$T_{\mathrm{PIE}}$ relation occurs for 
\ion{C}{iii} absorbers with a large temperature difference between
$T_{b}$ and $T_{\mathrm{PIE}}$. The higher $T_{b}$
than $T_{\mathrm{eq}}$ for all but one of these absorbers suggests that they have a
higher heating rate, i.e. experiencing a recent cooling 
from a hotter gas or exposed to 
additional radiation. 

The observed total (thermal and non-thermal) $b_{\mathrm{\ion{H}{i}}}$
increases weakly as [C/H]$_{\mathrm{temp}}$ decreases, as seen in
the third panel. Naturally, absorbers with a higher $T_{b}$ tend to have
a higher $b_{\mathrm{\ion{H}{i}}}$, as the non-thermal contribution to
the observed line width is not significant. 
There is no clear trend between
[C/H]$_{\mathrm{temp}}$ and non-thermal motion $b_{\mathrm{nt}}$ in
the right panel. 

\begin{figure}
\hspace{-0.2cm}
\includegraphics[width=88mm]{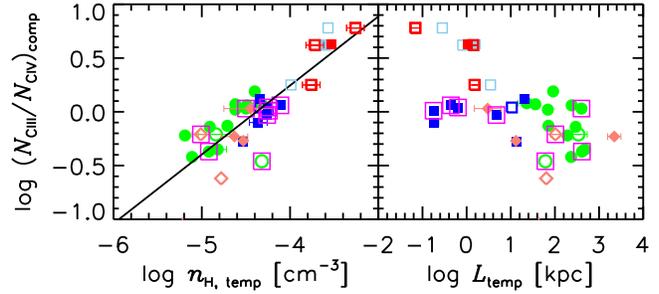}
\caption{The ratio of observed \ion{C}{iii} column
density and \ion{C}{iv} column density as a function of the
total hydrogen volume density (left panel) and  
$L_{\mathrm{temp}}$ (right panel). 
Symbols are the same as in Fig.~\ref{fig20}.
The least-square fit in the left panel is 
$\log N_{\mathrm{\ion{C}{iii}}}/N_{\mathrm{\ion{C}{iv}}} = 
2.81 + 0.64 \times \log n_{\mathrm{H, \, temp}}$.
Refer to the online version for clarity.
}
\label{fig21}
\end{figure}

The left panel of Fig.~\ref{fig21} implies that 
$N_{\mathrm{\ion{C}{iii}}}/N_{\mathrm{\ion{C}{iv}}}$ increases with 
the total hydrogen volume density. On the other hand, the right panel shows
that the line-of-sight length of high-metallicity branch absorbers is mostly
independent of $N_{\mathrm{\ion{C}{iii}}}/N_{\mathrm{\ion{C}{iv}}}$ at
$L_{\mathrm{temp}} \le 20$\,kpc. Low-metallicity branch
absorbers at $L_{\mathrm{temp}} \ge 20$\,kpc tend to be
$N_{\mathrm{\ion{C}{iii}}}/N_{\mathrm{\ion{C}{iv}}} \le 1$, simply because
that they have $\log N_{\mathrm{\ion{C}{iv}}} \le 13$
(Fig.~\ref{fig6}). When combined 
the left panel with the right one, low-metallicity branch absorbers tend to have
a lower $n_{\mathrm{H,\,temp}}$ at a larger $L_{\mathrm{temp}}$,
a good agreement with what simulations have predicted 
\citep{cen11, voort11, shen13}.

\subsection{Redshift evolution of carbon abundance}
\label{sec5.3}

\begin{figure}
\hspace{-0.2cm}
\includegraphics[width=88mm]{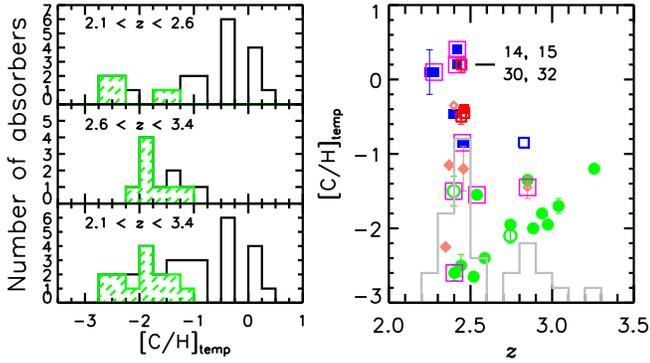}
\caption{{\it Left panel:} the number of absorbers as a function of 
[C/H]$_{\mathrm{temp}}$ at the three different redshift bins. 
The green-shaded region is for low-metallicity
branch \ion{C}{iii} absorbers. 
{\it Right panel:}
the redshift distribution of [C/H]$_{\mathrm{temp}}$. 
The overlaid gray histogram based at $-3$ in the y-axis
is the number
of absorbers in $\Delta z = 0.1$ bins.
Symbols are the same as in Fig.~\ref{fig21}.
}
\label{fig22}
\end{figure}

Despite the small number of absorbers at $z > 2.6$, we checked whether there is
noticeable redshift evolution in the [C/H]$_{\mathrm{temp}}$
distribution. 
Figure~\ref{fig22} shows the redshift distribution of 
[C/H]$_{\mathrm{temp}}$, and, to represent it in another way,
the number of \ion{C}{iii} absorbers as a function of [C/H]$_{\mathrm{temp}}$
for redshift ranges $2.1 < z < 2.6$ and $2.6 < z < 3.4$ as well as
for the whole sample. 

At $z \! \sim \! 2.4$ in the upper left panel, the \ion{C}{iii} absorbers display a
large range of the carbon abundance at 
[C/H]$_{\mathrm{temp}} \in [-2.7, 0.4]$. The mean 
[C/H]$_{\mathrm{temp}}$ is $-1.0\pm1.0$. 
On the other hand,
at $z \! \sim \! 2.9$, the \ion{C}{iii} absorbers show a smaller  
[C/H]$_{\mathrm{temp}}$ range at  
[C/H]$_{\mathrm{temp}} \in [-2.1, -0.9]$, 
with $<\!{\mathrm{[C/H]_{temp}}}\!> \,
= -1.6\pm0.4$. At both redshifts, the [C/H]$_{\mathrm{temp}}$ distribution
is not Gaussian. The two-sided Kolmogorov-Smirnov statistic gives
the significance level of 0.02, implying that the [C/H]$_{\mathrm{temp}}$
distribution at the two redshift ranges is significantly different.

At $z \sim 2.9$, a lack of data points at [C/H]$_{\mathrm{temp}} \le -2$
is probably due to lower-S/N data and a smaller number of
sightlines. However, a lack of absorbers at [C/H]$_{\mathrm{temp}} \in [-0.8, 0.4]$
is not an observational bias, since these absorbers have a stronger
\ion{C}{iv} compared to a typical absorber with a similar $N_{\mathrm{\ion{H}{i}}}$,
thus easy to detect. 
At $z \sim 2.4$, high-metallicity branch absorbers account for about 50\% of 
\ion{C}{iii} absorbers.
If a relative fraction of [C/H]$_{\mathrm{temp}}$ is
similar at both redshift ranges,  $5\pm2$ absorbers with
[C/H]$_{\mathrm{temp}} \in [-0.8, 0.4]$ are expected at $z \sim 2.9$.
Therefore, this lack of high-metallicity branch absorbers at $z \sim 2.9$ is probably real. 
In short, the [C/H]$_{\mathrm{temp}}$ distribution is more skewed toward
a higher [C/H]$_{\mathrm{temp}}$ at lower $z$, implying a metallicity
evolution.

The simulated temperature distribution by \citet{cen11} is similar to
the observed one shown in Section~\ref{sec3.3}. The
same simulation predicts that 
for $\log N_{\mathrm{\ion{C}{iv}}} \in [12, 14]$, the metallicity distribution
has a narrow (broad) Gaussian distribution covering 
$-3.0\! \sim \!-1.0$ ($-3.0 \!\sim \!-0.5$) with a peak at $-2.1$ ($-2.0$) at $z = 4$ 
($z=2.6$). This is roughly consistent with the Cloudy-predicted
[C/H] distribution,
except that the same simulation does not reproduce absorbers
with [C/H] $\ge -0.5$ at $\log N_{\mathrm{\ion{C}{iv}}} \in [12, 14]$.
A similar discrepancy is also seen in Figs. 1 and 3 by \citet{shen13}. 
Considering that the line-of-sight length of 
\ion{C}{iii} absorbers with [C/H]$_{\mathrm{temp}} \ge -0.5$ 
is less than 1\,kpc (Fig.~\ref{fig19}), 
the discrepancy may be
due to an insufficient resolution to resolve these absorbers in
simulations.

\begin{figure}
\includegraphics[width=86mm]{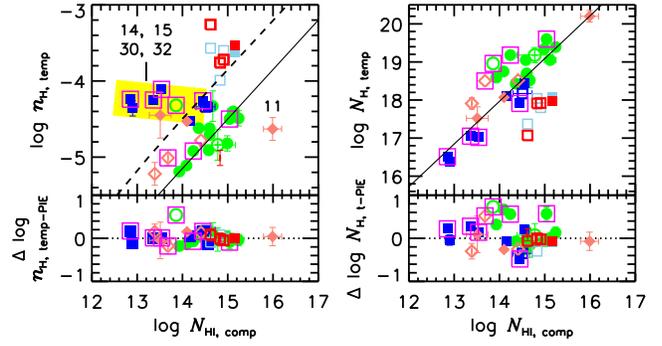}
\caption{{\it Upper panels:} the total hydrogen density $n_{\mathrm{H_{, \, temp}}}$
as a function of $N_{\mathrm{\ion{H}{i}}}$ (left panel) and
the total hydrogen column density ($N_{\mathrm{\ion{H}{i}}} + 
N_{\mathrm{\ion{H}{ii}}}$) as a function of
$N_{\mathrm{\ion{H}{i}}}$ (right panel), respectively.
All the symbols are the same as in Fig.~\ref{fig21}.
In the left panel, the dashed line indicates the Schaye 
$N_{\mathrm{\ion{H}{i}}}$--$n_{\mathrm{H}}$
relation at $z = 2.4$, 
$n_{\mathrm{H, \, temp}} = -13.65 + 0.65 \times N_{\mathrm{\ion{H}{i}}}$,
while the solid line is the Schaye relation shifted down by 0.6\,dex to match 
low-metallicity branch absorbers noted as green symbols.
The yellow-shaded region indicates a fit to the high-metallicity branch absorbers
when 4 absorbers with $\log n_{\mathrm{H}} \ge -4$ (\ion{C}{ii}-enriched
absorbers) are excluded,
$\log n_{\mathrm{H, \, temp}} = -3.49 - 0.06  \times \log N_{\mathrm{\ion{H}{i}}}$.
In the right panel, the solid line shows the least-square fit:
$\log N_{\mathrm{H, \, temp}} = 2.25 + 1.12  \times \log N_{\mathrm{\ion{H}{i}}}$.
{\it Lower panels:} the difference between the non-PIE $T_{b}$ 
and PIE models for the HM Q+G 2012 UVB.
Refer to the online version for clarity.
}
\label{fig23}
\end{figure}

\subsection{Total hydrogen volume density, total hydrogen column density
and $N_{\mathrm{\ion{H}{i}}}$}
\label{sec5.4}

Correlations between the observed \ion{H}{i} column density, the total
hydrogen volume density and the total hydrogen column density are 
presented in the upper panels of Fig.~\ref{fig23}. In the upper left panel, 
the dashed line 
shows the $N_{\mathrm{\ion{H}{i}}}$--$n_{\mathrm{H}}$ relation
derived by \citet{schaye00b} for a self-gravitating,
photoionised \ion{H}{i} gas in local hydrostatic equilibrium. 
The Schaye $N_{\mathrm{\ion{H}{i}}}$--$n_{\mathrm{H}}$ relation
requires the \ion{H}{i} photoionisation rate, the gas
temperature and the gas mass fraction. We used the \ion{H}{i} photoionisation 
rate $\Gamma = 9.575 \times 10^{-13}$ sec$^{-1}$
for our fiducial HM Q+G 2012 UVB at $z = 2.4$.
The gas temperature was calculated using Cloudy as a function
of $N_{\mathrm{\ion{H}{i}}}$ for the HM Q+G 2012 at $z = 2.4$, 
while the gas mass fraction was used as his default
value. 

Unlike the theoretical prediction, the observed and predicted 
parameters indicate 
no well-defined relation between $n_{\mathrm{H, \, temp}}$ and
$N_{\mathrm{\ion{H}{i}}}$ for the full sample. 
This departure from the Schaye relation is not due to the
fact that the observed data cover an extended redshift range
$2.1 < z < 3.4$, while the calculated Schaye relation
is at $z = 2.4$, since input parameters do not change significantly
at $2.1 < z < 3.4$. Nor is it due to the imposed non-PIE condition, since the
difference between the non-PIE and PIE models is negligible.

If the 4 \ion{C}{ii}-enriched 
absorbers at $\log n_{\mathrm{H}} \ge -4$ are excluded, the 
$N_{\mathrm{\ion{H}{i}}}$--$n_{\mathrm{H}}$ plane is reminiscent
of the $N_{\mathrm{\ion{C}{iv}}}$--[C/H]$_{\mathrm{temp}}$ plane
in Fig.~\ref{fig18}. The total hydrogen density $n_{\mathrm{H}}$
of low-metallicity branch absorbers (green symbols) 
increases with $N_{\mathrm{\ion{H}{i}}}$,
while $n_{\mathrm{H}}$ of high-metallicity branch absorbers in the 
yellow-shaded region is almost independent of $N_{\mathrm{\ion{H}{i}}}$. 
Interestingly, if the Schaye relation is shifted down by 0.6\,dex, 
it matches low-metallicity branch absorbers.
The shift in the normalisation from the Schaye equation 
can be caused by many factors, such as an incorrect temperature dependence. 
For example, the metal cooling in the gas could change the internal ionisation and
thermal structure, while the Schaye relation does not account for any
metals associated with the forest.

A group of absorbers at
$(\log N_{\mathrm{\ion{H}{i}}}, \log n_{\mathrm{H, \, temp}}) \sim (13.2, -4.3)$,
marked with an absorber number, is simple high-metallicity branch absorbers
characterised by a higher $N_{\mathrm{\ion{C}{iv}}}$ rather than typical \ion{H}{i} with
a similar $N_{\mathrm{\ion{H}{i}}}$.
If these absorbers are excluded, high-metallicity branch absorbers 
including 4 \ion{C}{ii} absorbers in red roughly follow
the Schaye relation. In fact, \ion{C}{iii} absorbers can 
be viewed to be confined by the dashed (the Schaye relation)
and solid lines on the 
$\log N_{\mathrm{\ion{H}{i}}}$--$\log n_{\mathrm{H, \, temp}}$ plane.
This suggests that overall the basic assumptions of the 
Schaye relation are not far-fetched despite being too simplistic and 
that the individual absorber differs from
each other, having an intrinsic scatter in the gas temperature, the gas density
and the metallicity even at the same overdensity.

A non-PIE condition introduces a large scatter on the
$N_{\mathrm{\ion{H}{i}}}$--$N_{\mathrm{H, \, temp}}$ plane in the
upper right panel, although a general trend exists. A large scatter 
at $\log N_{\mathrm{\ion{H}{i}}} \sim 14.5$ is due to the large scatter in 
$n_{\mathrm{H}}$ for the same $N_{\mathrm{\ion{H}{i}}}$ range.
Due to a higher $n_{\mathrm{H, \, temp}}$, the 4 \ion{C}{ii}-enriched 
absorbers
have a lower $N_{\mathrm{H}}$ compared to \ion{C}{ii}-free
absorbers with a similar $N_{\mathrm{\ion{H}{i}}}$. Again, there is a suggestion
that high-metallicity and low-metallicity branch absorbers follow a different
scaling relation.
The difference between
the non-PIE and PIE models 
is $<\!\Delta \log N_{\mathrm{H}}\!> \, = 0.07\pm0.30$, implying that
on average there is no significant systematic
difference in $N_{\mathrm{H}}$ between 
PIE and non-PIE models.

\subsection{Carbon abundance, line-of-sight length and $N_{\mathrm{\ion{H}{i}}}$}
\label{sec5.5}

\begin{figure}
\hspace{-0.2cm}
\includegraphics[width=86mm]{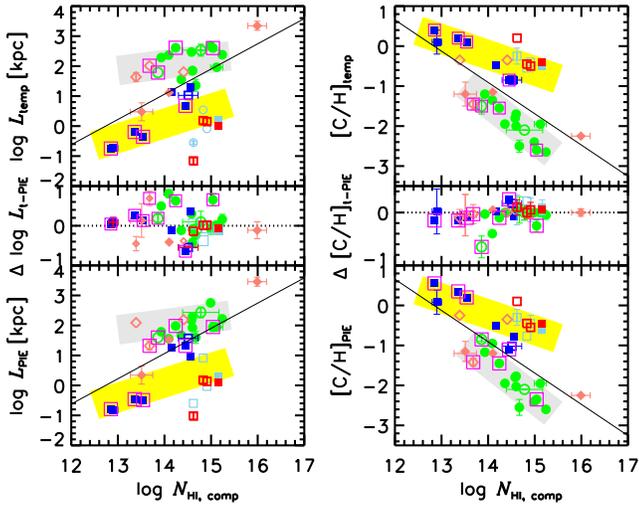}
\caption{{\it Upper left panel:} the line-of-sight length $L_{\mathrm{temp}}$ 
as a function of $N_{\mathrm{\ion{H}{i}}}$. 
All the symbols are the same as in Fig.~\ref{fig23}.  
The solid line, yellow-shaded and gray-shaded regions are 
the least-square fit to the full sample, high-metallicity branch
and low-metallicity branch absorbers, respectively.
{\it Upper right panel:} [C/H]$_{\mathrm{temp}}$ as a function of 
$N_{\mathrm{\ion{H}{i}}}$. 
{\it Middle panels:} the difference between the non-PIE $T_{b}$ and PIE models
for the HM Q+G 2012 UVB.
{\it Lower panels:} for the PIE model. 
The yellow-shaded and gray-shaded regions  
are the same ones as in the upper panels.
Refer to the online version for clarity.
}
\label{fig24}
\end{figure}

In Subsection~\ref{sec5.1}, we suggested that \ion{C}{iii} absorbers consist of 2 distinct
populations based on the $N_{\mathrm{\ion{C}{iv}}}$--[C/H]$_{\mathrm{temp}}$ 
and $N_{\mathrm{\ion{C}{iv}}}$--$L_{\mathrm{temp}}$ relations. In 
Subsections~\ref{sec5.2}--\ref{sec5.4}, these two populations are shown to 
follow their own 
scaling relations between various observed and predicted physical parameters. In
this Subsection, we present similar relations to those explored in Subsection~\ref{sec5.1},
but here for $N_{\mathrm{\ion{H}{i}}}$ instead of $N_{\mathrm{\ion{C}{iv}}}$.

In Fig.~\ref{fig24}, the entire $N_{\mathrm{\ion{H}{i}}}$--$L_{\mathrm{temp}}$ 
plane can be viewed as a scatter plot, with a spread of 2--3\,dex in $L_{\mathrm{temp}}$
for a given $N_{\mathrm{\ion{H}{i}}}$. However, only a few data points
are clustered around a least-square fit to the full sample, shown as a solid line. Indeed, 
except at $\log N_{\mathrm{\ion{H}{i}}} \sim 14.5$, no \ion{C}{iii} absorbers are
along the $\log L_{\mathrm{temp}} \sim 1.4$ line at other $N_{\mathrm{\ion{H}{i}}}$,
implying that the $N_{\mathrm{\ion{H}{i}}}$--$L_{\mathrm{temp}}$ plane
is also interpreted better in terms of two populations.
For both branch absorbers, $L_{\mathrm{temp}}$ increases as
$N_{\mathrm{\ion{H}{i}}}$ increases, but with a different scaling relation
as tabulated in Table~\ref{tab3}. 
The non-PIE nature disperses low-metallicity branch absorbers in to a wider 
$L_{\mathrm{temp}}$ range.

Similarly, tied \ion{C}{iii} absorbers 
on the $N_{\mathrm{\ion{H}{i}}}$--[C/H]$_{\mathrm{temp}}$ plane
tend to lie distinctly above or below the least-square fit line rather than on it,
with a division at [C/H]$_{\mathrm{temp}} \sim -1.0$. For both branch absorbers, 
[C/H]$_{\mathrm{temp}}$ decreases as $N_{\mathrm{\ion{H}{i}}}$ increases.
Table~\ref{tab4} lists the scaling relation for each population. 
Note that low-metallicity branch absorbers show
an anti-correlation between $N_{\mathrm{\ion{H}{i}}}$ and [C/H]$_{\mathrm{temp}}$,
an opposite trend from the $N_{\mathrm{\ion{C}{iv}}}$--[C/H]$_{\mathrm{temp}}$ relation.
Since only a few absorbers have a large difference from the PIE model shown in
the middle right panel, the non-PIE nature does not alter the relation.

\section{The origin of the bimodality of \ion{C}{iii} absorbers}
\label{sec6}

\subsection{Implication from the equivalent width and impact parameter relation}
\label{sec6.1}

Numerical simulations 
predict that the {\it projected} $N_{\mathrm{\ion{H}{i}}}$ around galaxies 
(the total $N_{\mathrm{\ion{H}{i}}}$
integrated along a sightline long enough to contain galaxies and their
surrounding in a simulation box) decreases
with a galactocentric distance in a well-defined way, depending on the
galaxy potential \citep{cen11, shen13, shull14a}. 
There is a large scatter, but the observed
\ion{H}{i} rest-frame equivalent width (REW -- used as a proxy for the
\ion{H}{i} column density)
roughly decreases with the galaxy impact parameter
both at low redshifts \citep{chen01, wakker09, liang14} and at high redshifts \citep{steidel10}.
The observed \ion{C}{iv} REW also decreases with the impact
parameters, but much more steeply 
so that \ion{C}{iv} reaches to its detection limit
at a smaller impact parameter than \ion{H}{i} \citep{chen01a, steidel10, liang14}.

A large scatter in the relation is reduced when the impact parameter is 
normalised with the galaxy luminosity \citep{chen01} or the virial radius
\citep{stocke13, liang14, shull14a}. Since 
the \ion{H}{i} distribution around galaxies is influenced by the galaxy potential, 
the projected $N_{\mathrm{\ion{H}{i}}}$ is a better
physical parameter than a geometrical impact parameter,  i.e. at the same
impact parameter, a more massive galaxy has a larger REW. In short,
the projected (or integrated)
$N_{\mathrm{\ion{H}{i}}}$ is a normalised impact parameter by the 
galaxy potential, as long as \ion{H}{i} gas is not significantly disturbed kinematically,
such as nearby supernova explosions. 
Compared to the impact parameter normalised by
a virial radius, the integrated $N_{\mathrm{\ion{H}{i}}}$ is
a simple and straightforward parameter, since 1) obtaining an impact parameter
requires a time-intensive deep galaxy survey around target QSOs, 2) the galaxy
survey is flux-limited, biasing toward brighter galaxies especially at high redshifts
and 3) assigning a single galaxy to a single absorber is not unambiguous due to the
galaxy clustering or no nearby galaxies above a detection limit
\citep{wakker09, steidel10, liang14, savage14}.

Simulations also predict that the projected gas metallicity 
(the metal mass density divided by the total mass density) 
distribution around a star-forming
galaxy is bipolar, since metals are transported to galactic halo and the surrounding
IGM by galactic outflows
which tend to propagate along the minor axis where the ISM density is low.
The higher-metallicity
gas at [Z/H]$\, \ge -0.5$ is close to the minor axis, while the gas
at [Z/H] $\in [-2.5, -1.0]$ is far from the galaxy along the minor axis (i.e. has a larger
impact parameter) or is more uniformly distributed along the major
axis from a small to large impact parameter beyond the virial radius, 
cf. Fig.~1 of \citet{shen13}. For an optically-thin gas at $\log T \sim 3.9$--5.1 exposed
to the HM Q+G 2012 UVB, the dominant ions of carbon is
\ion{C}{iii} and \ion{C}{iv}. Therefore, the projected $N_{\mathrm{\ion{C}{iv}}}$ is
roughly proportional to [C/H]$_{\mathrm{temp}}$, and the distribution
of observed $N_{\mathrm{\ion{C}{iv}}}$ is expected to follow 
the [Z/H] distribution predicted by simulations.

\subsection{Origin of high-metallicity branch absorbers}
\label{sec6.2}

Based on the observational and theoretical results in 
Subsection~\ref{sec6.1}, including an observational suggestion of 
bimodal metallicity distribution of \ion{Mg}{ii} absorbers at $z < 1$ \citep{bouche12,
lehner13}, it is
not unreasonable that \ion{C}{iii} absorbers could consist of two populations.

For high-metallicity absorbers, their predicted
line-of-sight length increases with observed 
$N_{\mathrm{\ion{C}{iv}}}$ and $N_{\mathrm{\ion{H}{i}}}$ up to
$\sim 20$\,kpc, roughly a
size of an extended optical disk or inner halo 
\citep{brooks11}.
The positive correlation between  
the line-of-sight length and the observed column density is reminiscent
of high ions, such as \ion{C}{iv}, \ion{N}{v} and \ion{O}{vi}, 
in the Milky Way, where the column density of high ions increases
with the distance to the background hot star in the Milky Way disk
\citep{jenkins78, lehner11}. This has been interpreted to imply that the number of 
intervening, randomly distributed, distinct high-ion gas clouds in the Milky Way
increases as the distance to the star increases. 

Our 12 high-metalliciy branch absorbers can be further classified into two subclasses.
About 67\% (8 out of 12)
of high-metallicity branch absorbers are part of a multi-component \ion{C}{iv} complex,
having
several, nearby \ion{C}{iv} clumps within $\sim 200$\,\kms\/. Therefore,
it is not implausible that a seemingly single \ion{C}{iv} component is in fact
a composite of several smaller gas clouds in extended disks, inner halos or outflowing
gas and our 6.7\,\kms\/ spectral resolution is not sufficient enough to resolve them.
All of them are associated with \ion{Si}{iv}, indicative of higher 
density and higher metallicity, thus closer to galaxies compared to \ion{Si}{iv}-free
\ion{C}{iv} absorbers \citep{shen13}.

Simple high-metallicity branch absorbers (the absorber number \#14, 15, 30 and 32)
are associated with strong \ion{C}{iv} and weak \ion{H}{i}. They
have $L_{\mathrm{temp}} \le 1$\,kpc and super-solar metallicity. 
Outliers in several well-characterised correlations
presented in Section~\ref{sec5} mostly belong to this subclass.
This subclass was studied extensively by 
\citet{schaye07}, which characterises this subclass as having a super-solar metallicity,
a size of about 100\,pc and a life time of $\sim 6 \times 10^{6}$ years, i.e. they are
not self-gravitating.
\citet{schaye07} conclude that these absorbers are 
caught in the process of transporting metals into the IGM.
Since there are only 4 absorbers at $L_{\mathrm{temp}} \le 0.35$\,kpc
in this subclass and since they are
usually outliers, it is not clear whether simple high-metallicity branch
absorbers are also unresolved multi-component
gas clouds in extended disks/inner halos
like complex high-metallicity branch absorbers.

From Figs~\ref{fig18} and \ref{fig24}, 
the carbon abundance of high-metallicity branch absorbers is anti-correlated
with $N_{\mathrm{\ion{C}{iv}}}$ and $N_{\mathrm{\ion{H}{i}}}$ with 

\begin{equation}
\label{eq2}
{\mathrm{[C/H]_{temp}}}  \propto -1.04 \times \log N_{\mathrm{\ion{C}{iv}}},
\end{equation}

\noindent and 

\begin{equation}
\label{eq3}
{\mathrm{[C/H]_{temp}}}  \propto -0.40 \times \log N_{\mathrm{\ion{H}{i}}}.
\end{equation}

This implies that $N_{\mathrm{\ion{C}{iv}}}$ and $N_{\mathrm{\ion{H}{i}}}$ of 
high-metallicity branch absorbers cannot be used as a proxy of impact parameter,
which seems to conflict with simulations and observations as discussed above.

This apparent discrepancy is mainly caused by the fact that 
$N_{\mathrm{\ion{C}{iv}}}$ and $N_{\mathrm{\ion{H}{i}}}$ plotted in 
Figs~\ref{fig18} and  \ref{fig24} 
are for an individual {\it component}. On the other hand,
simulations use the projected total column density of \ion{H}{i} or \ion{C}{iv},
not an individual component, as well as the averaged gas metallicity along the
sightline. To be a fair comparison, $N_{\mathrm{\ion{C}{iv}}}$
and $N_{\mathrm{\ion{H}{i}}}$ in Figs.~\ref{fig18} and \ref{fig24} should be
summed up all the components within a given velocity range appropriate for intervening
galaxies and ${\mathrm{[C/H]_{temp}}}$ should be an average of these individual
components. Unfortunately, due to the kinematically complicated component structure
of complex high-metallicity absorbers, there is no one-to-one correspondence
between \ion{H}{i} and \ion{C}{iv} components, and it is not possible to derive 
${\mathrm{[C/H]_{temp}}}$ of all the individual \ion{C}{iv} components. 

The anti-correlation between [C/H]$_{\mathrm{temp}}$ and column density
can be better understood, when combined with the 
[C/H]$_{\mathrm{temp}}$--$L_{\mathrm{temp}}$ anti-correlation.
Since there is a slight difference between \ion{H}{i} and \ion{C}{iv} distributions due to
non-uniform $n_{\mathrm{H}}$, peculiar velocity, UVB and metallicity \citep{cen11},
the $L_{\mathrm{temp}}$ dependence on the column density of \ion{H}{i} 
and \ion{C}{iv} is expected to be different. From Figs.~\ref{fig18} and \ref{fig24}, 
for high-metallicity branch absorbers,

\begin{equation}
{\mathrm{[C/H]_{temp}}}  \propto  
0.56 \times \log \left(\frac{N_{\mathrm{\ion{C}{iv}}}}{N_{\mathrm{\ion{H}{i}}}}\right), 
\end{equation}

\begin{equation}
N_{\mathrm{\ion{C}{iv}}} \propto L_{\mathrm{temp}}^{0.46},
\end{equation}

\begin{equation}
N_{\mathrm{\ion{H}{i}}} \propto L_{\mathrm{temp}}^{2.00},
\end{equation}

\noindent which leads to

\begin{equation}
{\mathrm{[C/H]_{temp}}} \propto -0.86 \times \log L_{\mathrm{temp}}.
\end{equation}

This is the anti-correlation between [C/H]$_{\mathrm{temp}}$ and $L_{\mathrm{temp}}$
displayed in Fig.~\ref{fig19}, caused by the fact that a higher amount of $N_{\mathrm{\ion{H}{i}}}$
is added up along the same line-of-sight than $N_{\mathrm{\ion{C}{iv}}}$. Due to a higher
$N_{\mathrm{\ion{H}{i}}}$ than $N_{\mathrm{\ion{C}{iv}}}$ for a larger-$L_{\mathrm{temp}}$
absorber, even if it is a unresolved component, [C/H]$_{\mathrm{temp}}$ decreases with
$L_{\mathrm{temp}}$.

Combined with an anti-correlation between $L_{\mathrm{temp}}$ and
[C/H]$_{\mathrm{temp}}$,
the anti-correlation between [C/H]$_{\mathrm{temp}}$ and $N_{\mathrm{\ion{C}{iv}}}$
or $N_{\mathrm{\ion{H}{i}}}$
suggests that [C/H]$_{\mathrm{temp}}$ decreases when
a \ion{C}{iv}-enriched gas is dispersed in to a larger distance or expands to be
mixed with an adjacent low-metallicity ISM/halo gas. 

\subsection{Origin of low-metallicity branch absorbers}
\label{sec6.4}

For low-metallicity branch absorbers, the median $L_{\mathrm{temp}}$ 
is $\sim$\,233\,kpc, ranged from 20 to 480 kpc. In addition to 
[C/H]$_{\mathrm{temp}} \le -1$ and the total hydrogen density at
$n_{\mathrm{H, \, temp}} \in [-5.2, -4.3]$, close to
the cosmic mean density,
the line-of-sight length argument is in favour of their origin as a galactic halo and the 
surrounding IGM filaments.

Their [C/H]$_{\mathrm{temp}}$ is 
$\propto 1.13 \times \log N_{\mathrm{\ion{C}{iv}}}$ and 
$\propto -0.91 \times \log N_{\mathrm{\ion{H}{i}}}$. 
This reflects that $N_{\mathrm{\ion{C}{iv}}}/N_{\mathrm{\ion{H}{i}}}$
(a proxy for [C/H]$_{\mathrm{temp}}$) decreases with 
$N_{\mathrm{\ion{H}{i}}}$, as $N_{\mathrm{\ion{C}{iv}}}$
is almost independent of $N_{\mathrm{\ion{H}{i}}}$ with a large scatter
(Fig.~\ref{fig9}). Being almost independent of 
$L_{\mathrm{temp}}$ on $N_{\mathrm{\ion{H}{i}}}$ and $N_{\mathrm{\ion{C}{iv}}}$,
the opposite trend of [C/H]$_{\mathrm{temp}}$ on $N_{\mathrm{\ion{C}{iv}}}$
and $N_{\mathrm{\ion{H}{i}}}$ implies that 1) \ion{C}{iv} (and hence \ion{C}{iii})
is not likely to be homogeneously mixed with \ion{H}{i} at the
scale of $\sim 200$\,kpc or 2) an expansion of \ion{C}{iv} gas 
into the ambient medium with time
increases a line-of-sight length for some gas clouds with a similar 
$N_{\mathrm{\ion{C}{iv}}}$.

\section{Summary}
\label{sec7}

We have analysed the physical properties of 53 intervening,
optically-thin \ion{C}{iv}+\ion{C}{iii} systems at $2.1 < z < 3.4$ in the
19 high-quality optical spectra taken with the UVES at the VLT and 
HIRES at Keck with $\sim 6.7$\,\kms\/ resolution.
The selected systems have the 
same velocity structure in both \ion{C}{iv} and \ion{C}{iii}
(also \ion{C}{ii} $\lambda\lambda$\,1334, 1036 if
detected), and 
were fitted with Voigt profiles
with the same fixed redshift and line widths (or $b$ parameters) 
for each corresponding component
when the velocity centroid difference between untied 
\ion{C}{iv} and \ion{C}{iii} is less than 1.5\,\kms\/.
When a clean, separable \ion{H}{i} component 
for each corresponding \ion{C}{iv}
exists within 15\,\kms\/,
the \ion{H}{i} component was fitted with the same redshift and
appropriate line width to build a sample of tied (aligned) 
\ion{H}{i}+\ion{C}{iv}+\ion{C}{iii} component trios, i.e.
kinematically simple absorbers.

The 53 \ion{C}{iii} systems consists of 155 \ion{C}{iv} 
components. Out of 155, 132 \ion{C}{iv} components are associated with
the tied \ion{C}{iii} 
(104 clean detections, 4 lower limits and 24 upper limits) and
23 \ion{C}{iv} components have either blended \ion{C}{iii} regions or 
shifted \ion{C}{iii}.
54 \ion{C}{iv} components (35\%, 54/155) have the tied \ion{H}{i},
while 33 components (21\%, 33/155) have both tied \ion{H}{i} and \ion{C}{iii}
with well-measured \ion{C}{iii} column densities.

For the 132 tied \ion{C}{iv}+\ion{C}{iv} components 
at $\log N_{\mathrm{\ion{C}{iv}}} \in [11.7, 14.1]$ and
$\log N_{\mathrm{\ion{C}{iii}}} \in [11.5, 14.0]$ for clean \ion{C}{iii}:

\begin{enumerate}

\item $N_{\mathrm{\ion{C}{iii}}} \propto
N_{\mathrm{\ion{C}{iv}}}^{1.42\pm0.11}$, with lower (higher)
$N_{\mathrm{\ion{C}{iii}}}$ at $\log N_{\mathrm{\ion{C}{iv}}} \le 13$ ($\ge 13$).  
There is a suggestion of the break-down of this linear relation 
at higher-$N_{\mathrm{\ion{C}{iv}}}$ ends.
The mean ratio of $N_{\mathrm{\ion{C}{iii}}}$ and  $N_{\mathrm{\ion{C}{iv}}}$
is $<\!N_{\mathrm{\ion{C}{iv}}}/N_{\mathrm{\ion{C}{iv}}}\!> \,= 1.0 \pm 0.3$,
with a negligible redshift evolution.  

\item For the \ion{C}{iv}+\ion{C}{iii} components with a tied \ion{H}{i},
no systematic trend is found in $N_{\mathrm{\ion{C}{iii}}}/N_{\mathrm{\ion{C}{iv}}}$
as a function of $N_{\mathrm{\ion{H}{i}}}$, implying that both
$N_{\mathrm{\ion{C}{iii}}}$ and $N_{\mathrm{\ion{C}{iv}}}$ behave in a similar
way to $N_{\mathrm{\ion{H}{i}}}$.

\end{enumerate}

From the 54 tied \ion{H}{i}+\ion{C}{iv} component
pairs at $\log N_{\mathrm{\ion{H}{i}}} \in [12.2, 16.0]$ and 
$\log N_{\mathrm{\ion{C}{iv}}} \in [11.8, 13.8]$:

\begin{enumerate}

\item The tied \ion{C}{iv} components cover the wide range of 
column densities
and $b$ parameters in the $N_{\mathrm{\ion{C}{iv}}}-b_{\mathrm{\ion{C}{iv}}}$
plane. On the other hand, most tied \ion{H}{i} 
components are located along the cutoff-$b$ line on
the $N_{\mathrm{\ion{H}{i}}}$--$b_{\mathrm{\ion{H}{i}}}$ plane. This suggests that the
tied pairs only sample an gas having a specific physical condition, not
a representative of the entire low-$N_{\mathrm{\ion{H}{i}}}$ absorber.

\item The temperature $T_{\mathrm{b}}$ 
estimated from the line widths of \ion{H}{i} and \ion{C}{iv}
is well-approximated to a Gaussian peaking at
$\log T_{b} \sim 4.4\pm0.3$ at 
a range of $\log T_{b} \sim 3.5$--5.5, implying
that the tied \ion{H}{i}+\ion{C}{iv} pairs sample both photoionised
and collisionally ionised \ion{C}{iv}. Excluding 4 pairs non-thermally
broadened, the mean and median gas temperatures are 
$<\!\log T\!> \, = 4.52 \pm 0.33$ and $\log T_{\mathrm{median}} = 4.48$. 

\item The median non-thermal line width and thermal line width are
$b_{\mathrm{nt, median}} = 7$ and $b_{\mathrm{th, median}} = 22$\,\kms\/.
About 59\% (32 out of 54) of the pairs has the non-thermal energy 
contribution to the total energy less than 10\%, meaning that the majority of the
pairs near the cutoff-$b_{\mathrm{\ion{H}{i}}}$ are thermally broadened.

\end{enumerate}

Including 4 \ion{C}{ii}-enriched \ion{C}{iii} absorbers, 
but excluding one non-thermal absorber,
32 \ion{C}{iii} absorbers 
have tied \ion{H}{i}, \ion{C}{iv} and \ion{C}{iii}, with
$\log N_{\mathrm{\ion{H}{i}}} \in [12.9, 16.0]$, 
$\log N_{\mathrm{\ion{C}{iv}}} \in [11.8, 13.8]$, 
$\log N_{\mathrm{\ion{C}{iii}}} \in [11.7, 13.8]$ and
$\log N_{\mathrm{\ion{C}{ii}}} \in [12.2, 12.7]$, respectively. We have built the 
Cloudy photoionisation models for three theoretical UV background
radiation models, the
Haardt-Madau (HM) 2005 UV background contributed both by QSOs and galaxies
(Q+G) and only by QSOs (Q) as well as our fiducial HM Q+G 2012 model.
As the Cloudy-predicted gas temperature for a 
photoionisation equilibrium (PIE) gas is different from the gas temperature $T_{b}$,
the non-PIE Cloudy model with the fixed temperature $T_{b}$ was also built
for the HM Q+G 2012, which is our fiducial model to derive the physical
parameters of 32 tied \ion{C}{iii} absorbers, subscripted with ``temp".

\begin{enumerate}

\item About 25\% of the absorbers has more than 50\% difference between
$T_{b}$ and $T_{\mathrm{PIE}}$, with a higher
$T_{b}$ for all the highly discrepant absorbers but one.
This  implies that the PIE assumption is 
not likely to hold in these absorbers, probably due to recent cooling
from a hotter gas
or additional radiation. 
There is a suggestion that 
more absorbers show a larger temperature difference at lower redshifts
and at $\log N_{\mathrm{\ion{H}{i}}} \le 14$. 

\item The Cloudy-predicted parameters do not show a strong discrepancy 
between the non-PIE $T_{b}$ and the PIE models for the same UVBs. 
However, the non-PIE condition increases a scatter in any PIE correlations.

\item There is evidence of two populations of kinematically simple, 
optically-thin \ion{C}{iii} absorbers,
occupied in a distinct region on the various parameter spaces with their own
scaling relation. 
High-metallicity (low-metallicity) branch absorbers have the carbon abundance 
[C/H]$_{\mathrm{temp}} \ge -1.0$ ($\le -1.0$) and a line-of-sight length
$L_{\mathrm{temp}} \le 20$\,kpc ($\ge 20$\,kpc). 

\item High-metallicity branch absorbers
are either complex with multi-component \ion{C}{iv} mostly associated with
\ion{Si}{iv}, or simple, sometimes having super-solar metallicity, which
are characterised by
stronger \ion{C}{iv} than the usual \ion{H}{i} absorbers with a similar 
$N_{\mathrm{\ion{H}{i}}}$. Considering their
higher $T_{b}$ than $T_{\mathrm{PIE}}$ and higher [C/H]$_{\mathrm{temp}}$,
simple high-metallicity branch absorbers are likely to be radiatively cooling gas from
a higher temperature and/or to be expanding.
Outliers from any well-characterised correlations
are mostly simple high-metallicity branch absorbers or 4 absorbers associated
with \ion{C}{ii}. 

\item High-metallicity branch absorbers have a lower gas temperature
than low-metallicity branch absorbers having $\log T_{b} \sim 4.5$, 
due to an increased cooling rate
by a higher [C/H]$_{\mathrm{temp}}$. 
The total (neutral and ionised) hydrogen volume density is
$\log n_{\mathrm{H, \, temp}} \in [-4.5, -3.3]$, about an order of magnitude higher
than low-metallicity branch absorbers. Compared to the warm ionised gas in the Milky
Way with a comparable temperature range, high-metallicity absorbers 
have two orders of magnitude lower $n_{\mathrm{H, \, temp}}$.

\item For high-metallicity branch absorbers,
$L_{\mathrm{temp}} \propto N_{\mathrm{\ion{C}{iv}}}^{2.17}$ and 
and $L_{\mathrm{temp}} \propto N_{\mathrm{\ion{H}{i}}}^{0.50}$ upto
$L_{\mathrm{temp}} \le 20$\,kpc, roughly a size of intervening disks or inner halos.
This trend is similar to what is observed
for high ions in the Milky Way. The observations imply that 1) a
seemingly single-component profile of \ion{C}{iv} and \ion{H}{i} is
in fact an ensemble average of many components unresolved in our spectra
and 2) a larger line-of-sight length passes through more numbers of
randomly distributed \ion{H}{i} and \ion{C}{iv} gas clouds
in extended disks or outflowing gas in inner halos.

\item For high-metallicity branch absorbers, [C/H]$_{\mathrm{temp}}$
is anti-correlated as $\propto -1.04 \times \log N_{\mathrm{\ion{C}{iv}}}$ and 
$\propto -0.40 \times \log N_{\mathrm{\ion{H}{i}}}$, implying that 
the total $N_{\mathrm{\ion{H}{i}}}$ integrated along the same
sightline is larger than the total $N_{\mathrm{\ion{C}{iv}}}$. Since
the relative
fraction between \ion{C}{iii} and \ion{C}{iv}
for optically-thin gas exposed to the HM Q+G 2012 UVB is similar as a function of
$N_{\mathrm{\ion{H}{i}}}$, a larger $N_{\mathrm{\ion{H}{i}}}$ decreases
[C/H]$_{\mathrm{temp}}$.

\item Low-metallicity branch absorbers do not have associated
\ion{Si}{iv} and have $\log n_{\mathrm{H, \, temp}} \in [-5.3, -4.2]$.
This corresponds to a cosmic over-density of 0.3--13, which is for
a typical low-density intergalactic \ion{H}{i} gas, not in a collapsed
object. Their gas temperature is
close to the PIE equilibrium temperature at $\log T_{b} \sim 4.5$
at the low-metallicity regime, except a few outliers with a higher
$T_{b}$, probably due to recent cooling.

\item For low-metallicity branch absorbers, 
$L_{\mathrm{temp}} \propto N_{\mathrm{\ion{C}{iv}}}^{0.55}$ and 
and $L_{\mathrm{temp}} \propto N_{\mathrm{\ion{H}{i}}}^{0.20}$ at 
$L_{\mathrm{temp}} \in [20, 480]$\,kpc, roughly a size of intervening halos
or the surrounding IGM filaments.

\item For low-metallicity branch absorbers, [C/H]$_{\mathrm{temp}} 
\propto 1.13 \times \log N_{\mathrm{\ion{C}{iv}}}$ and 
$\propto -0.91 \times \log N_{\mathrm{\ion{H}{i}}}$. 
This reflects that $N_{\mathrm{\ion{C}{iv}}}/N_{\mathrm{\ion{H}{i}}}$
(a proxy for [C/H]$_{\mathrm{temp}}$) decreases with 
$N_{\mathrm{\ion{H}{i}}}$, as $N_{\mathrm{\ion{C}{iv}}}$
is almost independent of $N_{\mathrm{\ion{H}{i}}}$ with a large scatter.
The opposite trend of [C/H]$_{\mathrm{temp}}$ on $N_{\mathrm{\ion{C}{iv}}}$
and $N_{\mathrm{\ion{H}{i}}}$ implies that 1) \ion{C}{iv} and \ion{C}{iii}
are not likely to be homogeneously mixed with \ion{H}{i} at the
scale of $\sim 200$\,kpc or 2) an expansion of \ion{C}{iv} gas 
into the ambient medium with time
increases a line-of-sight length for some gas clouds with a similar 
$N_{\mathrm{\ion{C}{iv}}}$.

\item At $2.1 < z < 2.6$, [C/H]$_{\mathrm{temp}}$ spans a wide range
from $-$2.7 to 0.4, with the mean value of  $-1.0\pm1.0$. Its distribution has a long
Gaussian tail at low carbon abundance. On the other hand,
at $2.6 < z < 3.4$, \ion{C}{iii} absorbers have a smaller range
of [C/H]$_{\mathrm{temp}}$ at $-$2.1 to $-$0.9, with the mean value of $-1.6\pm0.4$.
There is a lack of data points at [C/H]$_{\mathrm{temp}} \ge -0.8$, which is
not an observational bias as these absorbers are easy to detect. 

\item The total hydrogen density $n_{\mathrm{H_{, \, temp}}}$ 
does not follow the theoretical $N_{\mathrm{\ion{H}{i}}}$--$n_{\mathrm{H}}$ 
relation by \citet{schaye01} for a
self-gravitating, metal-free PIE gas in local hydrostatic equilibrium, since our \ion{C}{iii}
absorbers sample a gas in a wide range of physical conditions. However, when
excluded simple high-metallicity branch absorbers, the remaining high-metallicity
branch absorbers satisfy the Schaye relation.  
Low-metallicity branch absorbers follows the Schaye relation if shifted down
by 0.6\,dex.

\item The evidence for bimodal distributions in the observed and derived physical 
parameters
for optically-thin \ion{C}{iii} absorbers is largely independent of the choice of UVB, 
as long as all the
\ion{C}{iii} absorbers are exposed to a similar UVB.

\end{enumerate}

\section*{Acknowledgments}

TSK appreciates an insightful discussion with B.~D. Savage. 
TSK acknowledges funding support to 
the European Research Council Starting Grant ``Cosmology with the IGM"
through grant GA-257670. 
RFC is also supported by the same grant for his stay at Osservatorio Astronomico 
di Trieste to carry out part of this work.
TSK is also grateful to a travel support by the FP7 ERC Advanced Grant 
Emergence-320596 to IoA, Cambridge, where part of this work was done.
DR was partially supported by the National Science Foundation's Research 
Experience for Undergraduates program through NSF
Award AST-1004881 to the University of Wisconsin-Madison.

\bibliography{literature}

\appendix
\section{Predictions by Cloudy}


\begin{table*}
\caption{The tied \ion{H}{i}+\ion{C}{iv} components} 
\label{tab_a1}
\begin{tabular}{lcrrrrrrr}
\hline

\noalign{\smallskip}

 QSO & $z_{\mathrm{abs}}$ &  $b_{\mathrm{\ion{H}{i}}}$  & $\log N_{\mathrm{\ion{H}{i}}}$ & 
   $b_{\mathrm{\ion{C}{iv}}}$ & $\log N_{\mathrm{\ion{C}{iv}}}$ &  Temperature & 
   $b_{\mathrm{nt}}$  & Notes \\
    
\noalign{\smallskip}
   
&   &  (\kms\/) & [cm$^{-2}$]  &  (\kms\/)  & [cm$^{-2}$]   &  (K)   & (\kms\/)  & \\

\hline

    Q0055--269 &   3.257359 &   24.7{\scriptsize $\pm$0.9} &  13.93{\scriptsize $\pm$0.02} &   
           10.8{\scriptsize $\pm$0.8} &  12.76{\scriptsize $\pm$0.03} &  32860{\scriptsize $\pm$3012} &    
            8{\scriptsize $\pm$1} &     certain \\
    Q0055--269 &   3.038795 &   22.3{\scriptsize $\pm$1.2} &  14.60{\scriptsize $\pm$0.04} &   
           14.4{\scriptsize $\pm$1.4} &  12.72{\scriptsize $\pm$0.04} &  19220{\scriptsize $\pm$4373} &   
           13{\scriptsize $\pm$2} &     certain \\
    Q0055--269 &   2.744100 &   26.7{\scriptsize $\pm$4.8} &  15.12{\scriptsize $\pm$0.13} &   
           16.3{\scriptsize $\pm$1.9} &  12.79{\scriptsize $\pm$0.05} &  29830{\scriptsize $\pm$17130} &   
           15{\scriptsize $\pm$2} &     certain \\
    Q0055--269 &   2.743720 &   24.2{\scriptsize $\pm$10.1} &  14.78{\scriptsize $\pm$0.39} &    
           8.8{\scriptsize $\pm$2.3} &  12.48{\scriptsize $\pm$0.09} &  33490{\scriptsize $\pm$32600} &    
           6{\scriptsize $\pm$6} &   uncertain \\
    PKS2126--158 &   2.973015 &   23.8{\scriptsize $\pm$0.2} &  14.36{\scriptsize $\pm$0.01} &   
          12.5{\scriptsize $\pm$0.7} &  12.18{\scriptsize $\pm$0.03} &  27020{\scriptsize $\pm$1355} &   
          11{\scriptsize $\pm$1} &     certain \\
  HS1425$+$6099 &   3.024719 &   21.8{\scriptsize $\pm$1.0} &  14.75{\scriptsize $\pm$0.04} &   
          11.2{\scriptsize $\pm$2.0} &  12.11{\scriptsize $\pm$0.06} &  23210{\scriptsize $\pm$4140} &   
          10{\scriptsize $\pm$3} &     certain \\
   Q0636$+$6801 &   3.018604 &   28.9{\scriptsize $\pm$14.8} &  13.97{\scriptsize $\pm$0.31} &   
          14.7{\scriptsize $\pm$2.7} &  12.29{\scriptsize $\pm$0.06} &   41160{\scriptsize $\pm$56860} &   
          13{\scriptsize $\pm$5} &   uncertain \\
    Q0420--388 &   2.849598 &   27.4{\scriptsize $\pm$0.7} &  14.09{\scriptsize $\pm$0.02} &   
          15.0{\scriptsize $\pm$1.5} &  12.59{\scriptsize $\pm$0.04} &  34780{\scriptsize $\pm$3185} &   
          13{\scriptsize $\pm$2} &     certain \\
    Q0420--388 &   2.849229 &   32.7{\scriptsize $\pm$1.6} &  13.68{\scriptsize $\pm$0.06} &   
          12.3{\scriptsize $\pm$2.9} &  12.10{\scriptsize $\pm$0.10} &   60880{\scriptsize $\pm$7971} &    
          8{\scriptsize $\pm$5} &   uncertain \\
  HE0940--1050 &   2.937755 &   25.2{\scriptsize $\pm$0.3} &  14.58{\scriptsize $\pm$0.02} &    
          9.7{\scriptsize $\pm$0.4} &  12.65{\scriptsize $\pm$0.02} &  35740{\scriptsize $\pm$996} &    
          7{\scriptsize $\pm$1} &     certain \\
  HE0940--1050 &   2.937304 &   42.6{\scriptsize $\pm$1.5} &  14.23{\scriptsize $\pm$0.03} &   
          42.6{\scriptsize $\pm$4.3} &  12.66{\scriptsize $\pm$0.04} &   $\le 92$ &   
          43{\scriptsize $\pm$5} &     certain \\
  HE0940--1050 &   2.883509 &   26.1{\scriptsize $\pm$0.3} &  14.60{\scriptsize $\pm$0.01} &   
         15.8{\scriptsize $\pm$0.8} &  12.35{\scriptsize $\pm$0.03}  &  28560{\scriptsize $\pm$2009} &   
         15{\scriptsize $\pm$1} &     certain \\
  HE0940--1050 &   2.826555 &   17.2{\scriptsize $\pm$2.4} &  14.51{\scriptsize $\pm$0.21} &    
          6.4{\scriptsize $\pm$0.2} &  13.19{\scriptsize $\pm$0.02} &  16840{\scriptsize $\pm$5478} &    
          4{\scriptsize $\pm$1} &   uncertain \\
  HE2347--4342 &   2.787112 &   23.6{\scriptsize $\pm$0.1} &  14.58{\scriptsize $\pm$0.00} &   
          23.2{\scriptsize $\pm$4.2} &  12.22{\scriptsize $\pm$0.06} &  $\le 12880$ &   
          23{\scriptsize $\pm$5} &   uncertain \\
  HE2347--4342 &   2.572434 &   29.0{\scriptsize $\pm$0.2} &  15.02{\scriptsize $\pm$0.01} &    
          9.2{\scriptsize $\pm$3.4} &  11.88{\scriptsize $\pm$0.12} &   49890{\scriptsize $\pm$4252} &    
          4{\scriptsize $\pm$9} &   uncertain \\
  HE2347--4342 &   2.488605 &   34.4{\scriptsize $\pm$0.6} &  14.81{\scriptsize $\pm$0.01} &   
          11.9{\scriptsize $\pm$0.8} &  12.52{\scriptsize $\pm$0.02} &   68600{\scriptsize $\pm$3143} &    
          7{\scriptsize $\pm$2} &     certain \\
  HE2347--4342 &   2.347467 &   23.6{\scriptsize $\pm$1.3} &  15.99{\scriptsize $\pm$0.20} &    
          9.7{\scriptsize $\pm$0.1} &  13.49{\scriptsize $\pm$0.00} &  30490{\scriptsize $\pm$4035} &    
          7{\scriptsize $\pm$0} &     certain \\
  HE0151--4326 &   2.519825 &   26.1{\scriptsize $\pm$0.5} &  15.24{\scriptsize $\pm$0.02} &   
          10.2{\scriptsize $\pm$1.1} &  12.29{\scriptsize $\pm$0.04} &   38340{\scriptsize $\pm$2344} &    
           7{\scriptsize $\pm$2} &     certain \\
  HE0151--4326 &   2.468096 &   22.9{\scriptsize $\pm$1.0} &  12.84{\scriptsize $\pm$0.02} &    
           6.6{\scriptsize $\pm$0.1} &  12.96{\scriptsize $\pm$0.01} &  31730{\scriptsize $\pm$2820} &    
           0{\scriptsize $\pm$0} &     certain \\
  HE0151--4326 &   2.449902 &   13.6{\scriptsize $\pm$0.3} &  14.45{\scriptsize $\pm$0.02} &    
           5.5{\scriptsize $\pm$0.2} &  12.99{\scriptsize $\pm$0.01} &  10280{\scriptsize $\pm$599} &    
           4{\scriptsize $\pm$0} &     certain \\
  HE0151--4326 &   2.419676 &   19.4{\scriptsize $\pm$1.1} &  12.85{\scriptsize $\pm$0.04} &    
           8.0{\scriptsize $\pm$0.3} &  12.75{\scriptsize $\pm$0.01} &  20570{\scriptsize $\pm$2785} &    
           6{\scriptsize $\pm$1} &     certain \\
  HE0151--4326 &   2.415718 &   20.1{\scriptsize $\pm$0.4} &  13.36{\scriptsize $\pm$0.01} &    
           8.5{\scriptsize $\pm$0.2} &  13.03{\scriptsize $\pm$0.01} &  21850{\scriptsize $\pm$963} &    
           6{\scriptsize $\pm$0} &     certain \\
  HE0151--4326 &   2.401315 &   33.7{\scriptsize $\pm$0.7} &  15.04{\scriptsize $\pm$0.02} &   
          10.7{\scriptsize $\pm$1.4} &  12.34{\scriptsize $\pm$0.11} &  67770{\scriptsize $\pm$3127} &    
           4{\scriptsize $\pm$4} &     certain \\
    Q0002--422 &   2.539455 &   37.0{\scriptsize $\pm$1.0} &  14.24{\scriptsize $\pm$0.03} &   
           12.8{\scriptsize $\pm$0.8} &  12.55{\scriptsize $\pm$0.02} &  79970{\scriptsize $\pm$4910} &    
           7{\scriptsize $\pm$2} &     certain \\
    Q0002--422 &   2.463222 &   23.6{\scriptsize $\pm$2.8} &  14.56{\scriptsize $\pm$0.06} &   
           10.3{\scriptsize $\pm$0.5} &  13.26{\scriptsize $\pm$0.01} &  29670{\scriptsize $\pm$8781} &    
           8{\scriptsize $\pm$1} &     certain \\
    Q0002--422 &   2.462358 &   16.3{\scriptsize $\pm$0.5} &  15.16{\scriptsize $\pm$0.02} &    
           8.3{\scriptsize $\pm$0.3} &  13.20{\scriptsize $\pm$0.01} &  13020{\scriptsize $\pm$1186} &    
           7{\scriptsize $\pm$0} &     certain \\
     Q0002--422 &   2.462044 &   18.1{\scriptsize $\pm$0.3} &  14.83{\scriptsize $\pm$0.02} &   
           10.1{\scriptsize $\pm$0.7} &  13.34{\scriptsize $\pm$0.02} & 14800{\scriptsize $\pm$932} &    
            9{\scriptsize $\pm$0} &    uncertain \\
  PKS0329--255 &   2.587085 &   45.7{\scriptsize $\pm$1.9} &  14.82{\scriptsize $\pm$0.04} &   
          13.2{\scriptsize $\pm$6.1} &  12.00{\scriptsize $\pm$0.18} & 126700{\scriptsize $\pm$22160} &    
           0{\scriptsize $\pm$1} &     certain \\
  PKS0329--255 &   2.586757 &   22.5{\scriptsize $\pm$0.7} &  14.99{\scriptsize $\pm$0.03} &    
           7.3{\scriptsize $\pm$1.2} &  12.37{\scriptsize $\pm$0.06} &  29980{\scriptsize $\pm$2246} &    
           3{\scriptsize $\pm$3} &     certain \\
  PKS0329--255 &   2.456581 &   23.3{\scriptsize $\pm$2.1} &  13.51{\scriptsize $\pm$0.24} &    
           9.0{\scriptsize $\pm$4.7} &  11.97{\scriptsize $\pm$0.25} &  30640{\scriptsize $\pm$9385} &    
           6{\scriptsize $\pm$7} &     certain \\
  PKS0329--255 &   2.456208 &   25.4{\scriptsize $\pm$4.4} &  14.13{\scriptsize $\pm$0.15} &   
           25.4{\scriptsize $\pm$8.5} &  12.64{\scriptsize $\pm$0.09} &  $\le 538$ &   
           25{\scriptsize $\pm$8} &     certain \\
    Q0453--423 &   2.444109 &   13.0{\scriptsize $\pm$0.4} &  14.62{\scriptsize $\pm$0.01} &    
           5.2{\scriptsize $\pm$0.7} &  12.63{\scriptsize $\pm$0.06} & 9308{\scriptsize $\pm$566} &    
           4{\scriptsize $\pm$1} &   uncertain \\
    Q0453--423 &   2.443509 &   21.4{\scriptsize $\pm$4.2} &  14.14{\scriptsize $\pm$0.14} &    
           8.5{\scriptsize $\pm$0.2} &  13.38{\scriptsize $\pm$0.01} &  25450{\scriptsize $\pm$8891} &    
           6{\scriptsize $\pm$1} &     uncertain \\
    Q0453--423 &   2.442644 &   15.6{\scriptsize $\pm$0.5} &  14.91{\scriptsize $\pm$0.01} &    
           6.2{\scriptsize $\pm$0.4} &  13.04{\scriptsize $\pm$0.03} &  13500{\scriptsize $\pm$796} &    
           5{\scriptsize $\pm$1} &   uncertain \\
    Q0453--423 &   2.441813 &   22.1{\scriptsize $\pm$0.7} &  14.67{\scriptsize $\pm$0.01} &   
          8.4{\scriptsize $\pm$3.8} &  11.75{\scriptsize $\pm$0.13} &  27670{\scriptsize $\pm$3464} &    
          6{\scriptsize $\pm$5} &     certain \\
    Q0453--423 &   2.398159 &   19.0{\scriptsize $\pm$1.9} &  13.39{\scriptsize $\pm$0.07} &   
          11.7{\scriptsize $\pm$1.1} &  12.74{\scriptsize $\pm$0.04} &  14700{\scriptsize $\pm$4761} &   
          11{\scriptsize $\pm$1} &   uncertain \\
    Q0453--423 &   2.397801 &   18.3{\scriptsize $\pm$0.2} &  14.41{\scriptsize $\pm$0.01} &    
           9.7{\scriptsize $\pm$0.1} &  13.78{\scriptsize $\pm$0.01} & 15860{\scriptsize $\pm$595} &    
           9{\scriptsize $\pm$0} &   uncertain \\           
    Q0453--423 &   2.397447 &   47.0{\scriptsize $\pm$5.9} &  13.86{\scriptsize $\pm$0.05} &   
          16.7{\scriptsize $\pm$1.4} &  12.71{\scriptsize $\pm$0.03} &  127500{\scriptsize $\pm$36360} &   
          10{\scriptsize $\pm$3} &   uncertain \\           
    Q0453--423 &   2.396755 &   19.0{\scriptsize $\pm$0.6} &  14.16{\scriptsize $\pm$0.03} &    
          6.9{\scriptsize $\pm$0.1} &  13.34{\scriptsize $\pm$0.01} &  20640{\scriptsize $\pm$1491} &    
          4{\scriptsize $\pm$0} &   certain \\
    Q0453--423 &   2.277569 &   18.8{\scriptsize $\pm$0.9} &  13.54{\scriptsize $\pm$0.03} &    
          9.5{\scriptsize $\pm$0.2} &  12.99{\scriptsize $\pm$0.01} &  17540{\scriptsize $\pm$2380} &    
          8{\scriptsize $\pm$0} &     certain \\
    Q0453--423 &   2.169936 &   59.1{\scriptsize $\pm$1.9} &  13.30{\scriptsize $\pm$0.02} &   
         17.7{\scriptsize $\pm$2.2} &  12.27{\scriptsize $\pm$0.04} &  210300{\scriptsize $\pm$15510} &    
         5{\scriptsize $\pm$9} &     certain \\
    Q0453--423 &   2.169418 &   35.9{\scriptsize $\pm$3.9} &  13.36{\scriptsize $\pm$0.12} &   
         14.7{\scriptsize $\pm$1.1} &  12.45{\scriptsize $\pm$0.03} &  70900{\scriptsize $\pm$18650} &   
         11{\scriptsize $\pm$2} &     certain \\
 HE1347--2457 &   2.370003 &   20.6{\scriptsize $\pm$2.8} &  14.10{\scriptsize $\pm$0.09} &    
         9.6{\scriptsize $\pm$0.7} &  12.60{\scriptsize $\pm$0.03} &  21860{\scriptsize $\pm$7694} &    
         8{\scriptsize $\pm$1} &     certain \\
    Q0329--385 &   2.372835 &   75.7{\scriptsize $\pm$0.8} &  14.66{\scriptsize $\pm$0.02} &   
         21.9{\scriptsize $\pm$4.6} &  12.38{\scriptsize $\pm$0.09} &  347700{\scriptsize $\pm$26610} &    
         0{\scriptsize $\pm$0} &     certain \\
    Q0329--385 &   2.363795 &   25.9{\scriptsize $\pm$0.6} &  14.33{\scriptsize $\pm$0.05} &   
         10.8{\scriptsize $\pm$1.8} &  12.35{\scriptsize $\pm$0.05} &  36870{\scriptsize $\pm$3179} &    
         8{\scriptsize $\pm$3} &     certain \\
    Q0329--385 &   2.314255 &   22.7{\scriptsize $\pm$0.8} &  13.68{\scriptsize $\pm$0.03} &    
         6.6{\scriptsize $\pm$1.1} &  12.29{\scriptsize $\pm$0.05} &  31120{\scriptsize $\pm$2825} &    
         0{\scriptsize $\pm$0} &     certain \\
    Q0329--385 &   2.313933 &   21.9{\scriptsize $\pm$2.9} &  13.60{\scriptsize $\pm$0.10} &    
         9.2{\scriptsize $\pm$2.8} &  12.02{\scriptsize $\pm$0.10} &  26240{\scriptsize $\pm$8935} &    
         7{\scriptsize $\pm$4} &   uncertain \\
    Q0329--385 &   2.249389 &   18.0{\scriptsize $\pm$2.7} &  12.90{\scriptsize $\pm$0.10} &   
         14.1{\scriptsize $\pm$2.3} &  12.40{\scriptsize $\pm$0.06} &  8344{\scriptsize $\pm$7476} &   
         14{\scriptsize $\pm$3} &     certain \\
    Q0329--385 &   2.248824 &   52.2{\scriptsize $\pm$5.3} &  13.40{\scriptsize $\pm$0.05} &   
         18.8{\scriptsize $\pm$6.0} &  12.29{\scriptsize $\pm$0.12} &  157100{\scriptsize $\pm$42970} &   
         12{\scriptsize $\pm$11} &     certain \\
    Q0329--385 &   2.248307 &    8.5{\scriptsize $\pm$4.1} &  12.21{\scriptsize $\pm$0.31} &    
          8.5{\scriptsize $\pm$2.3} &  12.29{\scriptsize $\pm$0.12} &  $\le 385$ &    
          8{\scriptsize $\pm$3} &     certain \\
  HE1122--1648 &   2.339275 &   26.7{\scriptsize $\pm$0.1} &  14.26{\scriptsize $\pm$0.00} &    
          8.1{\scriptsize $\pm$1.5} &  11.81{\scriptsize $\pm$0.06} &  42870{\scriptsize $\pm$1645} &    
          3{\scriptsize $\pm$5} &     certain \\
  HE1122--1648 &   2.206364 &   23.4{\scriptsize $\pm$0.9} &  14.86{\scriptsize $\pm$0.04} &   
         11.1{\scriptsize $\pm$0.9} &  12.13{\scriptsize $\pm$0.03} &  27980{\scriptsize $\pm$3005} &    
         9{\scriptsize $\pm$1} &     certain \\
  HE0001--2340 &   2.163416 &   24.2{\scriptsize $\pm$0.5} &  14.64{\scriptsize $\pm$0.02} &    
         8.2{\scriptsize $\pm$6.0} &  11.89{\scriptsize $\pm$0.22} &  34320{\scriptsize $\pm$6670} &    
         4{\scriptsize $\pm$12} &     certain \\
  PKS1448--292 &   2.167959 &   24.2{\scriptsize $\pm$1.6} &  13.26{\scriptsize $\pm$0.05} &    
         7.0{\scriptsize $\pm$3.0} &  11.86{\scriptsize $\pm$0.14} &  35500{\scriptsize $\pm$6438} &    
         0{\scriptsize $\pm$1} &     certain \\

\noalign{\smallskip}
\hline
\end{tabular}

\end{table*}

\vfill


\newpage
\begin{landscape}
\begin{table}
\vspace{0.5cm}
\caption{Cloudy PIE prediction for the tied \ion{H}{i}+\ion{C}{iv}+\ion{C}{iii} components for
the HM Q+G 2012 UVB}
\label{tab_a2}
\scriptsize{
\begin{tabular}{rp{1.6cm}p{0.7cm}p{0.95cm}p{0.95cm}p{0.95cm}p{1.05cm}c || ccrccrrl}
\hline 

 & & & \multicolumn{5}{c}{Observed} & \multicolumn{7}{c}{Predicted} & \\

\noalign{\smallskip}

\cline{4-7} \cline{9-15} \\[-0.2cm]


\# &  QSO & $z_{\mathrm{abs}}$ & $\log N_{\mathrm{\ion{H}{i}}}$ & $\log N_{\mathrm{\ion{C}{iv}}}$ &
   $\log N_{\mathrm{\ion{C}{iii}}}$ & $\log N_{\mathrm{\ion{C}{ii}}}$ & $\log T_{b}$ &
   $\log U$ & $\log n_{\mathrm{\ion{H}}}$ & [C/H] & 
   $\log N_{\mathrm{\ion{H}}}$ & $\log T_{\mathrm{PIE}}$ & $\log L$ & $\log P/k$ & Notes$^{\mathrm{a}}$ \\
   
\noalign{\smallskip}
   
& &   &  [cm$^{-2}$]  &  [cm$^{-2}$]   &  [cm$^{-2}$]   &  [cm$^{-2}$]  & [K] &  &
 [cm$^{-3}$] &   &  [cm$^{-2}$] & [K] &  [kpc] &  [K cm$^{-3}$] &  \\
 
\hline

  1 &      Q0055--269 &   3.257359 &  13.93{\tiny $\pm$0.02} &  12.76{\tiny $\pm$0.03} &  12.54{\tiny $\pm$0.06} &
$\le  12.30$ &   4.517{\tiny $\pm$0.040} &  -1.10{\tiny $\pm$0.05} &   -5.01 &
 -1.17{\tiny $\pm$0.05} &   18.27 &   4.341 &   1.788 &   -0.306 &       Low \\
  2 &      Q0055--269 &   3.038795 &  14.60{\tiny $\pm$0.04} &  12.72{\tiny $\pm$0.04} &  12.59{\tiny $\pm$0.15} &
$\le  12.70$ &   4.284{\tiny $\pm$0.099} &  -1.20{\tiny $\pm$0.10} &   -4.87 &
 -1.77{\tiny $\pm$0.08} &   18.90 &   4.427 &   2.283 &   -0.080 &       Low \\
  3 &      Q0055--269 &   2.744100 &  15.12{\tiny $\pm$0.13} &  12.79{\tiny $\pm$0.05} &  12.98{\tiny $\pm$0.06} &
bl$^{\mathrm{b}}$ &   4.475{\tiny $\pm$0.249} &  -1.63{\tiny $\pm$0.08} &   -4.39 &
 -1.95{\tiny $\pm$0.05} &   19.04 &   4.481 &   1.946 &    0.450 &       Low \\
  4 &      Q0055--269 &   2.743720 &  14.78{\tiny $\pm$0.39} &  12.48{\tiny $\pm$0.09} &  12.27{\tiny $\pm$0.19} &
bl$^{\mathrm{b}}$ &   4.525{\tiny $\pm$0.422} &  -1.21{\tiny $\pm$0.18} &   -4.81 &
 -2.10{\tiny $\pm$0.10} &   19.11 &   4.482 &   2.435 &    0.031 &       Low \\
  5 &    PKS2126--158 &   2.973015 &  14.36{\tiny $\pm$0.01} &  12.18{\tiny $\pm$0.03} &  12.25{\tiny $\pm$0.03} &
$\le  11.45$ &   4.432{\tiny $\pm$0.022} &  -1.40{\tiny $\pm$0.05} &   -4.66 &
 -1.95{\tiny $\pm$0.05} &   18.50 &   4.474 &   1.672 &    0.178 &       Low \\
  6 &      Q0420--388 &   2.849598 &  14.09{\tiny $\pm$0.02} &  12.59{\tiny $\pm$0.04} &  12.17{\tiny $\pm$0.06} &
  bl$^{\mathrm{c}}$ &   4.541{\tiny $\pm$0.040} &  -1.07{\tiny $\pm$0.08} &   -4.97 &
 -0.95{\tiny $\pm$0.05} &   18.08 &   4.393 &   1.558 &   -0.213 &       Low \\
  7 &      Q0420--388 &   2.849229 &  13.68{\tiny $\pm$0.06} &  12.10{\tiny $\pm$0.10} &  11.89{\tiny $\pm$0.14} &
 bl$^{\mathrm{c}}$ &   4.784{\tiny $\pm$0.057} &  -1.21{\tiny $\pm$0.15} &   -4.83 &
 -1.42{\tiny $\pm$0.10} &   17.98 &   4.450 &   1.322 &   -0.016 &        \\
  8 &    HE0940--1050 &   2.937755 &  14.58{\tiny $\pm$0.02} &  12.65{\tiny $\pm$0.02} &  12.51{\tiny $\pm$0.01} & 
$\le  11.95$ &   4.553{\tiny $\pm$0.012} &  -1.25{\tiny $\pm$0.05} &   -4.80 &
 -1.80{\tiny $\pm$0.05} &   18.86 &   4.458 &   2.169 &    0.016 &       Low \\
  9 &    HE0940--1050 &   2.883509 &  14.60{\tiny $\pm$0.01} &  12.35{\tiny $\pm$0.03} &  12.37{\tiny $\pm$0.02} &
$\le  12.25$ &   4.456{\tiny $\pm$0.031} &  -1.39{\tiny $\pm$0.05} &   -4.65 &
 -2.03{\tiny $\pm$0.05} &   18.76 &   4.483 &   1.923 &    0.192 &       Low \\
 10 &    HE0940--1050 &   2.826555 &  14.51{\tiny $\pm$0.21} &  13.19{\tiny $\pm$0.02} &  13.23{\tiny $\pm$0.03} &
$\le  12.00$ &   4.226{\tiny $\pm$0.141} &  -1.49{\tiny $\pm$0.05} &   -4.54 &
 -1.02{\tiny $\pm$0.05} &   18.52 &   4.419 &   1.570 &    0.238 &      High, \ion{Si}{iv}, comp \\
 11 &    HE2347--4342 &   2.347467 &  15.99{\tiny $\pm$0.20} &  13.49{\tiny $\pm$0.00} &  13.26{\tiny $\pm$0.22} &
$\le  11.75$ &   4.484{\tiny $\pm$0.057} &  -1.31{\tiny $\pm$0.15} &   -4.67 &
 -2.25{\tiny $\pm$0.05} &   20.27 &   4.521 &   3.458 &    0.215 &        bl \\
 12 &    HE0151--4326 &   2.519825 &  15.24{\tiny $\pm$0.02} &  12.29{\tiny $\pm$0.04} &  12.35{\tiny $\pm$0.06} &
$\le  11.70$ &   4.584{\tiny $\pm$0.027} &  -1.55{\tiny $\pm$0.08} &   -4.45 &
 -2.60{\tiny $\pm$0.05} &   19.26 &   4.507 &   2.219 &    0.423 &        Low \\
 13 &    HE0151--4326 &   2.449902 &  14.45{\tiny $\pm$0.02} &  12.99{\tiny $\pm$0.01} &  12.96{\tiny $\pm$0.04}  &
$\le  12.20$ &   4.012{\tiny $\pm$0.025} &  -1.57{\tiny $\pm$0.05} &   -4.42 &
 -1.10{\tiny $\pm$0.05} &   18.39 &   4.442 &   1.318 &    0.384 &      High, \ion{Si}{iv}, comp \\
 14 &    HE0151--4326 &   2.419676 &  12.85{\tiny $\pm$0.04} &  12.75{\tiny $\pm$0.01} &  12.76{\tiny $\pm$0.02} &
bl$^{\mathrm{d}}$  &   4.313{\tiny $\pm$0.059} &  -1.58{\tiny $\pm$0.05} &   -4.41 &
  0.55{\tiny $\pm$0.05} &   16.29 &   3.785 &  -0.790 &   -0.260 &      High, simp \\
 15 &    HE0151--4326 &   2.415718 &  13.36{\tiny $\pm$0.01} &  13.03{\tiny $\pm$0.01} &  13.07{\tiny $\pm$0.02} &
bl$^{\mathrm{d}}$  &   4.339{\tiny $\pm$0.019} &  -1.73{\tiny $\pm$0.05} &   -4.26 &
  0.35{\tiny $\pm$0.05} &   16.77 &   3.955 &  -0.459 &    0.057 &      High, simp \\
 16 &    HE0151--4326 &   2.401315 &  15.04{\tiny $\pm$0.02} &  12.34{\tiny $\pm$0.11} &  12.37{\tiny $\pm$0.06} &
bl$^{\mathrm{e}}$  &   4.831{\tiny $\pm$0.020} &  -1.58{\tiny $\pm$0.15} &   -4.41 &
 -2.35{\tiny $\pm$0.05} &   19.03 &   4.507 &   1.947 &    0.463 &       Low \\
 17 &      Q0002--422 &   2.539455 &  14.24{\tiny $\pm$0.03} &  12.55{\tiny $\pm$0.02} &  12.18{\tiny $\pm$0.06} &
$\le  11.95$ &   4.903{\tiny $\pm$0.027} &  -1.15{\tiny $\pm$0.05} &   -4.85 &
 -1.45{\tiny $\pm$0.05} &   18.62 &   4.480 &   1.983 &   -0.006 &       Low \\
 18 &      Q0002--422 &   2.463222 &  14.56{\tiny $\pm$0.06} &  13.26{\tiny $\pm$0.01} &  13.38{\tiny $\pm$0.02} &
$\le  11.75$ &   4.472{\tiny $\pm$0.128} &  -1.79{\tiny $\pm$0.05} &   -4.20 &
 -0.77{\tiny $\pm$0.05} &   18.24 &   4.364 &   0.949 &    0.524 &      High, \ion{Si}{iv}, comp \\
 19 &      Q0002--422 &   2.462358 &  15.16{\tiny $\pm$0.02} &  13.20{\tiny $\pm$0.01} &  13.83{\tiny $\pm$0.03} &
12.60{\tiny $\pm$0.03} &   4.115{\tiny $\pm$0.040} &  -2.46{\tiny $\pm$0.05} &   -3.53 &
 -0.45{\tiny $\pm$0.05} &   18.04 &   4.193 &   0.084 &    1.016 &      High, \ion{Si}{iv}, comp, error \\
 20 &      Q0002--422 &   2.462044 &  14.83{\tiny $\pm$0.02} &  13.34{\tiny $\pm$0.02} &  13.59{\tiny $\pm$0.02} &
12.19{\tiny $\pm$0.08} &   4.170{\tiny $\pm$0.027} &  -2.25{\tiny $\pm$0.10} &   -3.74 &
 -0.45{\tiny $\pm$0.10} &   17.92 &   4.228 &   0.172 &    0.844 &      High, \ion{Si}{iv}, comp, no \\
 21 &    PKS0329--255 &   2.586757 &  14.99{\tiny $\pm$0.03} &  12.37{\tiny $\pm$0.06} &  12.02{\tiny $\pm$0.11} &
$\le  11.90$ &   4.477{\tiny $\pm$0.033} &  -1.15{\tiny $\pm$0.10} &   -4.85 &
 -2.40{\tiny $\pm$0.05} &   19.39 &   4.493 &   2.753 &    0.003 &       Low \\
 22 &    PKS0329--255 &   2.456581 &  13.51{\tiny $\pm$0.24} &  11.97{\tiny $\pm$0.25} &  12.00{\tiny $\pm$0.29} &
$\le  12.30$ &   4.486{\tiny $\pm$0.133} &  -1.59{\tiny $\pm$0.30} &   -4.40 &
 -1.15{\tiny $\pm$0.25} &   17.43 &   4.447 &   0.342 &    0.409 &         \\
23 &      Q0453--423 &   2.444109 &  14.62{\tiny $\pm$0.01} &  12.63{\tiny $\pm$0.06} &  13.41{\tiny $\pm$0.08} 
       &   12.66{\tiny $\pm0.11^{\mathrm{f}}$ }&   3.969{\tiny $\pm$0.026} &  -2.65{\tiny $\pm$0.10} &   -3.34 &
       0.10{\tiny $\pm$0.10} &   17.13 &   3.932 &  -1.022 &    0.944 &      High, \ion{Si}{iv}, comp, no\\
24 &      Q0453--423 &   2.442644 &  14.91{\tiny $\pm$0.01} &  13.04{\tiny $\pm$0.03} &  13.66{\tiny $\pm$0.07} 
     &  12.18{\tiny $\pm$0.07} &   4.130{\tiny $\pm$0.026} &  -2.32{\tiny $\pm$0.05} &   -3.67 &
     -0.55{\tiny $\pm$0.10} &   17.95 &   4.243 &   0.132 &    0.930 &      High, \ion{Si}{iv}, comp, error \\
25 &      Q0453--423 &   2.441813 &  14.67{\tiny $\pm$0.01} &  11.75{\tiny $\pm$0.13} &  11.83{\tiny $\pm$0.14} 
      &   $\le  11.65$ &   4.442{\tiny $\pm$0.054} &  -1.58{\tiny $\pm$0.20} &   -4.41 &
     -2.55{\tiny $\pm$0.20} &   18.68 &   4.510 &   1.600 &    0.462 &       Low \\
 26 &      Q0453--423 &   2.398159 &  13.39{\tiny $\pm$0.07} &  12.74{\tiny $\pm$0.04} &  11.69{\tiny $\pm$0.25} &
$\le  12.05$ &   4.167{\tiny $\pm$0.141} &  -0.62{\tiny $\pm$0.10} &   -5.37 &
 -0.25{\tiny $\pm$0.15} &   18.21 &   4.370 &   2.088 &   -0.634 &         \\
 27 &      Q0453--423 &   2.397801 &  14.41{\tiny $\pm$0.01} &  13.78{\tiny $\pm$0.01} &  13.16{\tiny $\pm$0.02} &
$\le  11.70$ &   4.200{\tiny $\pm$0.016} &  -1.08{\tiny $\pm$0.05} &   -4.91 &
 -0.35{\tiny $\pm$0.05} &   18.77 &   4.367 &   2.188 &   -0.177 &         \\
 28 &      Q0453--423 &   2.397447 &  13.86{\tiny $\pm$0.05} &  12.71{\tiny $\pm$0.03} &  12.25{\tiny $\pm$0.09} &
$\le  11.80$ &   5.106{\tiny $\pm$0.124} &  -1.13{\tiny $\pm$0.08} &   -4.86 &
 -0.85{\tiny $\pm$0.05} &   18.23 &   4.441 &   1.596 &   -0.053 &       Low \\
 29 &      Q0453--423 &   2.396755 &  14.16{\tiny $\pm$0.03} &  13.34{\tiny $\pm$0.01} &  13.06{\tiny $\pm$0.04} &
$\le  11.60$ &   4.315{\tiny $\pm$0.031} &  -1.42{\tiny $\pm$0.05} &   -4.57 &
 -0.50{\tiny $\pm$0.05} &   18.17 &   4.359 &   1.254 &    0.154 &      High, \ion{Si}{iv}, comp\\
 30 &      Q0453--423 &   2.277569 &  13.54{\tiny $\pm$0.03} &  12.99{\tiny $\pm$0.01} &  13.05{\tiny $\pm$0.05} &
bl$^{\mathrm{g}}$ &   4.244{\tiny $\pm$0.059} &  -1.86{\tiny $\pm$0.05} &   -4.12 &
  0.20{\tiny $\pm$0.05} &   16.88 &   4.038 &  -0.488 &    0.277 &      High, \ion{Si}{iv}, simp/comp \\
 31 &    HE1347--2457 &   2.370003 &  14.10{\tiny $\pm$0.09} &  12.60{\tiny $\pm$0.03} &  12.33{\tiny $\pm$0.04} &
$\le  11.55$ &   4.340{\tiny $\pm$0.153} &  -1.30{\tiny $\pm$0.05} &   -4.68 &
 -1.20{\tiny $\pm$0.05} &   18.33 &   4.480 &   1.528 &    0.157 &       Low \\
 32 &     Q0329--385 &   2.249389 &  12.90{\tiny $\pm$0.10} &  12.40{\tiny $\pm$0.06} &  12.30{\tiny $\pm$0.34} &
$\le  12.55$ &   3.921{\tiny $\pm$0.389} &  -1.74{\tiny $\pm$0.07} &   -4.24 &
  0.08{\tiny $\pm$0.30} &   16.43 &   4.136 &  -0.820 &    0.257 &      High, simp/comp \\

\noalign{\smallskip}
\hline
\end{tabular}

\begin{list}{}{}
\item[$^{\mathrm{a}}$]
{\it High:} high-metallicity branch absorbers. {\it Low:} low-metallicity
branch absorbers. {\it blank:} not classified. \ion{Si}{iv}: associated 
\ion{Si}{iv} is detected. {\it bl:} \ion{Si}{iv} region
is blended. {\it comp:} part of a multi-component \ion{C}{iv} complex. 
{\it simp:} high-metallicity branch absorbers with a simple, isolated \ion{C}{iv}.
{\it simp/comp:} simple high-metallicity branch absorbers 
close to a separate \ion{C}{iv} complex.
{\it error:} the predicted column densities are within the observed errors.
{\it no:} no single $U$ matches all the observed column densities
within errors. 
\item[$^{\mathrm{b}}$]
A weak \ion{H}{i} Ly$\alpha$ absorption at $z = 3.109959$ 
exists at $-16$\,\kms\/, while
\ion{C}{ii} $\lambda$ 1334 is expected at $-30$\,\kms\/ and $0$\,\kms\/.
At $-30$\,\kms\/ and $0$\,\kms\/, a negligible absorption is due to the wing
of the aforementioned \ion{H}{i}. Therefore, \ion{C}{ii} $\lambda$ 1334, if
exists, would be very weak.
\item[$^{\mathrm{c}}$]
At the expected \ion{C}{ii} $\lambda$\,1334 position, an absorption feature by
\ion{Si}{iv} $\lambda$ 1402 at $z = 2.662123$ and at $z = 2.662294$ exists,
while \ion{C}{ii} $\lambda$ 1036 is blended. Since the \ion{Si}{iv}
$\lambda\lambda$\,1393, 1402 doublet falls on the absorption-free region
without any hint of additional absorption features, \ion{C}{ii} $\lambda$ 1334,
if exists, would be very weak.
\item[$^{\mathrm{d}}$] Despite the normalised flux being less than 1, 
the expected \ion{C}{ii} $\lambda$\.1334 spectral region
is flat, which does not suggesting any additional absorption feature.
\item[$^{\mathrm{e}}$] 
Both \ion{C}{ii} $\lambda\lambda$\,1334, 1036 regions are severely blended.
\item[$^{\mathrm{f}}$] 
The column density of \ion{C}{ii} is from \ion{C}{ii} $\lambda$\,1036 in
an absorption-free region.
\item[$^{\mathrm{g}}$] 
At the expected position of \ion{C}{ii} $\lambda$\,1334, a narrow absorption
feature exists, but it over-produces a \ion{C}{ii} $\lambda$\,1036 absorption which
is located in a region deblended reliably. In addition, 
this narrow feature occurs at $+3$\,\kms\/ away from its supposed position, 
it is likely to be a unidentified metal line.
\end{list}
}
\end{table}
\end{landscape}

\vfill


\newpage

\begin{landscape}
\begin{table}
\vspace{0.5cm}
\caption{Non-PIE Cloudy-prediction for the tied \ion{H}{i}+\ion{C}{iv}+\ion{C}{iii} components for
the HM Q+G 2012 UVB at a fixed temperature}
\label{tab_a3}
\scriptsize{
\begin{tabular}{rp{1.6cm}p{0.7cm}p{0.95cm}p{0.95cm}p{0.95cm}p{0.95cm}c || ccrcrrl}
\hline 

 & & & \multicolumn{4}{c}{Observed} & \multicolumn{7}{c}{Predicted} & \\

\noalign{\smallskip}

\cline{4-8}\\[-0.2cm]  


\# & QSO & $z_{\mathrm{abs}}$ & $\log N_{\mathrm{\ion{H}{i}}}$ & $\log N_{\mathrm{\ion{C}{iv}}}$ &
   $\log N_{\mathrm{\ion{C}{iii}}}$ & $\log N_{\mathrm{\ion{C}{ii}}}^{\mathrm{a}}$ & $\log T_{b}$ &
   $\log U$ & $\log n_{\mathrm{\ion{H}}}$ & [C/H] & 
   $\log N_{\mathrm{\ion{H}}}$ & $\log L$ & $\log P/k$ & Notes$^{\mathrm{b}}$ \\
   
\noalign{\smallskip}
   
&  &   &  [cm$^{-2}$]  &  [cm$^{-2}$]   &  [cm$^{-2}$]   &  [cm$^{-2}$]  &  [K] &  &
 [cm$^{-3}$] &   &  [cm$^{-2}$] &  [kpc] &  [K cm$^{-3}$] &  \\
 
\hline

  1  &      Q0055--269 &   3.257359 &  13.93{\tiny $\pm$0.02} &  12.76{\tiny $\pm$0.03} &  12.54{\tiny $\pm$0.06} &
$\le  12.30$ &   4.517{\tiny $\pm$0.040} &  -0.92{\tiny $\pm$0.05} &   -5.19 &
 -1.20{\tiny $\pm$0.05} &  18.59 &    2.284 &   -0.309 &                                Low \\
  2  &      Q0055--269 &   3.038795 &  14.60{\tiny $\pm$0.04} &  12.72{\tiny $\pm$0.04} &  12.59{\tiny $\pm$0.15} &
$\le  12.70$ &   4.284{\tiny $\pm$0.099} &  -1.36{\tiny $\pm$0.05} &   -4.71 &
 -1.70{\tiny $\pm$0.10} &  18.64 &    1.854 &   -0.065 &                                Low \\
  3  &      Q0055--269 &   2.744100 &  15.12{\tiny $\pm$0.13} &  12.79{\tiny $\pm$0.05} &  12.98{\tiny $\pm$0.06} &
bl  &   4.475{\tiny $\pm$0.249} &  -1.62{\tiny $\pm$0.05} &   -4.40 &
 -1.95{\tiny $\pm$0.05} &  19.05 &    1.960 &    0.434 &                                Low \\
  4  &      Q0055--269 &   2.743720 &  14.78{\tiny $\pm$0.39} &  12.48{\tiny $\pm$0.09} &  12.27{\tiny $\pm$0.19} &
bl  &   4.525{\tiny $\pm$0.422} &  -1.18{\tiny $\pm$0.20} &   -4.84 &
 -2.10{\tiny $\pm$0.10} &  19.18 &    2.531 &    0.045 &                                Low \\
  5  &    PKS2126--158 &   2.973015 &  14.36{\tiny $\pm$0.01} &  12.18{\tiny $\pm$0.03} &  12.25{\tiny $\pm$0.03} &
$\le  11.45$ &   4.432{\tiny $\pm$0.022} &  -1.44{\tiny $\pm$0.05} &   -4.62 &
 -1.95{\tiny $\pm$0.05} &  18.43 &    1.557 &    0.174 &                                Low \\
  6  &      Q0420--388 &   2.849598 &  14.09{\tiny $\pm$0.02} &  12.59{\tiny $\pm$0.04} &  12.17{\tiny $\pm$0.06} &
bl  &   4.541{\tiny $\pm$0.040} &  -0.93{\tiny $\pm$0.05} &   -5.11 &
 -1.35{\tiny $\pm$0.05} &  18.75 &    2.367 &   -0.205 &                                Low \\
  7  &      Q0420--388 &   2.849229 &  13.68{\tiny $\pm$0.06} &  12.10{\tiny $\pm$0.10} &  11.89{\tiny $\pm$0.14} &
bl  &   4.784{\tiny $\pm$0.057} &  -1.03{\tiny $\pm$0.10} &   -5.01 &
 -1.45{\tiny $\pm$0.15} &  18.50 &    2.012 &    0.140 &                                 \\
  8  &    HE0940--1050 &   2.937755 &  14.58{\tiny $\pm$0.02} &  12.65{\tiny $\pm$0.02} &  12.51{\tiny $\pm$0.01} &
$\le  11.95$ &   4.553{\tiny $\pm$0.012} &  -1.14{\tiny $\pm$0.05} &   -4.91 &
 -1.80{\tiny $\pm$0.05} &  19.05 &    2.470 &    0.003 &                                Low \\
  9  &    HE0940--1050 &   2.883509 &  14.60{\tiny $\pm$0.01} &  12.35{\tiny $\pm$0.03} &  12.37{\tiny $\pm$0.02} &
$\le  12.25$ &   4.456{\tiny $\pm$0.031} &  -1.42{\tiny $\pm$0.05} &   -4.62 &
 -2.00{\tiny $\pm$0.05} &  18.70 &    1.840 &    0.193 &                                Low \\
 10  &    HE0940--1050 &   2.826555 &  14.51{\tiny $\pm$0.21} &  13.19{\tiny $\pm$0.02} &  13.23{\tiny $\pm$0.03} &
$\le  12.00$ &   4.226{\tiny $\pm$0.141} &  -1.69{\tiny $\pm$0.05} &   -4.34 &
 -0.85{\tiny $\pm$0.05} &  18.17 &    1.024 &    0.241 &  High, \ion{Si}{iv}, comp \\
 11  &    HE2347--4342 &   2.347467 &  15.99{\tiny $\pm$0.20} &  13.49{\tiny $\pm$0.00} &  13.26{\tiny $\pm$0.22} &
$\le  11.75$ &   4.484{\tiny $\pm$0.057} &  -1.35{\tiny $\pm$0.15} &   -4.63 &
 -2.25{\tiny $\pm$0.05} &  20.20 &    3.345 &    0.212 &  bl                              \\
 12  &    HE0151--4326 &   2.519825 &  15.24{\tiny $\pm$0.02} &  12.29{\tiny $\pm$0.04} &  12.35{\tiny $\pm$0.06} &
$\le  11.70$ &   4.584{\tiny $\pm$0.027} &  -1.51{\tiny $\pm$0.10} &   -4.49 &
 -2.65{\tiny $\pm$0.05} &  19.39 &    2.380 &    0.459 &  Low                             \\
 13  &    HE0151--4326 &   2.449902 &  14.45{\tiny $\pm$0.02} &  12.99{\tiny $\pm$0.01} &  12.96{\tiny $\pm$0.04} &
$\le  12.20$ &   4.012{\tiny $\pm$0.025} &  -1.73{\tiny $\pm$0.05} &   -4.26 &
 -0.85{\tiny $\pm$0.05} &  17.91 &    0.675 &    0.112 & High, \ion{Si}{iv}, comp \\
 14  &    HE0151--4326 &   2.419676 &  12.85{\tiny $\pm$0.04} &  12.75{\tiny $\pm$0.01} &  12.76{\tiny $\pm$0.02} &
bl  &   4.313{\tiny $\pm$0.059} &  -1.75{\tiny $\pm$0.05} &   -4.24 &
  0.40{\tiny $\pm$0.05} &  16.51 &   -0.748 &    0.438 &  High, simp \\
 15  &    HE0151--4326 &   2.415718 &  13.36{\tiny $\pm$0.01} &  13.03{\tiny $\pm$0.01} &  13.07{\tiny $\pm$0.02} &
bl  &   4.339{\tiny $\pm$0.019} &  -1.74{\tiny $\pm$0.05} &   -4.25 &
  0.20{\tiny $\pm$0.05} &  17.05 &   -0.197 &    0.454 &  High, simp \\
 16  &    HE0151--4326 &   2.401315  &  15.04{\tiny $\pm$0.02} &   12.34{\tiny $\pm$0.11} &  12.37{\tiny $\pm$0.06} &
  bl   &   4.831{\tiny $\pm$0.020}  &  -1.49{\tiny $\pm$0.05}  &  -4.50  & 
  -2.60{\tiny $\pm$0.10}  & 19.60  &   2.602  &  0.696  &                        Low \\
 17  &      Q0002--422 &   2.539455 &  14.24{\tiny $\pm$0.03} &  12.55{\tiny $\pm$0.02} &  12.18{\tiny $\pm$0.06} &
$\le  11.95$ &   4.903{\tiny $\pm$0.027} &  -1.08{\tiny $\pm$0.05} &   -4.92 &
 -1.55{\tiny $\pm$0.05} &  19.18 &    2.612 &    0.347 &                                Low \\
 18  &      Q0002--422 &   2.463222 &  14.56{\tiny $\pm$0.06} &  13.26{\tiny $\pm$0.01} &  13.38{\tiny $\pm$0.02} &
$\le  11.75$ &   4.472{\tiny $\pm$0.128} &  -1.65{\tiny $\pm$0.05} &   -4.34 &
 -0.85{\tiny $\pm$0.05} &  18.45 &    1.300 &    0.493 & High, \ion{Si}{iv}, comp \\
 19  &      Q0002--422 &   2.462358 &  15.16{\tiny $\pm$0.02} &  13.20{\tiny $\pm$0.01} &  13.83{\tiny $\pm$0.03} &
12.60{\tiny $\pm$0.03} &   4.115{\tiny $\pm$0.040} &  -2.46{\tiny $\pm$0.05} &   -3.53 &
 -0.40{\tiny $\pm$0.05} &  17.97 &    0.017 &    0.937 & High, \ion{Si}{iv}, comp, error \\
 20  &      Q0002--422 &   2.462044 &  14.83{\tiny $\pm$0.02} &  13.34{\tiny $\pm$0.02} &  13.59{\tiny $\pm$0.02} &
12.19{\tiny $\pm$0.08} &   4.170{\tiny $\pm$0.027} &  -2.23{\tiny $\pm$0.10} &   -3.76 &
 -0.45{\tiny $\pm$0.05} &  17.91 &    0.180 &    0.766 & High, \ion{Si}{iv}, comp, no \\
 21  &    PKS0329--255 &   2.586757 &  14.99{\tiny $\pm$0.03} &  12.37{\tiny $\pm$0.06} &  12.02{\tiny $\pm$0.11} &
$\le  11.90$ &   4.477{\tiny $\pm$0.033} &  -1.18{\tiny $\pm$0.10} &   -4.82 &
 -2.40{\tiny $\pm$0.05} &  19.34 &    2.679 &    0.015 &                                Low \\
 22  &    PKS0329--255 &   2.456581 &  13.51{\tiny $\pm$0.24} &  11.97{\tiny $\pm$0.25} &  12.00{\tiny $\pm$0.29} &
$\le  12.30$ &   4.486{\tiny $\pm$0.133} &  -1.54{\tiny $\pm$0.30} &   -4.45 &
 -1.20{\tiny $\pm$0.30} &  17.52 &    0.476 &    0.398 &                                 \\
 23  &      Q0453--423 &   2.444109 &  14.62{\tiny $\pm$0.01} &  12.63{\tiny $\pm$0.06} &  13.41{\tiny $\pm$0.08} &
12.66{\tiny $\pm$0.11} &   3.969{\tiny $\pm$0.026} &  -2.73{\tiny $\pm$0.10} &   -3.26 &
  0.20{\tiny $\pm$0.10} &  17.07 &   -1.162 &    1.060 & High, \ion{Si}{iv}, comp, no \\
 24  &      Q0453--423 &   2.442644 &  14.91{\tiny $\pm$0.01} &  13.04{\tiny $\pm$0.03} &  13.66{\tiny $\pm$0.07} &
  12.18{\tiny $\pm$0.07} &   4.130{\tiny $\pm$0.026} &  -2.27{\tiny $\pm$0.10} &   -3.72 &
 -0.50{\tiny $\pm$0.10} &  17.92 &    0.149 &    0.768 &  High, \ion{Si}{iv}, comp, error \\
 25  &      Q0453--423 &   2.441813 &  14.67{\tiny $\pm$0.01} &  11.75{\tiny $\pm$0.13} &  11.83{\tiny $\pm$0.14} &
$\le  11.65$ &   4.442{\tiny $\pm$0.054} &  -1.66{\tiny $\pm$0.20} &   -4.33 &
 -2.50{\tiny $\pm$0.15} &  18.52 &    1.359 &    0.475 &                                Low \\
 26  &      Q0453--423 &   2.398159 &  13.39{\tiny $\pm$0.07} &  12.74{\tiny $\pm$0.04} &  11.69{\tiny $\pm$0.25} &
$\le  12.05$ &   4.167{\tiny $\pm$0.141} &  -0.77{\tiny $\pm$0.15} &   -5.22 &
 -0.35{\tiny $\pm$0.05} &  17.91 &    1.637 &   -0.687 &                                  \\
 27  &      Q0453--423 &   2.397801 &  14.41{\tiny $\pm$0.01} &  13.78{\tiny $\pm$0.01} &  13.16{\tiny $\pm$0.02} &
$\le  11.70$ &   4.200{\tiny $\pm$0.016} &  -1.21{\tiny $\pm$0.05} &   -4.78 &
 -0.35{\tiny $\pm$0.05} &  18.52 &    1.804 &   -0.214 &                                 \\
 28  &      Q0453--423 &   2.397447 &  13.86{\tiny $\pm$0.05} &  12.71{\tiny $\pm$0.03} &  12.25{\tiny $\pm$0.09} &
$\le  11.80$ &   5.106{\tiny $\pm$0.124} &  -1.67{\tiny $\pm$0.05} &   -4.32 &
 -1.50{\tiny $\pm$0.20} &  18.96 &    1.784 &    1.152 &                                Low \\
 29  &      Q0453--423 &   2.396755 &  14.16{\tiny $\pm$0.03} &  13.34{\tiny $\pm$0.01} &  13.06{\tiny $\pm$0.04} &
$\le  11.60$ &   4.315{\tiny $\pm$0.031} &  -1.46{\tiny $\pm$0.05} &   -4.53 &
 -0.47{\tiny $\pm$0.05} &  18.11 &    1.140 &    0.149 &  High, \ion{Si}{iv}, comp \\
 30  &      Q0453--423 &   2.277569 &  13.54{\tiny $\pm$0.03} &  12.99{\tiny $\pm$0.01} &  13.05{\tiny $\pm$0.05} &
 bl  &   4.244{\tiny $\pm$0.059} &  -1.87{\tiny $\pm$0.05} &   -4.11 &
  0.10{\tiny $\pm$0.05} &  17.02 &   -0.358 &    0.494 & High, \ion{Si}{iv}, simp/comp \\
 31  &    HE1347--2457 &   2.370003 &  14.10{\tiny $\pm$0.09} &  12.60{\tiny $\pm$0.03} &  12.33{\tiny $\pm$0.04} &
$\le  11.55$ &   4.340{\tiny $\pm$0.153} &  -1.45{\tiny $\pm$0.05} &   -4.53 &
 -1.15{\tiny $\pm$0.05} &  18.07 &    1.116 &    0.167 &                                Low \\
 32  &    Q0329--385 &   2.249389 &  12.90{\tiny $\pm$0.10} &  12.40{\tiny $\pm$0.06} &  12.30{\tiny $\pm$0.34} &
$\le  12.55$ &   3.921{\tiny $\pm$0.389} &  -1.62{\tiny $\pm$0.10} &   -4.36 &
  0.10{\tiny $\pm$0.30} &  16.39 &   -0.734 &   -0.079 &  High, simp/comp \\

\noalign{\smallskip}
\hline
\end{tabular}

\begin{list}{}{}
\item[$^{\mathrm{a}}$]
See Table~\ref{tab_a2} on $N_{\mathrm{\ion{C}{ii}}}$.
\item[$^{\mathrm{b}}$]
See Table~\ref{tab_a2} for notes.
\end{list}
}
\end{table}
\end{landscape}

\end{document}